\documentclass[%
superscriptaddress,
preprint,
amsmath,amssymb,
aps,
]{revtex4-1}
\usepackage{graphicx}
\usepackage{epstopdf, epsfig}

\usepackage{floatrow}
\usepackage{subfig}

\setcitestyle{authoryear,round}

\begin{document}
	\title[Draft]{Effects of the quiescent core in turbulent channel flow on transport and clustering of inertial particles }
	
	\author{Yucheng Jie}
	\affiliation{AML, Department of Engineering Mechanics,Tsinghua University,100084 Beijing,China}
	\author{Helge I. Andersson}%
	\affiliation{Department of Energy and Process Engineering, Norwegian University of Science and Technology,7491 Trondheim, Norway}
	\author{Lihao Zhao}
	\altaffiliation[Corresponding author:]{zhaolihao@tsinghua.edu.cn}
	\affiliation{AML, Department of Engineering Mechanics,Tsinghua University,100084 Beijing,China}

\begin{abstract}
	The existence of a quiescent core (QC) in the center of turbulent channel flows was  demonstrated in recent experimental and numerical studies. The QC-region, which is characterized by relatively uniform velocity magnitude and weak turbulence levels, occupies about $40\%$ of the cross-section at Reynolds numbers $Re_\tau$ ranging from $1000$ to $4000$. The influence of the QC region and its boundaries on transport and accumulation of inertial particles has never been investigated before. Here, we first demonstrate that a QC is unidentifiable at $Re_\tau = 180$, before an in-depth exploration of particle-laden turbulent channel flow at $Re_\tau = 600$ is performed. The inertial spheres exhibited a tendency to accumulate preferentially in high-speed regions within the QC, i.e. contrary to the well-known concentration in low-speed streaks in the near-wall region. The particle wall-normal distribution, quantified by means of Vorono\"i volumes and particle number concentrations, varied abruptly across the QC-boundary and vortical flow structures appeared as void areas due to the centrifugal mechanism. The QC-boundary, characterized by a localized strong shear layer, appeared as a \emph{barrier}, across which transport of inertial particles is hindered. Nevertheless, the statistics conditioned in QC-frame show that the mean velocity of particles outside of the QC was towards the core, whereas particles within the QC tended to migrate towards the wall. Such upward and downward particle motions are driven by similar motions of fluid parcels. The present results show that the QC exerts a substantial influence on transport and accumulation of inertial particles, which is of practical relevance in high-Reynolds number channel flow.
\end{abstract}
	
	\pacs{Valid PACS appear here}
	\maketitle

\section{Introduction}\label{Sec:Intro}
Suspensions of particles with different inertia in wall-bounded turbulent flows are of fundamental significance in both industrial processes and natural phenomena. Particle-laden flows are common, such as flows in a biomass combustor with pellets, oceanic flows with planktons and air flows with contaminant dust. The effects of the carrying fluid on the suspended particles are different due to various size, shape and inertia of the particles, giving rise to a wide range of phenomena needed to be understood. 
One of key questions, which is not fully understood, is how inertial particles distribute and transport in inhomogeneous and anisotropic wall turbulence \citep{eaton_preferential_1994,balachandar_turbulent_2010}. An inhomogeneous spatial distribution of particles in wall-bounded turbulence has been observed in both the near-wall region and the core region. The \emph{concentration} in the vicinity of solid walls, reflecting the inhomogeneous distribution in the wall-normal direction, is caused by the turbophoresis effect and sweep/ejection events \citep{reeks_transport_1983,marchioli_mechanisms_2002,guha_transport_2008}, and has been widely observed and studied both in laboratory experiments and numerical simulations \citep{fessler_preferential_1994,rouson_preferential_2001,fong_velocity_2019}. The uneven particle distribution in the core of wall-bounded turbulence (referred to as preferential \emph{clustering}), mainly caused by centrifuging of particles away from rotation-dominant regions \citep{fessler_preferential_1994}, has received considerably less attention.

During the last decade, numerous studies have been devoted to exploitation and understanding of the mechanisms of preferential particle concentration in near-wall turbulence. \citet{picciotto_characterization_2005} examined the distribution of inertial particles in the vicinity of the wall in turbulent boundary layer flow and discovered that a large proportion of the particles tends to stay in convergence regions. \citet{sardina_wall_2012} studied turbophoresis in wall-turbulence using the particle pair correlation function and found that particles tend to accumulate in strongly directional and elongated flow structures. In recent years, preferential concentration in the near-wall region has been explored the effects of gravity \citep{nilsen_voronoi_2013,yuan_dynamics_2017} and particle shape \citep{zhao_slip_2014,njobuenwu_effect_2014,rabencov_voronoi_2015,yuan_three-dimensional_2018}. The role played by coherent flow structures for particle dispersion has been studied. \citet{bernardini_effect_2013} investigated the effect of large-scale structures on dynamics of particles in a turbulent Couette flow and a turbulent Poiseuille flow, observing different spatial distributions due to the underlying flow structures. \citet{yang_preferential_2017,yang_particle_2018} studied particle segregation in a turbulent Couette-Poiseuille flow and found that particles of large inertia segregated near both the stationary wall and the shearless moving wall and exhibited different patterns. 

The particle transport is affected by the spatial distribution of particles in the near-wall region. For instance, particles in wall-turbulence tend to accumulate in low-speed streaks, which would hinder the transport of particles in the streamwise direction \citep{eaton_preferential_1994}. \citet{marchioli_mechanisms_2002} investigated particle transfer and segregation in a turbulent boundary layer flow and revealed that particles are driven by sweep and ejection events. The offspring quasi-streamwise vortices trap the particles near the wall, resulting in preferential concentration. A systematic review of the physics and modeling of particle deposition and entrainment was summarized by \citet{soldati_physics_2009}. Similar particle translation mechanisms in a turbulent pipe flow were studied by \citet{picano_spatial_2009}. Moreover, the translational motion of non-spherical particles in turbulent channel flows was studied \citep{marchioli_orientation_2010, challabotla_orientation_2015}, indicating that the shape of an inertial spheroid has only a weak influence on the translational motion. Gravity, however, has a non-negligible effect in vertical channel flow for the particles with large inertia, for instance to increase the flux of particles towards the walls in an upward flow \citep{nilsen_voronoi_2013}.

Direct numerical simulations (DNSs) of particle-laden wall turbulence become unaffordable expensive at higher Reynolds numbers. Therefore, the vast majority of earlier numerical studies of particle-laden flows have been performed at Reynolds numbers way below those encountered in most practical applications. However, the characteristic flow structures in wall-bounded turbulence have been observed at moderate and high Reynolds numbers, such as large-scale motions (LSMs) and very-large-scale motions (VLSMs) \citep{smits_highreynolds_2011,jimenez_cascades_2012}. The formation and breakdown of LSMs are of significance in the self-sustaining process of wall-bounded turbulence \citep{hamilton_regeneration_1995}. In addition to LSMs and VLSMs, another kind of typical flow structure named a uniform momentum zone (UMZ) was discovered by \citet{meinhart_existence_1995} and further discussed by \citet{de_silva_uniform_2016} in turbulent boundary layer flows at high Reynolds number. \citet{meinhart_existence_1995}, utilizing particle image velocimetry (PIV), measured the instantaneous velocity field and discovered two new structural features. One is large zones with nearly uniform streamwise momentum and time-varying shape throughout the turbulent boundary flow, namely UMZs. The other is viscous-inertial layers, which separate these UMZs. Recently, \citet{chen_uniform-momentum_2020} have detected the UMZs in a turbulent pipe flow at $Re_\tau=500$, suggesting that the innermost UMZs show different properties to the others. By applying a similar method to identify UMZs in turbulent channel flows, \citet{kwon_quiescent_2014} revealed that a large core, with high and uniform velocity magnitude, settled in the channel center and named it \textit{quiescent core} (QC). A sharp jump in velocity is evident at the QC-boundary and the peak vorticity at the meandering boundary is therefore much larger than the local mean vorticity. Subsequently, the structural organization of the QC in a turbulent channel flow was explored using DNS at $Re_\tau=930$ by \citet{yang_structural_2016}. Prograde vortices were found to dominantly contribute to the total mean shear near the core. Their influence on the dynamics of inertial particles is likely to be of importance, but not well understood. \citet{yang_influence_2019} studied the entrainment phenomena of the QC in a turbulent pipe flow at $Re_\tau=926$ and suggested that the shape of the QC-boundary is related to large-scale structures of streamwise velocity. 

Due to demanding cost of computational resources, only a few numerical studies of inertial particles in wall turbulence at medium Reynolds numbers have been carried out in recent years. \citet{bernardini_reynolds_2014} performed one-way coupled simulations of particle-laden channel flow and examined the Reynolds number effect and Stokes number effect on the statistical particle distribution in the wall-normal direction. The turbophoretic drift for particles with intermediate inertia turned out to be independent of the large-scale outer motions, whereas highly inertial particles responded mainly to the action of the outer-layer structures. \citet{wang_two_2019} recently found that both mean concentration profiles and particle clustering are influenced by LSMs and VLSMs in an open channel flow at a moderate Reynolds number. In addition, \citet{jie_influence_2019} have recently studied the rotational motion of tracer spheroidal particles inside and outside of the QC at $Re_\tau=1000$.

Above all, in a traditional Reynolds-averaged view, a turbulent channel flow is divided in an inner and an outer region, where the inner region obeys the law-of-the wall and the streamwise velocity is unconditionally time-averaged. However, the aforementioned QC region identified by \citet{kwon_quiescent_2014} has instantaneous and dynamic features. The boundary of the QC is formed by a particularly high shear rate and associated with large-scale flow structures, which show a significantly greater complexity than the conventional inner/outer-region description. Therefore, one may naturally ask what kind of roles the typical flow structures, such as the QC, observed in high-Re wall-bounded turbulence play in particle behaviour. \citet{kwon_quiescent_2014} found that the streamwise velocity changes abruptly near the QC-boundary, which inevitably results in a strong velocity gradient. Therefore, the distribution and transport of particles may be influenced. Moreover, since the flow inside the QC is quiescent, the behaviour of particles inside the QC can be expected to be different compared to the particles in the near-wall region where the flow is characterized by strong mean shear and highly anisotropic turbulence. More importantly, in practice, the region of QC is significant in high Reynolds number flows (occupying close to $40\%$ of the flow field in the present case) and the dominance of the QC further increases at higher Reynolds numbers. Therefore, it is essential to understand how the particles distribute and transport in this typical flow region and in the vicinity of this region. In particular, we are curious to address the following questions: how do particles cluster in the QC and what is the physical mechanism? Does the presence of a QC affect the particle transport and, if so, how? Motivated by these yet unanswered questions, a DNS of turbulent channel flow at $Re_\tau=600$ laden with inertial particles was performed. The 3D Vorono\"i method \citep{monchaux_preferential_2010,monchaux_analyzing_2012,nilsen_voronoi_2013} was adopted to investigate the clustering of particles, especially around the QC-boundary. Conditionally sampled statistics both inside and outside of QC are computed to analyze particle motion in a QC-frame.

\section{Methodology}\label{Sec:Meth}
Spherical particles suspended in a turbulent channel flow are simulated by coupling Eulerian DNS with a Lagrangian point-particle approach. 
\subsection{Eulerian fluid phase}
In the present work, the fluid inside the channel is an incompressible Newtonian fluid with density $\rho$ and kinematic viscosity $\nu$. The flow is obtained by solving the mass conservation and the Navier-Stokes equations:
\begin{equation}\label{eq:ns1}
\nabla \cdot {{\mathbf{U}}_{f}}=\frac{\partial {{u}_{f}}}{\partial x}+\frac{\partial {{v}_{f}}}{\partial y}+\frac{\partial {{w}_{f}}}{\partial z}=0 ,
\end{equation}
\begin{equation}\label{eq:ns2}
\frac{\partial {{\mathbf{U}}_{f}}}{\partial t}+{{\mathbf{U}}_{f}}\cdot \nabla {{\mathbf{U}}_{f}}=-\frac{1}{\rho }\nabla p+\nu {{\nabla }^{2}}{{\mathbf{U}}_{f}}  ,
\end{equation}
with periodic boundary conditions in the streamwise ($x$) and spanwise ($y$) direction and no-slip conditions at both walls ($z$ denotes the wall-normal direction). In the above equations, ${{\mathbf{U}}_{f}}=\left( {{u}_{f}},{{v}_{f}},{{w}_{f}} \right)$ is the instantaneous fluid velocity, $p$ denotes pressure and $\mathbf{x}=\left( x,y,z \right)$ stands for coordinates in three different directions. The present study considers a sufficiently dilute suspension of sub-Kolmogorov particles so that one-way coupling can be adopted without including feedback force term in the momentum equation (\ref{eq:ns2}).
In the present channel flow, the friction Reynolds number is defined as $Re_\tau=u_\tau h/ \nu = 600$, where $h$ stands for the half-channel height and the wall-friction velocity $u_\tau$ is defined by $u_\tau = \sqrt{\tau_w/\rho}$ using wall shear stress $\tau_w$. In the present study, the fluid phase is considered as air and its kinematic viscosity $\nu$ and density $\rho$ are $1.4638\times10^{-5} m^2/s$ and $1.225 kg/m^3$, respectively. The corresponding friction velocity $u_\tau$ and channel height $2h$ are $0.2 m/s$ and $0.088 m$, respectively. As a result, the viscous length and time scales are $\delta_\nu = \nu / u_\tau$ and $\tau_\nu = \nu / u_\tau^2$, respectively, which are used to nondimensionalize physical variables. Hereinafter, the superscript $+$ denotes variables nondimensionalized by viscous scales.

In this simulation, the computational domain size is $6h\times 3h \times 2h$ in the streamwise, spanwise and wall-normal directions, respectively, with 384 grid points in each direction. Grid spacings are uniform in homogeneous directions, namely $x$ and $y$ direction, which gives $\Delta{x}^+ = 9.375$ and $\Delta{y}^+ = 4.6875$. The wall-normal grid spacing is refined near both walls and ranges from $\Delta{z}_1^+ = 0.143$ next to the wall and $\Delta{z}_c^+ = 5.2876$ in the channel center. The time step is $\Delta{t}^+ = 0.018$. A second-order finite difference scheme is utilized in the wall-normal direction and a pseudo-spectral method is employed in the streamwise and spanwise directions. In addition, a second-order Adams-Bashforth scheme is used for time evolution. The same DNS solver was used in previous studies, e.g. \citet{gillissen_performance_2007}, \citet{mortensen_dynamics_2008} and \citet{nilsen_voronoi_2013}.

\subsection{Lagrangian particle phase}
A Lagrangian particle tracking approach is employed in the present investigation to calculate the location and velocity of spherical particles in the turbulent channel flow. The governing equations of particle motion are:
\begin{equation}\label{eq:p1}
\frac{\mathrm{d}{{\mathbf{x}}_{p}}}{\mathrm{d}t}={{\mathbf{U}}_{p}}   ,
\end{equation}
\begin{equation}\label{eq:p2}
\frac{\mathrm{d}{{\mathbf{U}}_{p}}}{\mathrm{d}t}=\frac{1}{{{\tau }_{p}}}\left( {{\mathbf{U}}_{f@p}}-{{\mathbf{U}}_{p}} \right)\left( 1+0.15 {Re}_{p}^{0.687} \right) ,
\end{equation}
where ${{\mathbf{x}}_{p}}=\left( {{x}_{p}},{{y}_{p}},{{z}_{p}} \right)$ denotes the location of a particle, ${{\mathbf{U}}_{p}}=\left( {{u}_{p}},{{v}_{p}},{{w}_{p}} \right)$ stands for the velocity of the particle while ${{\mathbf{U}}_{f@p}}=\left( {{u}_{f@p}},{{v}_{f@p}},{{w}_{f@p}} \right) $ represents the local fluid velocity at particle position. The particle inertia is parameterized as a Stokes number $St = \tau_p / \tau_\nu$, in which $\tau_p = \rho_{p} d_p^2 /18 \nu \rho_f$ is the particle response time. $Re_p$ is the particle Reynolds number based on particle size $d_p$ and relative velocity between local fluid and particle. Here in equation (\ref{eq:p2}), only the Stokes drag force is included, considering that the particles are much heavier than fluid. In addition, a semi-empirical correction to Stokes drag \citep{schiller_ueber_1933} is used to ensure a reasonable drag force when $Re_p$ is greater than 1. Note that gravity is neglected in the present study to highlight the effects of particle inertia and turbulence on particle dynamics. The gravity, if included, does not influence the qualitative particle behaviour in a vertical channel flow \citep{nilsen_voronoi_2013,marchioli_influence_2007,yuan_dynamics_2017}. Periodic boundary conditions are used in both the streamwise and the spanwise direction. Whereas, collisions between particles and the solid wall are fully elastic. 

After the turbulent channel flow is statistically fully developed, eight sorts of inertial particles with Stokes numbers $St=0.5,1,5,10,30,100,200$ and $500$ are released into the flow with random distribution at $t^+=0$. The particle properties are given in  Table \ref{tab:PartPara}. The number of particles is $N_p  = 6 \times 10^5$ for each kind. The statistical results in Section \ref{Sec:Result} are obtained by averaging over a time-window from $t^+=2610$ to $3492$.

\begin{table}
	\begin{center}
		\def~{\hphantom{0}}
		\setlength{\tabcolsep}{7mm}{
		\begin{tabular}{cccc}
			\hline
			$St$    &   Density ratio $\rho_p/\rho_f$   & Radius $a [\mu m]$   & Radius $a^+$ \\ [3pt]
			$0.5$   &   $2700$  &   $2.12$    &   $2.89\times 10^{-2}$    \\
			$  1$   &   $2700$  &   $2.99$    &   $4.08\times 10^{-2}$    \\
			$  5$   &   $2700$  &   $6.68$    &   $9.13\times 10^{-2}$    \\
			$ 10$   &   $2700$  &   $9.49$    &   $1.29\times 10^{-1}$    \\
			$ 30$   &   $2700$  &   $16.3$    &   $2.24\times 10^{-1}$    \\
			$100$   &   $2700$  &   $29.9$    &   $4.08\times 10^{-1}$    \\
			$200$   &   $2700$  &   $42.2$    &   $5.77\times 10^{-1}$    \\
			$500$   &   $2700$  &   $66.8$    &   $9.13\times 10^{-1}$    \\
			\hline	 
		\end{tabular}
		\caption{Particle parameters in the present study.}
		\label{tab:PartPara}}
	\end{center}
\end{table}

\subsection{Vorono\"i volume analysis}
\citet{monchaux_analyzing_2012} summarized various methods for quantifying preferential particle distributions, such as visualizations, clustering index, box counting methods, correlation dimension, pair correlation, Vorono\"i diagrams and so on. Among these methods, the Vorono\"i volume field reflects the local clustering field directly and allows researchers to access statistics along Lagrangian trajectories. Moreover, the variance of the distribution of Vorono\"i volumes shows the degree of clustering. A Vorono\"i cell contains the space which is closer to the particle than to any other and it is a unique decomposition of 3D space into a 3D Vorono\"i diagram. \citet{monchaux_preferential_2010} first introduced this method to identify particle clusters and the method has subsequently been widely used in analyzing particle clustering \citep{nilsen_voronoi_2013,baker_coherent_2017,yuan_three-dimensional_2018}.

To calculate the Vorono\"i volumes of particles, mirrored ghost-particles are added into the domain to guarantee that all Vorono\"i cells of particles near the boundary have closed boundaries \citep{nilsen_voronoi_2013}. At each time step, all $N_p$ particles inside the computational domain $L_x \times L_y \times L_z=V_d$ are considered. First, the domain is extended by $2L$ in each direction to $\left[-L,L_x+L\right] \times \left[-L,L_y+L\right] \times \left[-L,L_z+L\right]$. Mirrored ghost-particles outside of the original $\left[0,L_x \right] \times \left[0,L_y \right] \times \left[0,L_z \right]$ domain are added in the extended domain. Second, the Vorono\"i volume of each and every particle in the extended domain is computed (using the Vorono\"i function in MATLAB). Finally, the Vorono\"i volumes of each particle inside of the original domain are saved for subsequent analyses. The mean Vorono\"i volume is $V_m=V_d/N_p$. Hereafter, Vorono\"i volumes of particles are scaled by the mean volume $V_m$. As a consequence, the local distribution of particles is close to a random distribution when the dimensionless Vorono\"i volume is about one. Any difference between the mean Vorono\"i volume in different flow regions, such as the mean Vorono\"i volume in different $x$-$y$ planes, reflects a non-uniform distribution of particles. While the variance $\sigma^2$ of the Vorono\"i volumes represents the degree of clustering.

\section{Results and Discussion}\label{Sec:Result}

\subsection{Detection of the quiescent core}
\floatsetup[figure]{style=plain,subcapbesideposition=top}
\begin{figure}
	\centering
	\sidesubfloat[]{\includegraphics[width=0.45\linewidth]{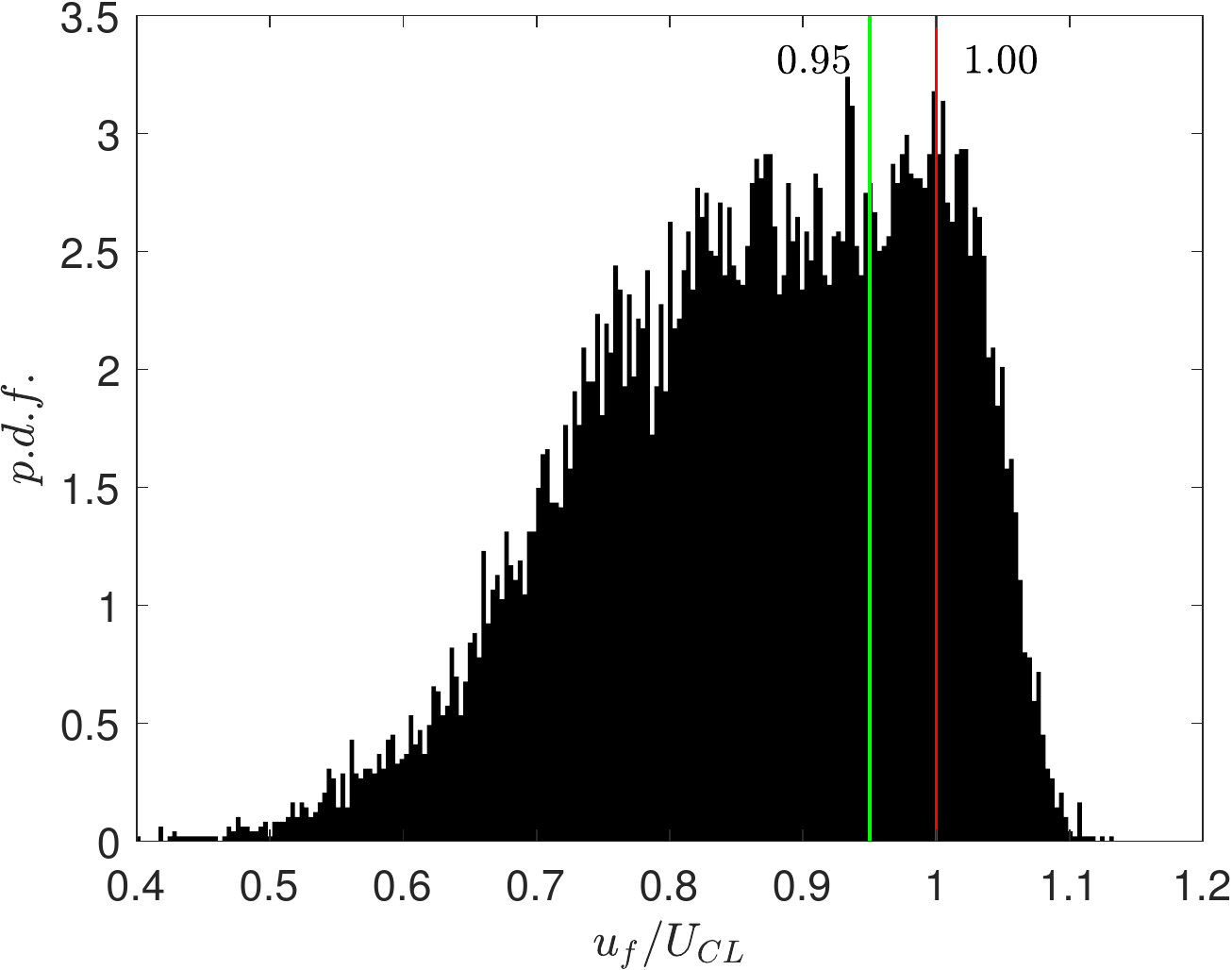}}
	\sidesubfloat[]{\includegraphics[width=0.45\linewidth]{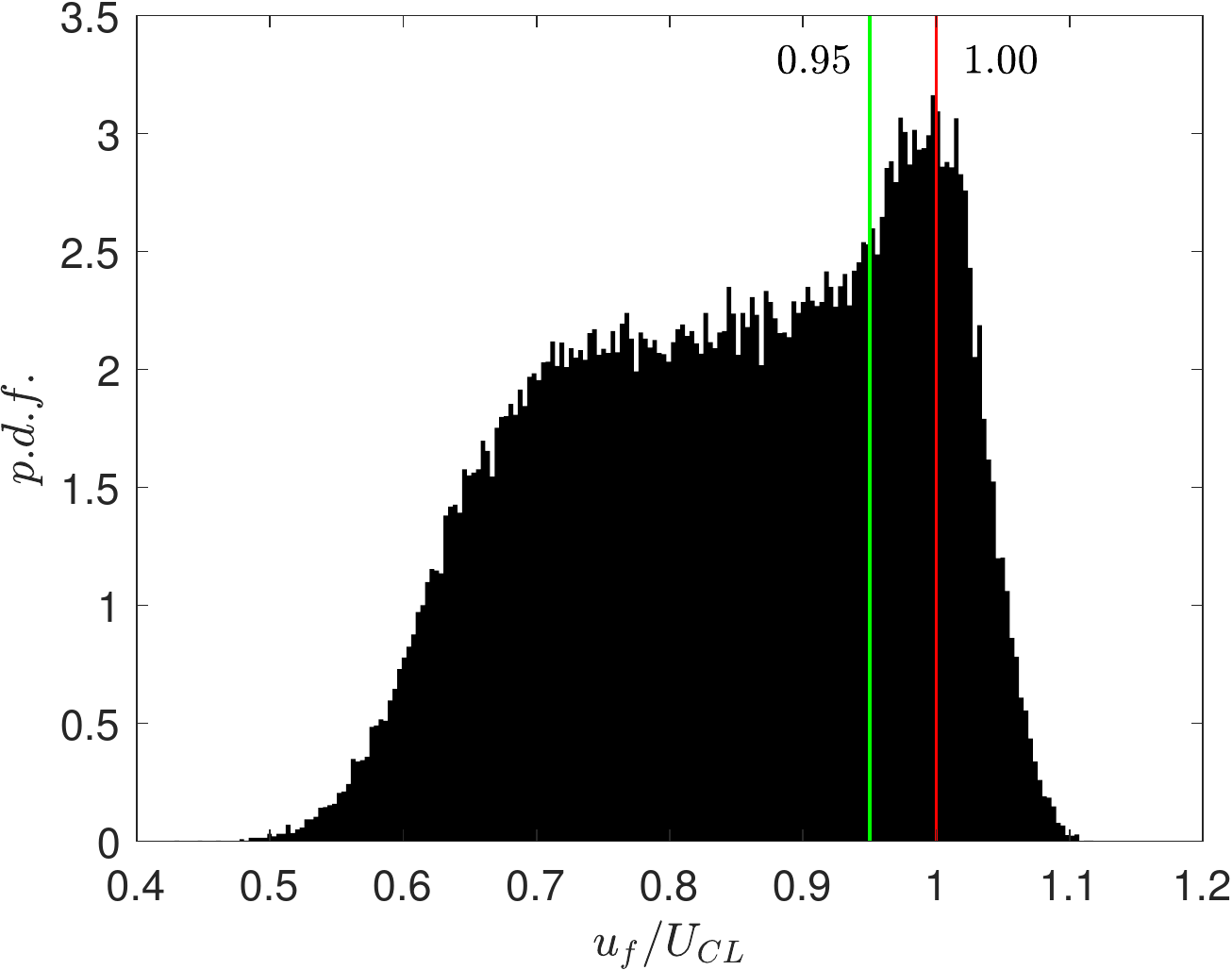}}  \hfill
	\caption{P.d.f.s of the streamwise modal velocity in turbulent channel flows at (a) $Re_\tau=180$; (b) $Re_\tau=600$.}
	\label{fig:UMZ}
\end{figure}

\begin{figure}
	\centering
	\sidesubfloat[]{\includegraphics[width=0.45\linewidth]{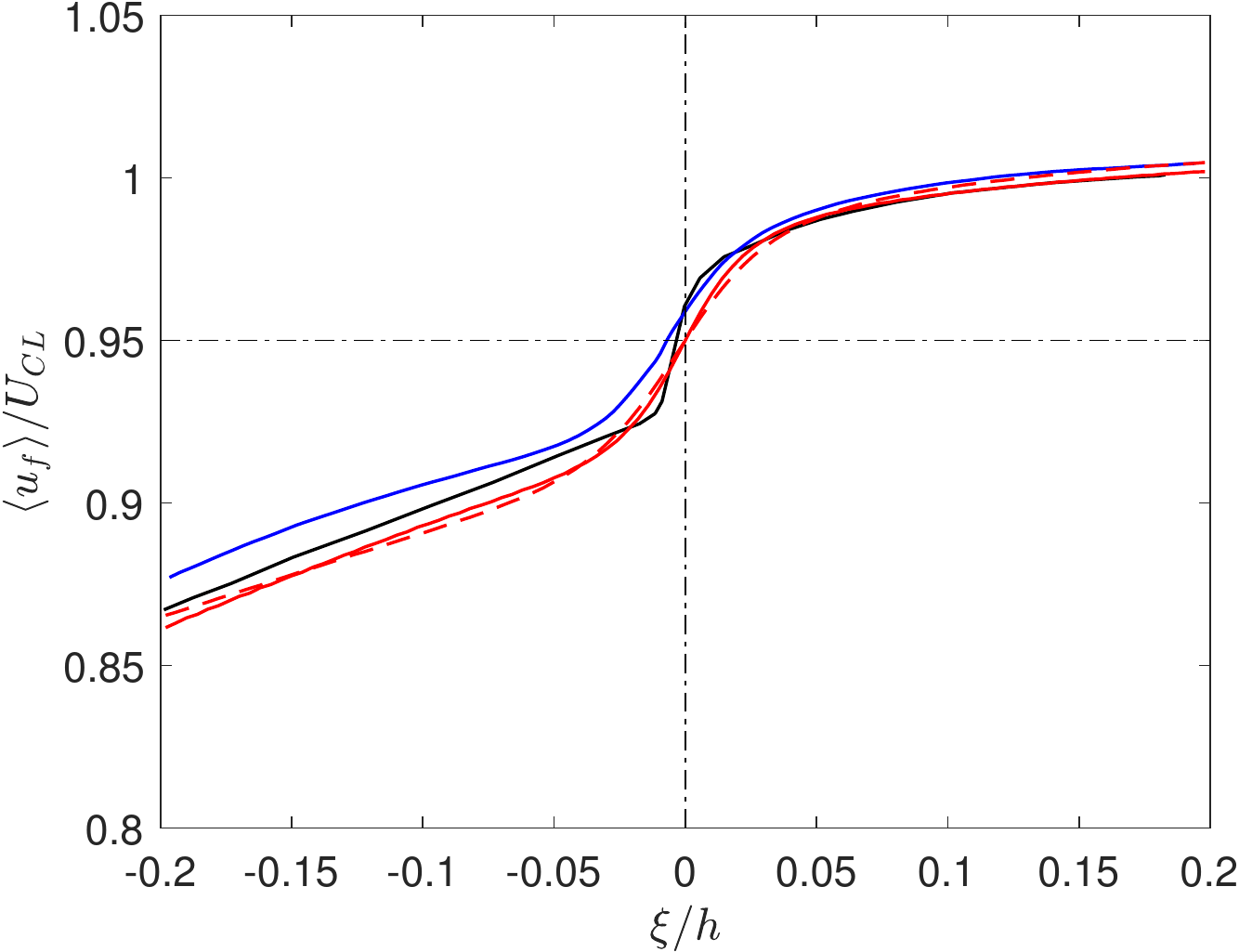}}
	\sidesubfloat[]{\includegraphics[width=0.45\linewidth]{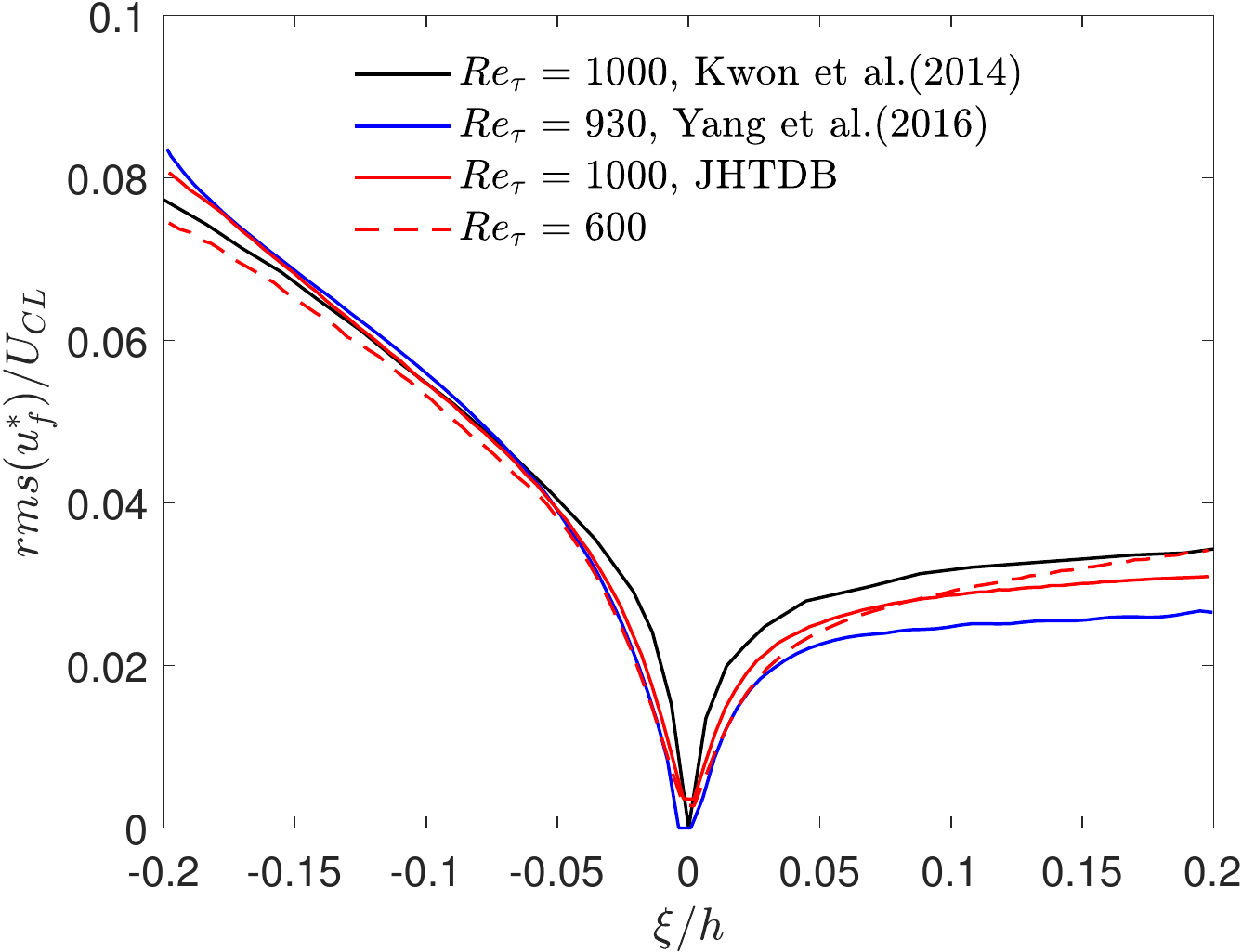}}  \hfill
	\caption{Comparison of (a) conditional mean streamwise velocity, and (b) conditional r.m.s. of streamwise velocity fluctuation around the QC boundary in turbulent channel flows at some different $Re_\tau$.}
	\label{fig:mU_rmsUF}
\end{figure}

\begin{figure}[t]
	\sidesubfloat[]{\includegraphics[width=0.90\linewidth]{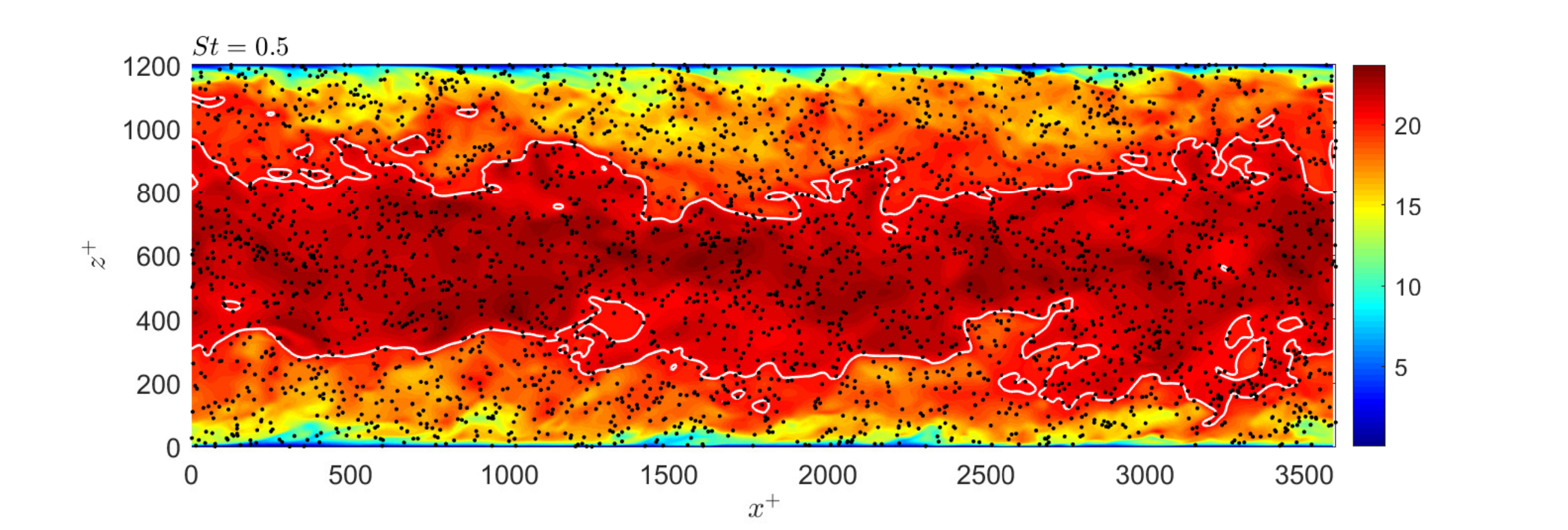}} \\
	\sidesubfloat[]{\includegraphics[width=0.90\linewidth]{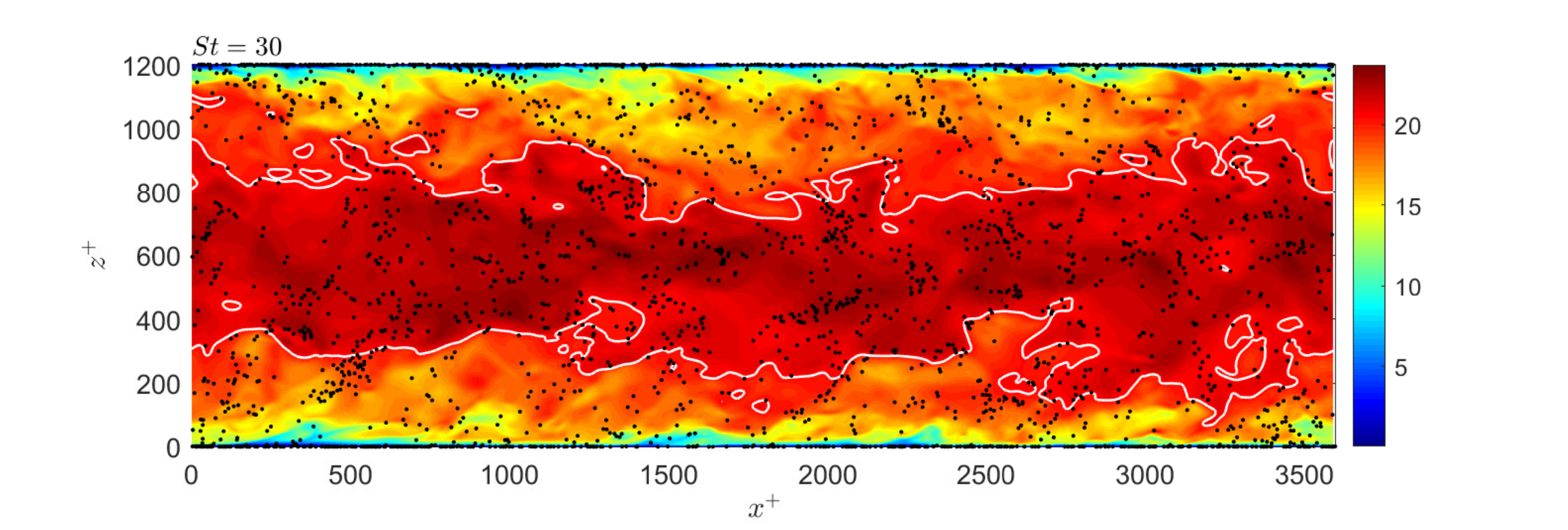}} \\
	\sidesubfloat[]{\includegraphics[width=0.90\linewidth]{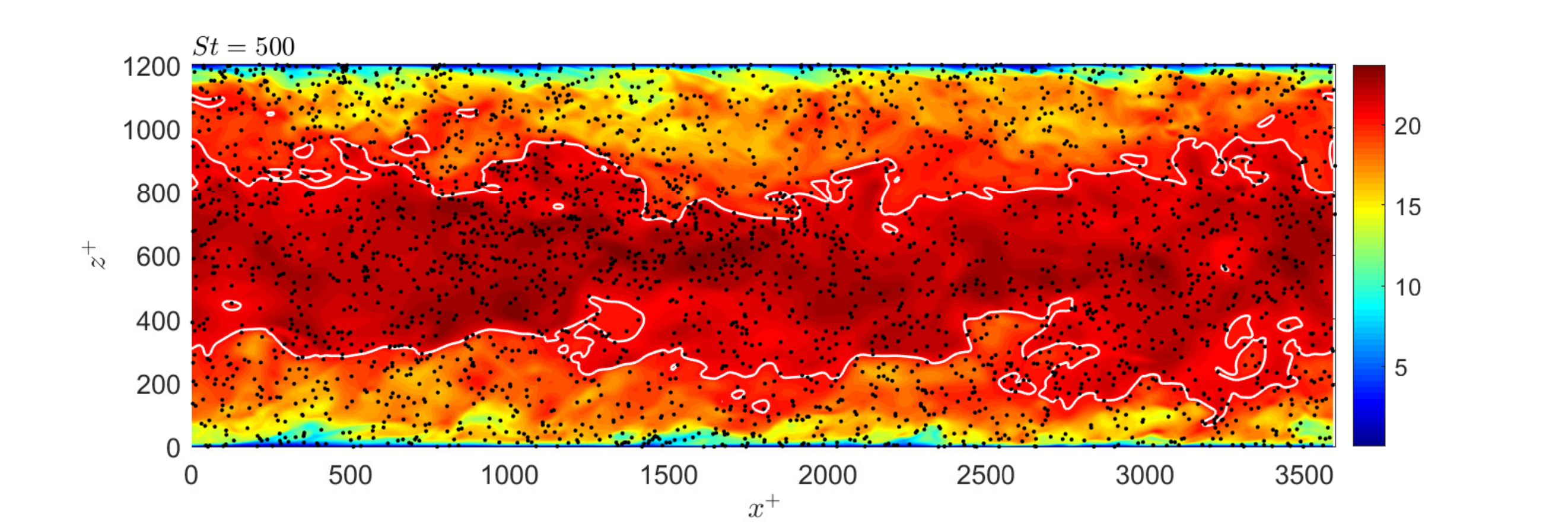}}
	\caption{Instantaneous distribution of particles in a $x$-$z$ plane at $y^+ \approx 708$ with (a)$St=0.5$; (b)$St=30$; (c)$St=500$ in turbulent channel flow at $Re_\tau=600$. The background color represents the fluid streamwise velocity and the white lines show the QC-boundary. Note that the flow field is the same in all plots.}
	\label{fig:xz}
\end{figure}

In the present study, the quiescent core of the turbulent channel flow was detected using the same method as given by \citet{kwon_quiescent_2014}. 
Firstly, the \emph{probability density function} (p.d.f.) of the instantaneous streamwise velocity in an $x$-$z$ plane is constructed and the size of the $x$-$z$ plane is $1.2h \times 2h$, similar to \citet{kwon_quiescent_2014}. $1.2h$ is one fifth of the streamwise length of the present channel. Secondly, the distinct local peaks of the p.d.f. are identified as \emph{modal velocities}, which were also used to detect the regions with relatively uniform momentum in turbulent boundary flows by \citet{meinhart_existence_1995}. Thirdly, by repeating the above procedure for different $x$-$z$ planes, a collection of modal velocities is obtained. 
The p.d.f.s of the streamwise modal velocity in turbulent channel flows at $Re_\tau=180$ and $Re_\tau=600$ are calculated and shown in figure \ref{fig:UMZ} for comparison. A prominent peak is present in the p.d.f. of the modal velocity at $Re_\tau=600$ while there is no obvious peak in the p.d.f. at $Re_\tau=180$. We have verified that the absence of a prominent peak in the p.d.f. of the modal velocity at $Re_\tau = 180$ is independent of the chosen streamwise length $1.2h$ used in the present analysis. As pointed out by \citet{kwon_quiescent_2014}, the thickness of the QC is inversely proportional to the Reynolds number and the discontinuity of the QC grows as Reynolds number reduces. Therefore, it is reasonable that the QC is insignificant and difficult to identify at a fairly low Reynolds number, for instance $Re_\tau = 180$. According to former studies by \citet{kwon_quiescent_2014} and \citet{yang_structural_2016}, the peak represents a large zone with relatively uniform streamwise momentum and low turbulence intensity, namely the quiescent core (QC). Therefore, Figure \ref{fig:UMZ} indicates the presence of a quiescent core region in the channel flow at $Re_\tau=600$ while no apparent QC exists at $Re_\tau=180$. The threshold velocity used to detect the boundary of the QC is $u_f=0.95U_{CL}$, the same as that in \citet{kwon_quiescent_2014} and \citet{yang_structural_2016} in channel flows and in \citet{yang_influence_2019} in pipe flows, where $U_{CL}$ denotes the mean streamwise fluid velocity at the center line. The wall-normal location of the velocity threshold $u_f=0.95U_{CL}$ may occasionally be a multi-valued function due to the local folding of the contour line, as evidenced by figure \ref{fig:xz}. Following \citet{kwon_quiescent_2014} we adopt the concept of ‘enveloped’ boundaries used earlier by \citet{westerweel_turbulentnon-turbulent_2002} and \citet{chauhan_turbulent/non-turbulent_2014} to explore the turbulent/non-turbulent interfaces (TNTIs). The ‘inner’ enveloped core boundary is assigned the wall-normal position of the boundary that is furthest away from the wall. In addition, the conditional-averaged streamwise velocity and conditional \emph{root-mean-square} (r.m.s.) of the streamwise velocity fluctuations around the QC-boundary in the turbulent channel flows at $Re_\tau=600$ and at $Re_\tau=1000$ were calculated and shown in figure \ref{fig:mU_rmsUF} so as to verify the detection method of QC in the present investigation. Data of turbulent channel flow at $Re_\tau=1000$ comes from the Johns Hopkins Turbulence Databases \citep{perlman_data_2007,li_public_2008,graham_web_2016}.
Hereinafter, $\xi$ stands for the coordinate axis attached to the QC-boundary, parallel to the wall-normal direction and pointing to the QC. $\xi=0$ represents the location of the QC-boundary and where $\xi>0$ indicates the QC region. The total velocity is decomposed in the reference frame of the core boundary (QC frame) as in \citet{kwon_quiescent_2014}, namely,
\begin{equation}
\centering
u_{f}^{*}\left( x, y, \xi, t \right)={{u}_{f}}\left( x, y, \xi, t \right)-\left\langle {{u}_{f}}\left( x, y, \xi, t \right) \right\rangle	.
\end{equation}
In this decomposition, $\langle u_f(x, y, \xi, t)\rangle$ stands for the conditional-averaged streamwise velocity in the QC-frame and the angle brackets represent an averaging operation in the streamwise, the spanwise direction and in time. 
$u_f^*$ represents the instantaneous fluctuation of $u_f$.
As illustrated in figure \ref{fig:mU_rmsUF}, the conditional-averaged streamwise velocity is approximately uniform in the QC, i.e. $\xi/h > 0$, while a localized shear layer, whose thickness is smaller than $0.1h$, exists around the QC-boundary. In the right panel, conditional r.m.s. of the streamwise velocity fluctuation remains almost the same and low inside the QC when $\xi/h > 0.05$, while it rapidly increases with increasing distance $\left| \xi\right| $ from the QC-boundary outside the QC region. The four lines almost collapse in both figure \ref{fig:mU_rmsUF}(a) and (b), indicating that the method used to identify the QC-boundary is adequate also in present study. Moreover, the almost collapse between the red solid and dashed lines indicates a negligible Reynolds number dependence of the conditional-averaged streamwise velocity and r.m.s. of streamwise velocity fluctuation around the QC-boundary in the $Re_\tau$ range between 600 and 1000.

\begin{table}
	\begin{center}
		\def~{\hphantom{0}}
		\setlength{\tabcolsep}{7mm}{
		\begin{tabular}{ccccccc}
			\hline
			$Re_\tau$			&  	600	&	1030	&	2100	&	3090	&	3965  \\ [3pt]
			$\mu(t_{core})/h$	&	0.695 &	0.778  	&  	0.830	&  	0.846	&	0.876 \\
			$\sigma(t_{core})/h$&	0.291 &	0.260  	& 	0.272 	&	0.278	&	0.280 \\
			\hline	 
		\end{tabular}
		\caption{The thickness $t_{core}$ of the QC changes with the friction Reynolds number $Re_\tau$. Data at $Re_\tau=1030,2100,3090$ and $3965$ are from \citet{kwon_quiescent_2014}. $\mu(t_{core})$ is the mean thickness and $\sigma(t_{core})$ is the standard deviation of $t_{core}$.}
		\label{tab:t}}
	\end{center}
\end{table}

The thickness of the QC is defined as $t_{core} = z_{upper} - z_{lower}$, where $z_{upper}$ and $z_{lower}$ denote the upper and lower QC-boundaries, respectively. It was first found by \citet{kwon_quiescent_2014} that the mean thickness of the QC is close to $h$, i.e. the half-channel height, and increases with increasing $Re_\tau$. Table \ref{tab:t} shows the thickness of the QC at different Reynolds numbers. The results at $Re_\tau=1030, 2100, 3090$ and $3965$ are from \citet{kwon_quiescent_2014}. Comparing to the present result at $Re_\tau=600$, the tendency of thickness versus $Re_\tau$ remains consistent. The thickness $t_{core}$ increases monotonically from about $0.7$ at $Re_\tau=600$ to $0.876$ at $Re_\tau=3965$, while the standard deviations appear to be $Re_\tau$-independent. This indicates that the QC in turbulent channel flow becomes of greater significance at high Reynolds number since it occupies a gradually larger flow volume. Moreover, the mean distance between the center of the QC and the channel centerline is approximately zero. Therefore, the mean wall-normal distance of the lower QC-boundary is $\mu(z_{lower})/h = (h-\mu(t_{core})/2)/h \approx 0.65$, namely $\mu^+(z_{lower}) \approx 390$, at $Re_\tau=600$.

\begin{figure}
	\sidesubfloat[]{\includegraphics[width=0.90\linewidth]{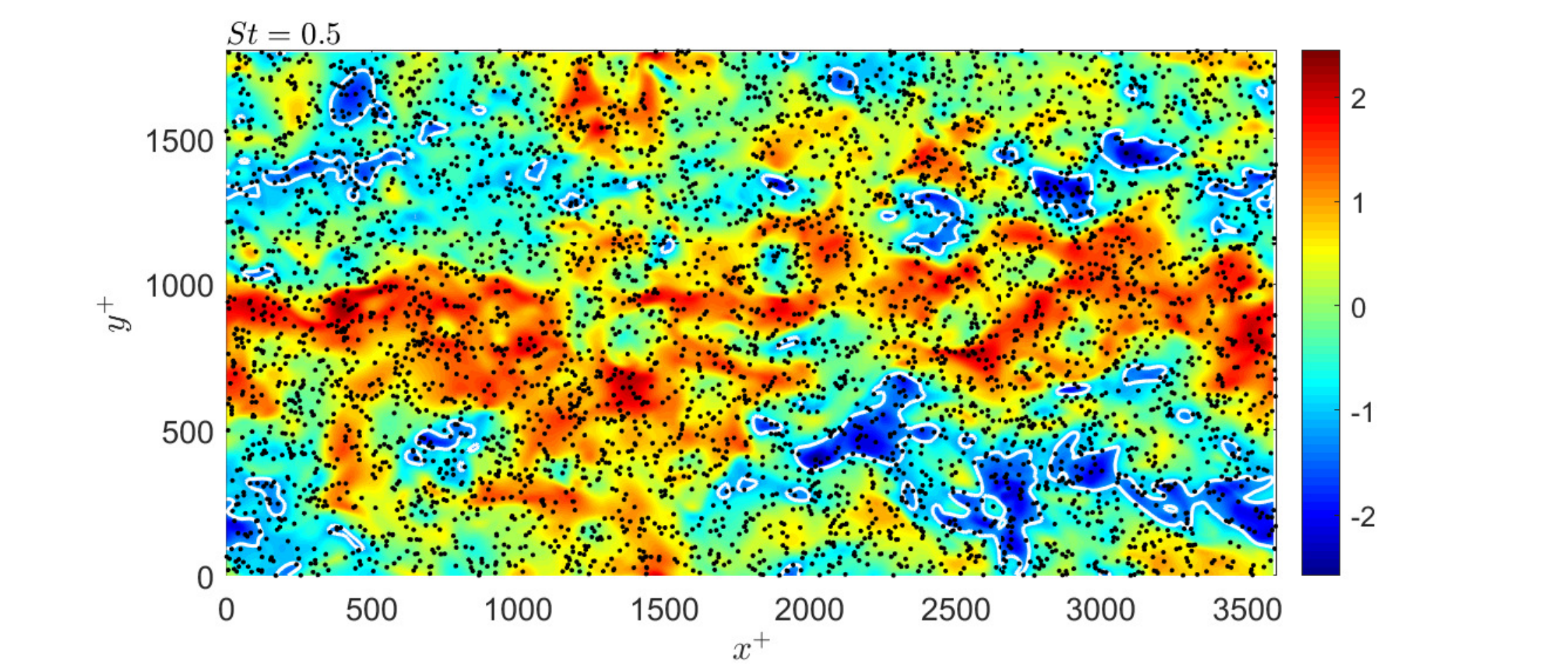}} \\
	\sidesubfloat[]{\includegraphics[width=0.90\linewidth]{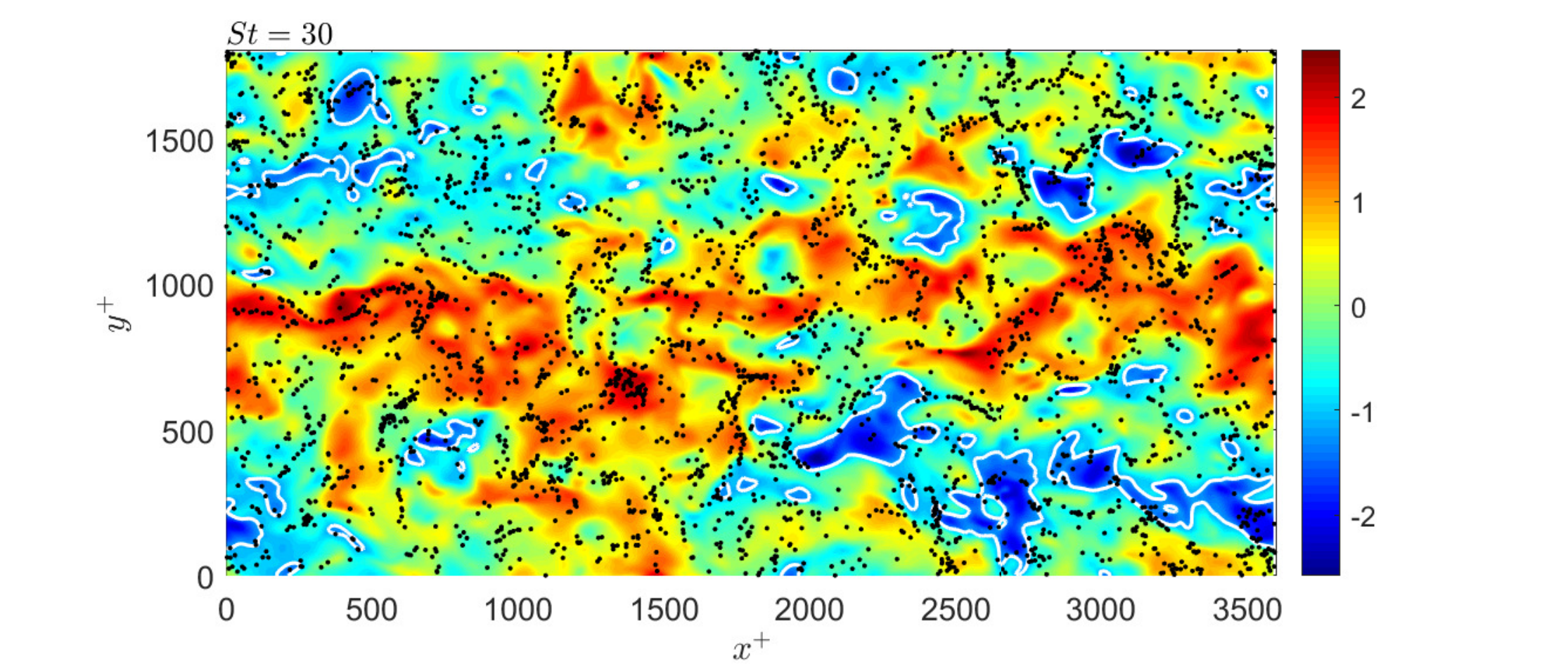}} \\
	\sidesubfloat[]{\includegraphics[width=0.90\linewidth]{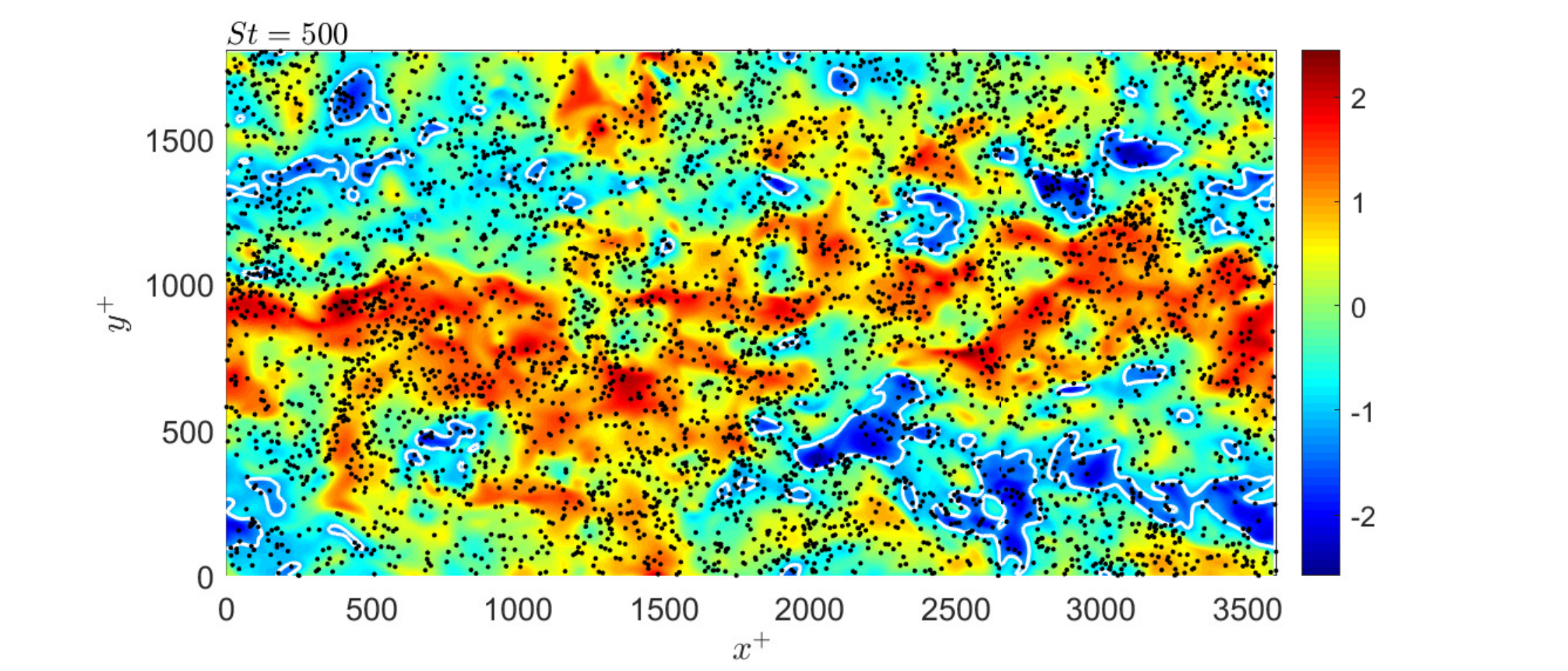}}
	\caption{Instantaneous distribution of particles with (a)$St=0.5$; (b)$St=30$; (c)$St=500$ in the mid plane of turbulent channel flow at $Re_\tau=600$. The background color represents the streamwise fluid velocity fluctuation and the white lines show the QC-boundary.}
	\label{fig:xy}
\end{figure}

\subsection{Distribution of particles} 
\subsubsection{Instantaneous distribution}
Instantaneous distributions of particles with different inertia in a streamwise--wall-normal plane of the channel flow are shown in figure \ref{fig:xz}, where the white lines stand for the boundary of the QC while black points denote particles. It can be observed that the particle distribution is more uniform when the Stokes number is either sufficiently small or large, while clustering of particles is more prominent at intermediate $St=30$. Figure \ref{fig:xz} also shows that the QC-boundary occasionally can be surprisingly close to the wall (for instance at about $x^+=3200$). This observation suggests that the fluid and particle dynamics also in the near-wall region can be influenced by the existence of the QC, which according to the detection criterion contains high-speed fluid with $u_f > 0.95U_{CL}$. There is also an early observation by \citet{johansson_structure_1982} that the streamwise velocity $u_f$ varies widely about its mean $\langle u_f \rangle$ near a solid wall. In addition, instantaneous distributions of particles in the channel mid plane at $z^+=600$ are presented in figure \ref{fig:xy}, where the white lines denote the QC-boundary. In this plane, the flow outside the QC (blue regions) is recognized with negative streamwise velocity fluctuation. Obviously the particles distributed more randomly when the Stokes number is either sufficiently small or large, similar to the observations in figure \ref{fig:xz}. It should be pointed out that $St=30$ particles tend to accumulate in regions with positive streamwise velocity fluctuations, namely higher streamwise velocity $u_f$, i.e. opposite of the phenomenon that particles prefer to concentrate in low-speed streaks in the near-wall region \citep{marchioli_mechanisms_2002}.
\subsubsection{Particle distribution in the quiescent core}	
\begin{figure}
	\centering
	\sidesubfloat[]{\includegraphics[width=0.45\linewidth]{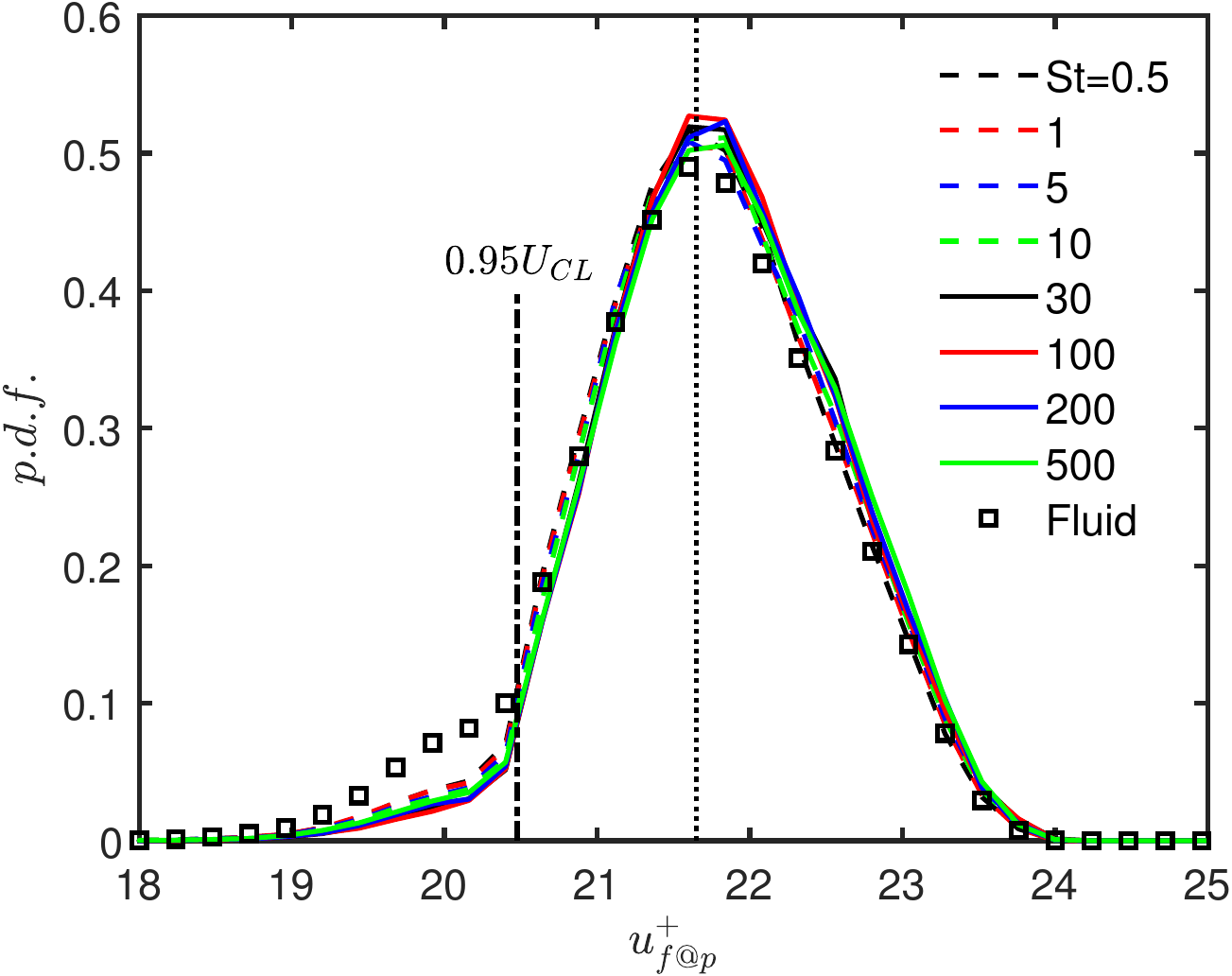}}
	\sidesubfloat[]{\includegraphics[width=0.45\linewidth]{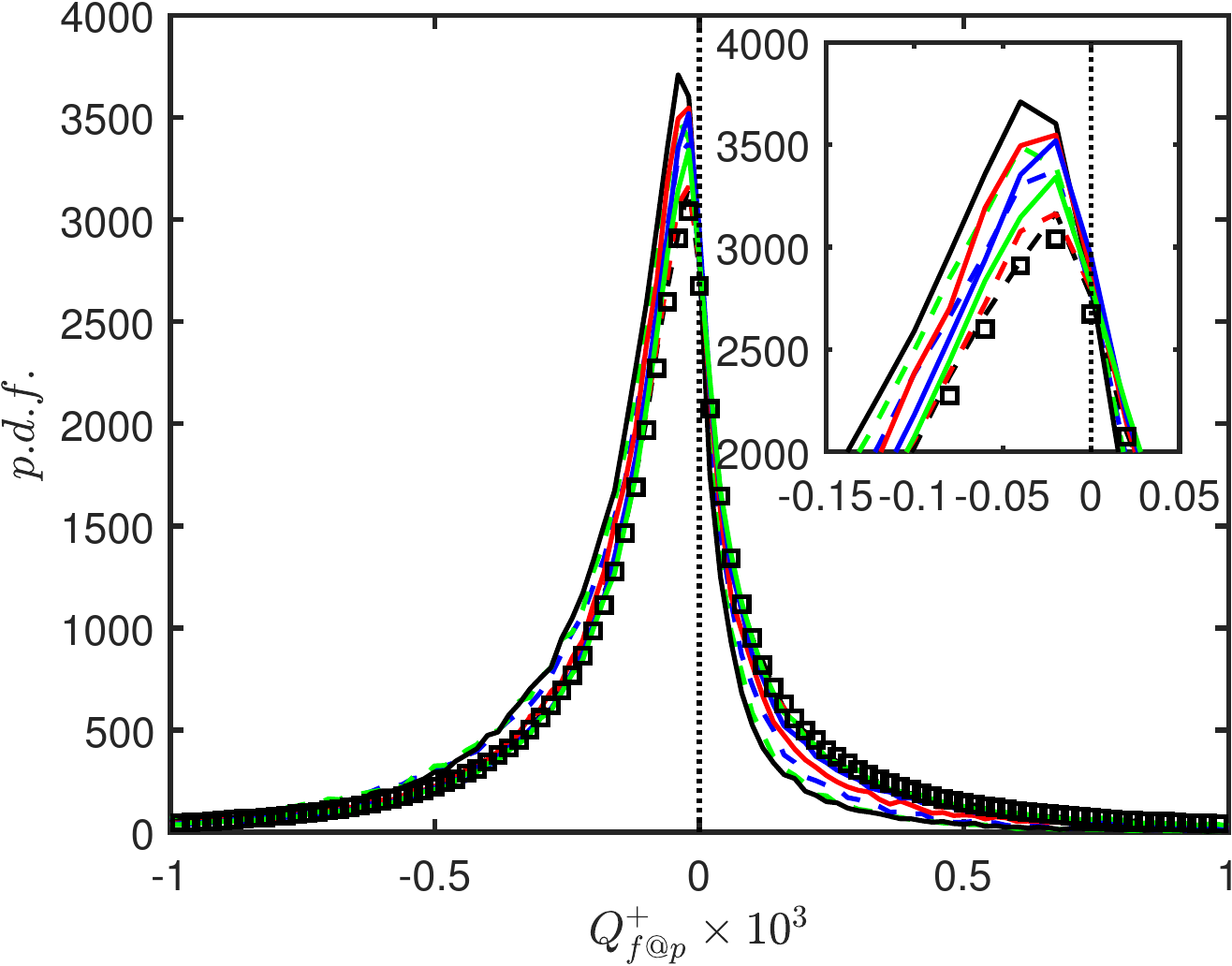}}  \hfill
	\caption{P.d.f.s of (a) streamwise fluid velocity at particle position, and (b) $Q$ of the fluid at particle position in the QC regions in figure \ref{fig:xy}. The squares denotes (a) the mean streamwise velocity, and (b) the $Q$ value of fluid in the QC in the mid plane. The black-dash-dotted vertical line stands for the value of the velocity threshold $0.95U_{CL}$.}
	\label{fig:CenterPdf}
\end{figure}

It is interesting to observe the accumulation of particles with intermediate $St$ in high-speed regions in the QC in the channel center. To quantitatively analyze this phenomenon, p.d.f.s of streamwise fluid velocity fluctuations at particle position are presented in figure \ref{fig:CenterPdf}(a). The squares denote the mean streamwise velocity of fluid in the QC in the channel mid plane ($z^+=600$). The curves representing p.d.f.s of fluid at particle position are all above the squares. This shows that the particles tend to locate in regions with higher streamwise velocity and the maximum peak is at $St =100$. This reveals that particles with $St = 100$ have strongest tendency to locate in regions with high streamwise velocity, although this trend is not very prominent. However, the tendency to stay in high-speed regions at channel center is most pronounced for $St = 30$ particles at $Re_\tau = 180$ (not shown), which reflects an implicit Reynolds number dependence due to the enlarged Kolmogorov length scale at high Reynolds numbers. 

In order to better interpret the mechanism of the particles’ preferential accumulation in regions of high streamwise velocity, we further examined the second invariant $Q$ of the velocity gradient tensor, which is defined as:
\begin{equation}
\centering
Q=\frac{1}{2}\left( {{\left\| \boldsymbol{\Omega } \right\|}^{2}}-{{\left\| \mathbf{S} \right\|}^{2}} \right),
\end{equation}
where $\mathbf{S}=\frac{1}{2}\left[ \nabla {{\mathbf{U}}_{f}} + (\nabla {{\mathbf{U}}_{f}})^\intercal \right]$ is the strain-rate tensor and $\boldsymbol{\Omega }=\frac{1}{2}\left[ \nabla {{\mathbf{U}}_{f}} - (\nabla {{\mathbf{U}}_{f}})^\intercal \right]$ is the rotation-rate tensor.
Considering particles inside the QC in the channel mid plane, figure \ref{fig:CenterPdf}(b) shows the p.d.f.s of $Q$ at particle position, where negative $Q$ represents local high strain rate and low vorticity. As shown in figure \ref{fig:CenterPdf}(b), the area below a solid or dashed line and with negative $Q$ is larger than that below the squares. We can therefore infer that particles tend to reside in the regions with negative $Q$ in the flow field. Moreover, the magnified figure shows that an inertia effect is present and the trend of curves is non-monotonic in $St$. The topmost black-solid line denotes fluid at particles with $St=30$, revealing that intermediate inertial particles are most prone to locate in regions with negative $Q$ than any other inertial particles. This also suggests that the clustering of particles with $St = 30$ is the most prominent. Therefore, p.d.f.s of $Q$ reveal that small particles in QC tend to stay in the regions with high strain rate and low vorticity, similarly to earlier findings in HIT \citep{squires_preferential_1991}. This observation also suggests that the physical mechanism of the preferential clustering in the QC is similar to that in HIT and caused by centrifuging of particles away from high vorticity regions. More importantly, however, we found that negative $Q$-regions correlated with high streamwise fluid velocity in the QC so that the preferential accumulation in the high-speed regions around the core of channel accelerates the particle transport in the streamwise direction.

\begin{figure}[t]
	\centering
	\sidesubfloat[]{\includegraphics[width=0.45\linewidth]{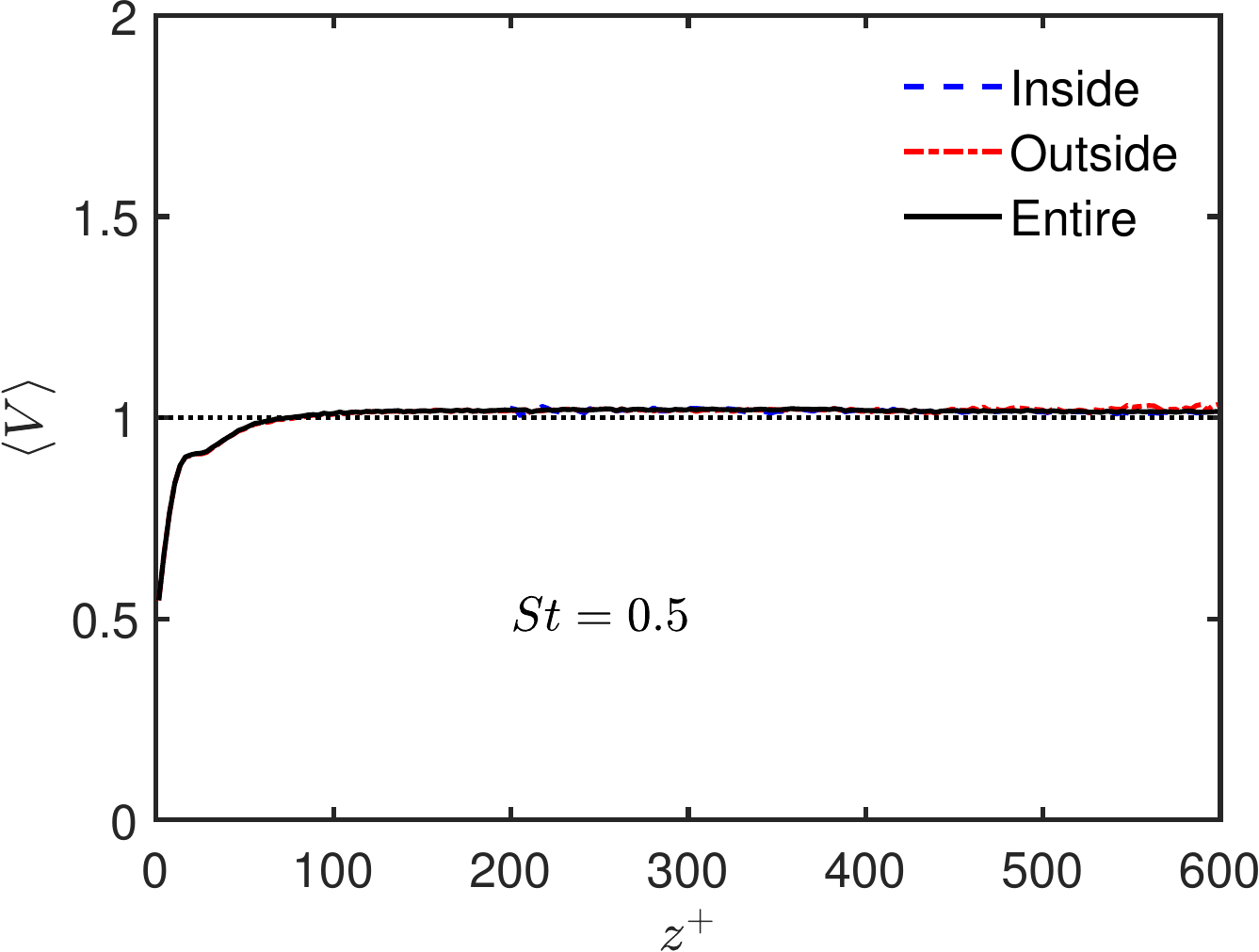}}
	\sidesubfloat[]{\includegraphics[width=0.45\linewidth]{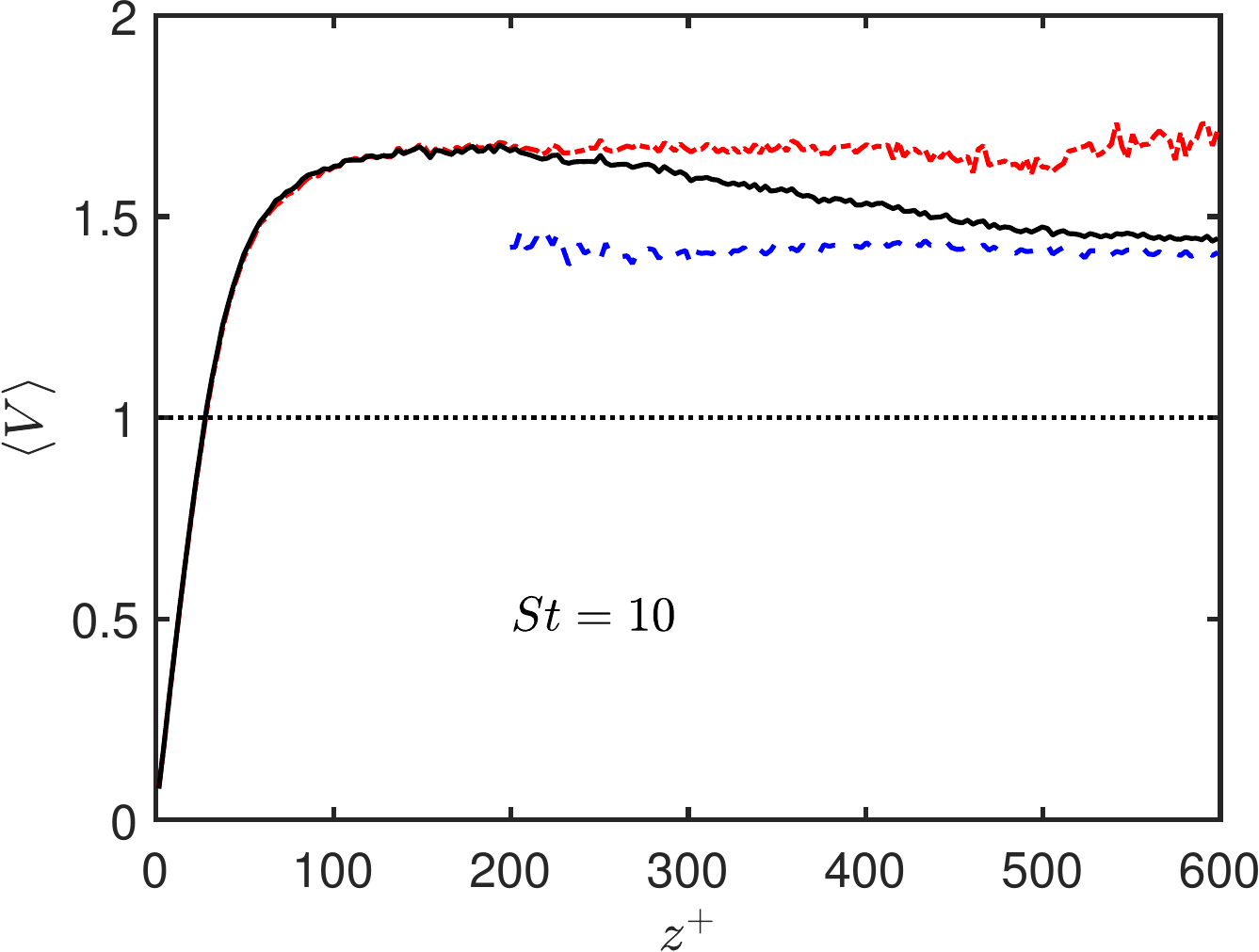}}  \hfill
	\\
	\sidesubfloat[]{\includegraphics[width=0.45\linewidth]{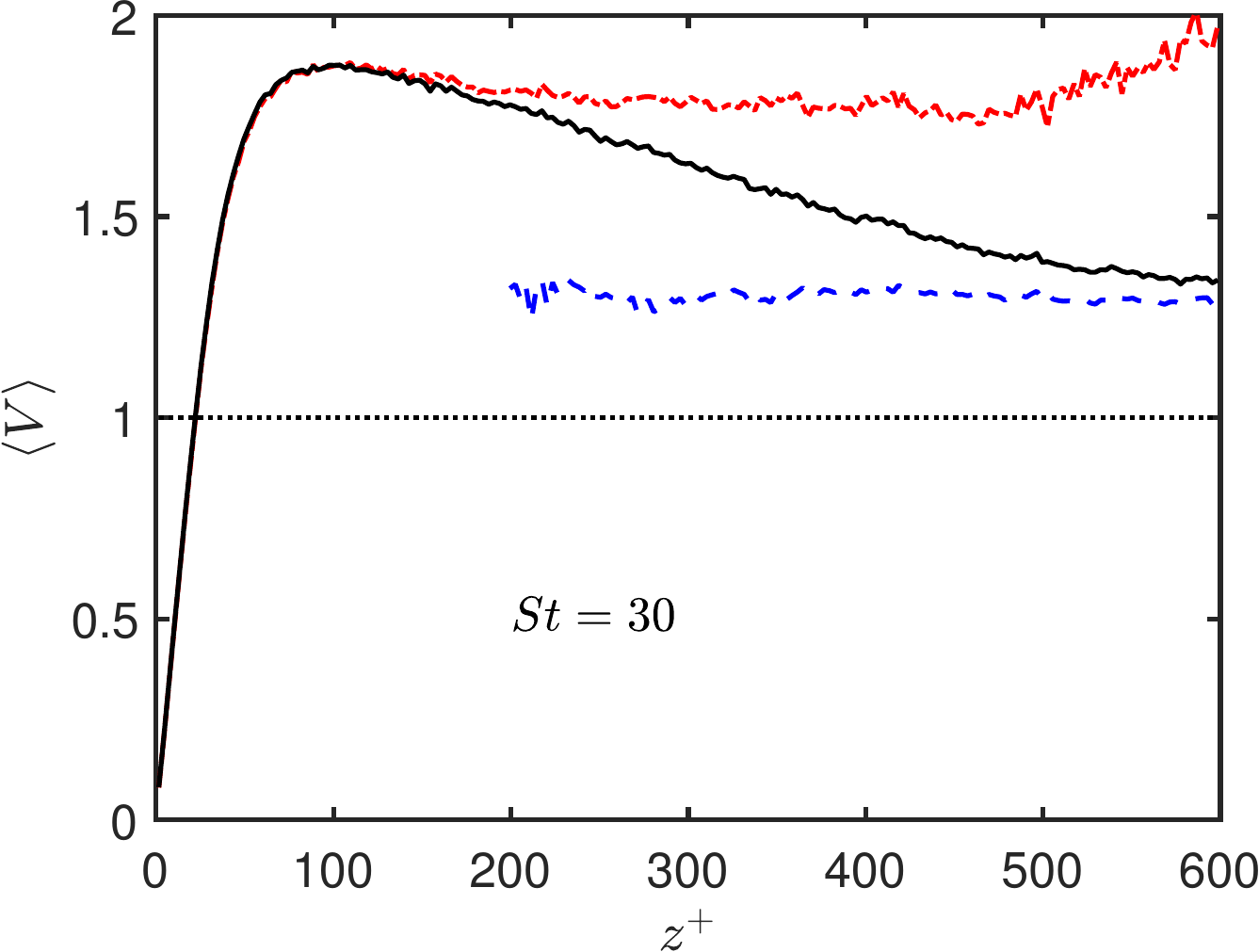}}
	\sidesubfloat[]{\includegraphics[width=0.45\linewidth]{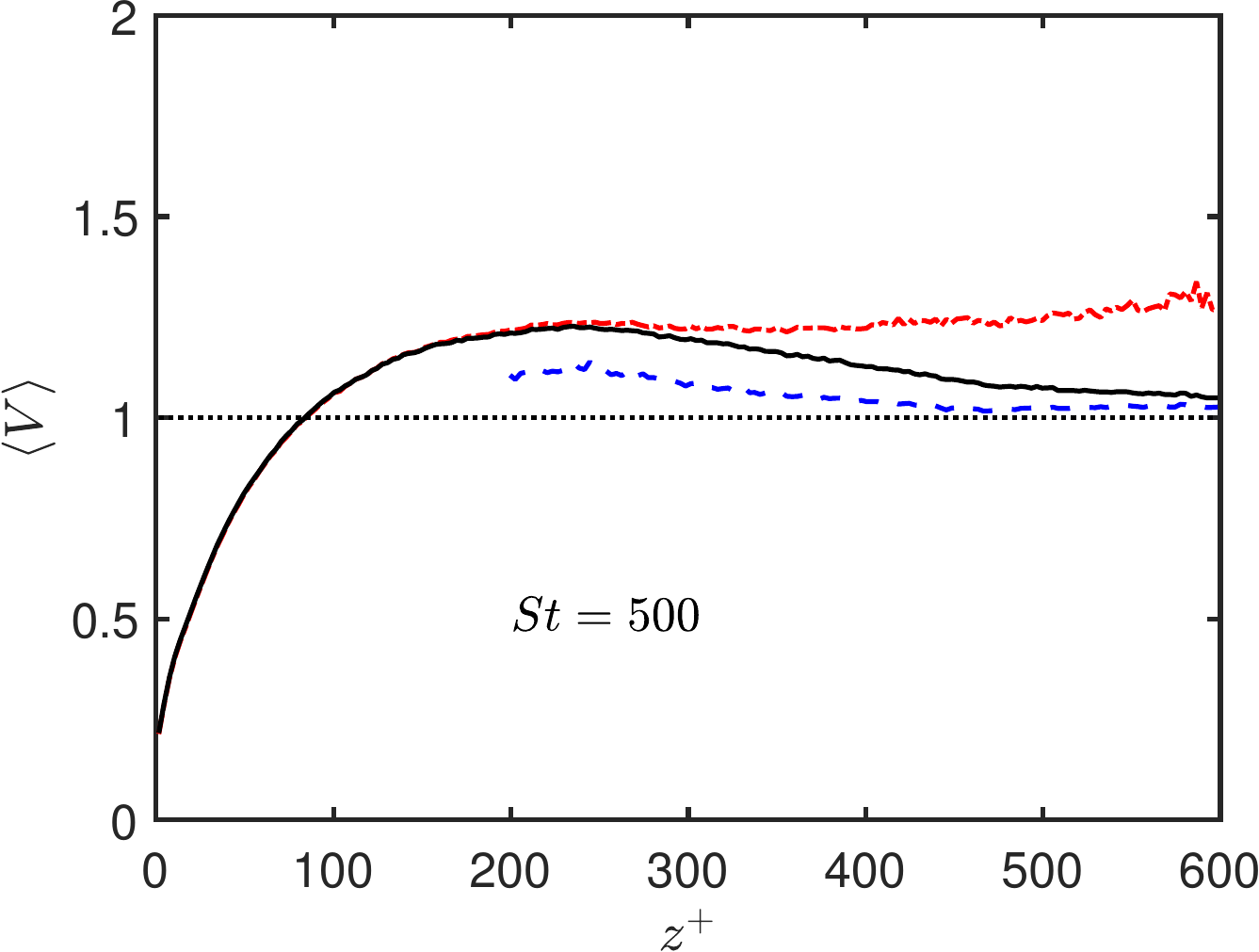}}  \hfill
	\caption{Conditional-averaged Vorono\"i volume $\langle V \rangle$ for particles in different regions. (a)$St=0.5$; (b)$St=10$; (c)$St=30$; (d)$St=500$. The blue-dashed lines stand for conditional-averaged Vorono\"i volumes of particles inside the QC while the red-dash-dotted lines represent particles outside the QC. The black-solid lines denote mean volumes of particles in the entire streamwise-spanwise plane, i.e. an unconditional average.}
	\label{fig:V_IO}
\end{figure}

\begin{figure}[t]
	\centering
	\sidesubfloat[]{\includegraphics[width=0.45\linewidth]{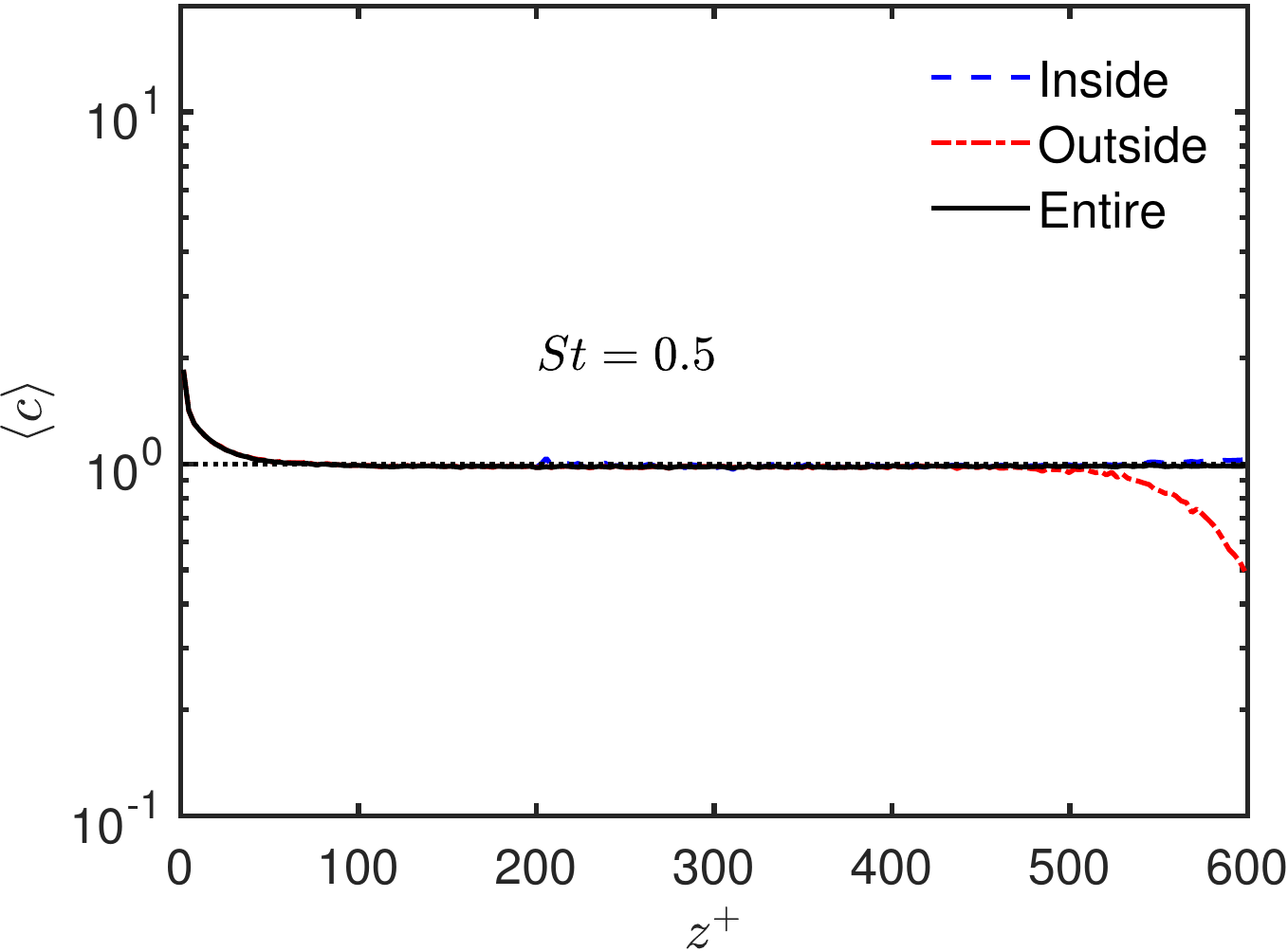}}
	\sidesubfloat[]{\includegraphics[width=0.45\linewidth]{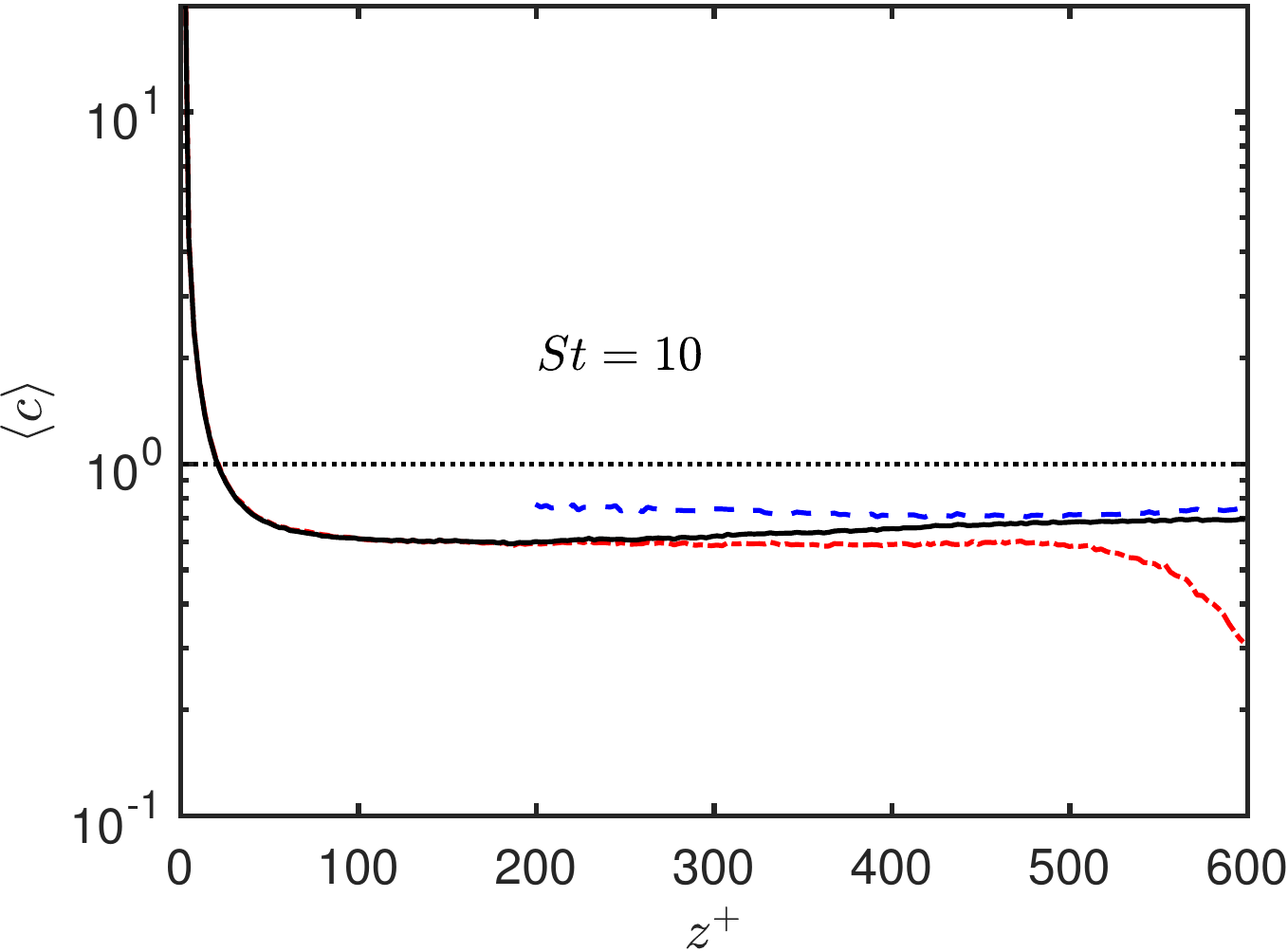}}  \hfill
	\\
	\sidesubfloat[]{\includegraphics[width=0.45\linewidth]{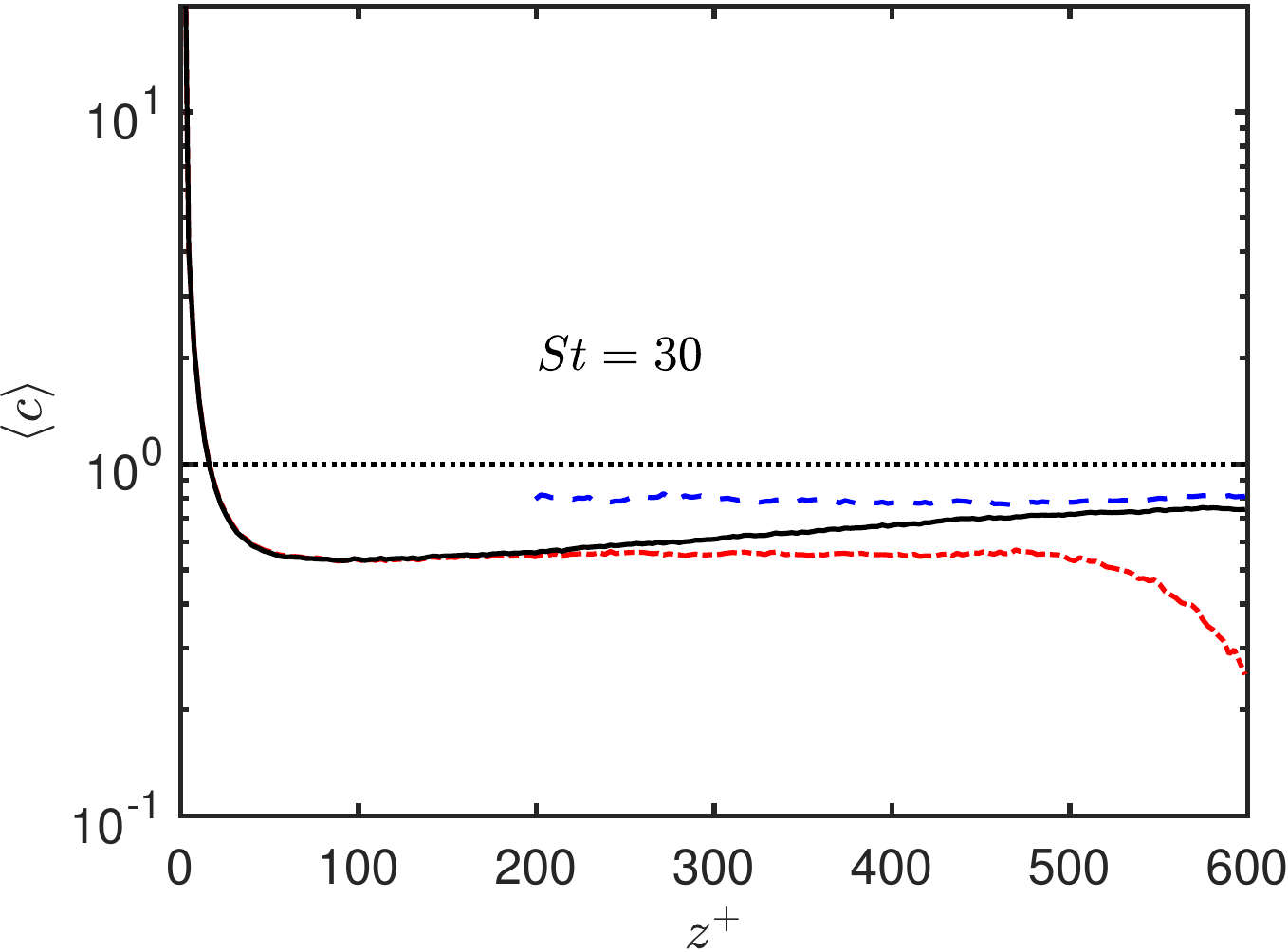}}
	\sidesubfloat[]{\includegraphics[width=0.45\linewidth]{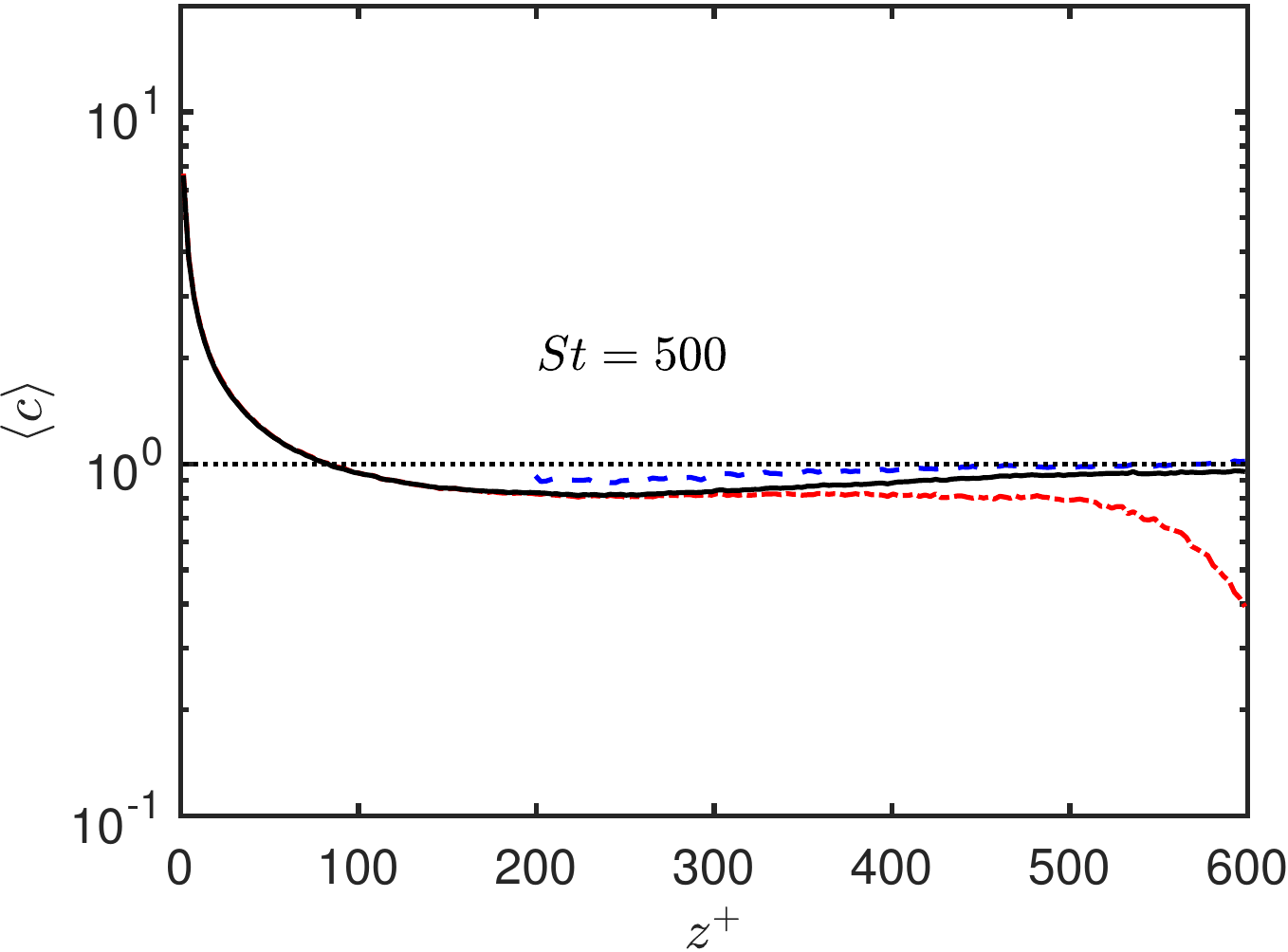}}  \hfill
	\caption{Concentration profiles $\langle c\rangle$ for four different Stokes numbers (a)$St=0.5$; (b)$St=10$; (c)$St=30$; (d)$St=500$. The continuous black line represents the unconditioned concentration.}
	\label{fig:c_IO}
\end{figure}

To study the role of the QC on the particle distribution in the wall-normal direction, conditional-averaged Vorono\"i volumes for particles in and outside of the QC are shown in figure \ref{fig:V_IO}. Hereinafter, we adopt the term \emph{concentration} for describing the non-uniform distribution of particles in the wall-normal direction, for instance the preferential near-wall concentration, and the term \emph{clustering} refers to the local spatial distribution in the channel core region. The inverse of the mean Vorono\"i volume, averaged over a wall-parallel slab at a certain wall distance, is proportional to the number of particles in this slab. In the vicinity of the wall, all black lines are lower than the dotted line $\langle V \rangle =1$, which denotes a random distribution. This means that inertial particles prefer to concentrate in the viscous layer in wall-bounded turbulence \citep{mclaughlin_aerosol_1989}. Away from the wall, the wall-normal concentration of particles with low $St$ is close to a random distribution (see figure \ref{fig:V_IO}a), while the mean Vorono\"i volume of particles with intermediate and large $St$ is larger than a random distribution, thus indicating a lower particle accumulation. The concentration in the logarithmic region (e.g. at about $z^+=100$) is lower than that in the outer layer \citep{nilsen_voronoi_2013}. The blue-dashed line is lower than the red-dash-dotted line in figure \ref{fig:V_IO}(b-d), revealing that the number of particles inside the QC is higher than that outside the QC at the same wall distance. Therefore, at a given wall-normal location, particles prefer to reside in the QC rather than outside the QC. Since particles in the QC travel faster in the streamwise direction than those outside the QC, the presence of the QC enhances the streamwise transport of particles in the channel flow. Moreover, both the mean Vorono\"i volume in and outside the QC are relatively uniform from $z^+ \approx 200$ to the channel center. The probability to locate in the QC attenuates from the channel center to the wall \citep{kwon_quiescent_2014}. The black-solid line accordingly approaches the blue-dashed line at the channel center while the black-solid line and the red-dash-dotted line are almost overlapping below $z^+ \approx 200$. Note that particle concentration in the near-wall region has not yet reached a statistically steady state, especially for those particles with intermediate inertia \citep{sardina_wall_2012,bernardini_reynolds_2014}. A quantitative difference of particle number in the vicinity of the wall may be expected in the comparison with the case of a steady distribution. However, because of the relatively large local Kolmogorov time scale near the QC, which results in low local Stokes numbers, the particle spatial distribution in the central region reaches a statistically steady state faster than in the near-wall region.

We furthermore examined the wall-normal distribution of the particle concentration in wall-parallel slabs (as shown in figure \ref{fig:c_IO}). The concentration profiles are computed by using 400 uniform slabs in the wall-normal direction. The concentration index $c$ is defined as $c(z^+) = N_s(z^+)/N_{uni}$, where $N_s(z^+)$ is the number of particles in the slab at wall-normal location $z^+$, while $N_{uni}$ represents the number of uniformly distributed particles with the same $N_p$. Similarly for the concentration of particles in the QC, $c(z^+)=N_{s,in}/N_{uni,in}$. In each slab, $N_{s,in}(z^+)$ is the number of particles in the QC while $N_{uni,in}$ denotes the number of particles in the QC in a uniformly distributed case. The statistics are consistent with the observations made from figure \ref{fig:V_IO} that the number of particles inside the QC is higher than that outside of the QC for particles with intermediate inertia. Note that the red and blue profiles are conditional-averaged by QC. The red profile, which represents the particle distribution outside the QC, extends all the way to the center, thereby revealing that the discontinuity of QC occurs even in the channel center.

\begin{figure}[t]
	\centering
	\sidesubfloat[]{\includegraphics[width=0.45\linewidth]{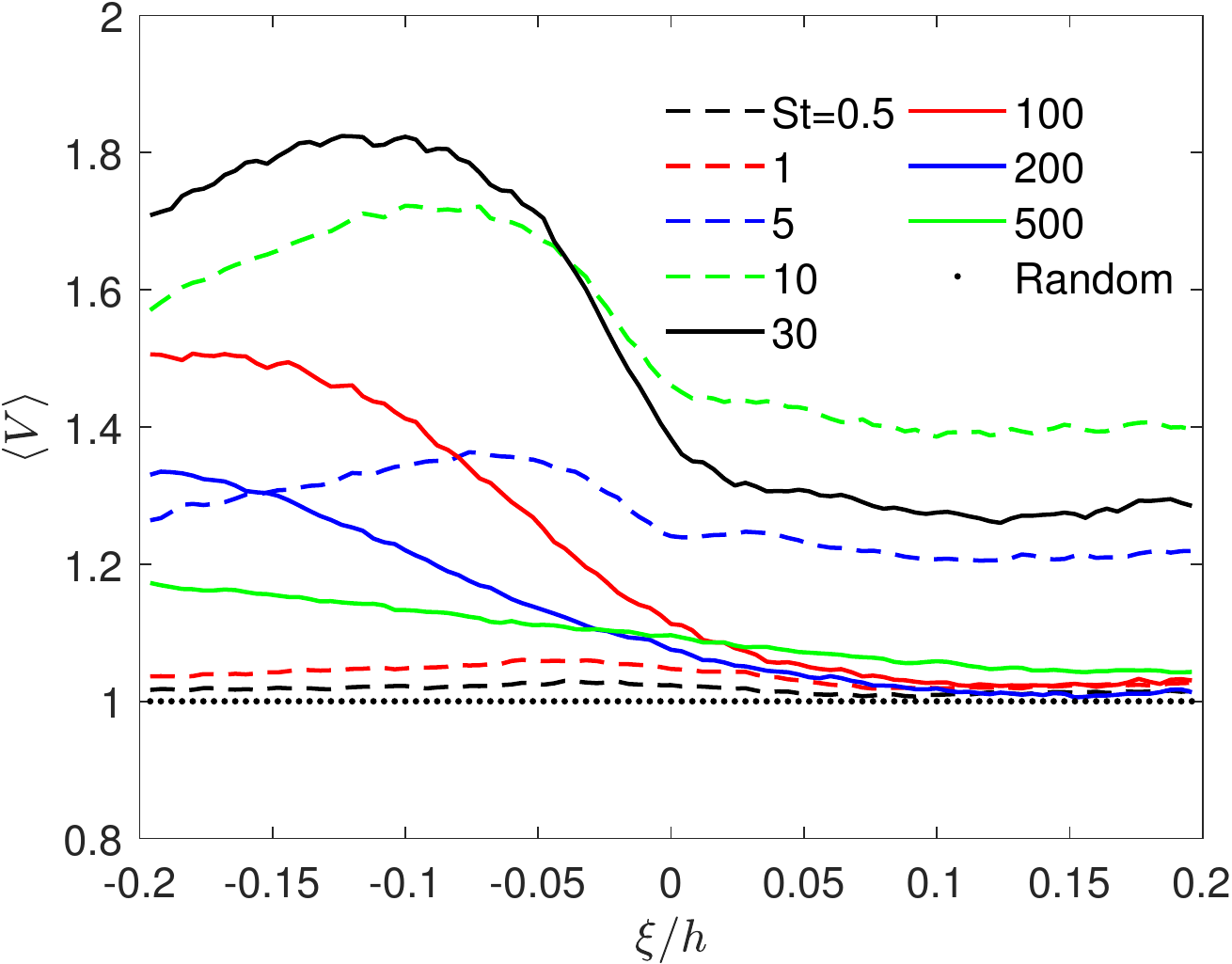}}
	\sidesubfloat[]{\includegraphics[width=0.45\linewidth]{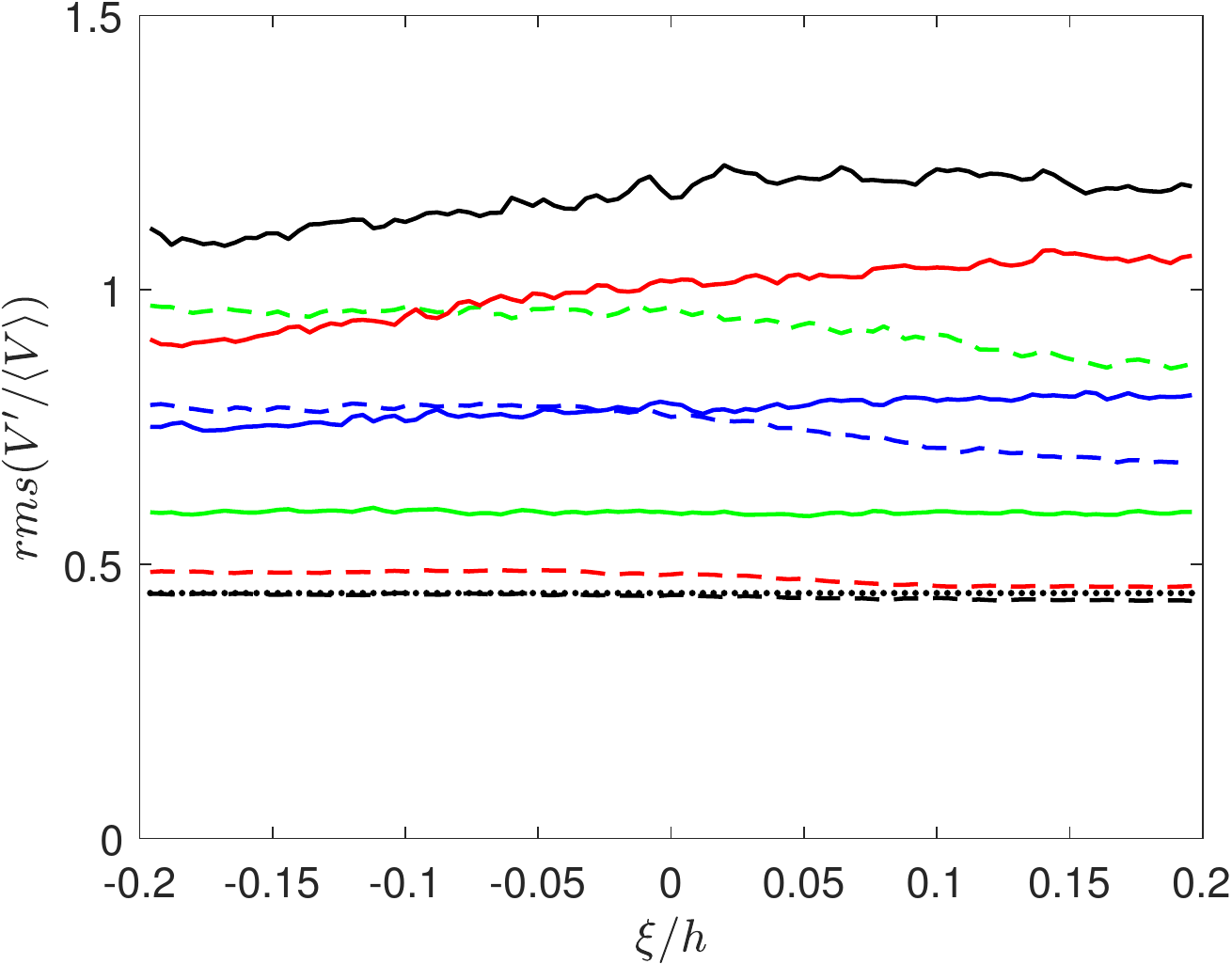}}  \hfill
	\caption{(a) Conditional-averaged Vorono\"i volume of particle Vorono\"i cells around the QC-boundary. (b) Conditional r.m.s. of non-dimensional Vorono\"i volume fluctuation around the QC-boundary. Here, the Vorono\"i volume is nondimensionalized so that $V=1$ represents a random distribution (the same as the particles' initial condition).}	
	\label{fig:VoroV}
\end{figure}

The flow features in and outside the QC region are quite different and an abrupt change of flow features is observed near the QC-boundary \citep{kwon_quiescent_2014,yang_structural_2016}. For example, the streamwise velocity $\langle u_f \rangle$ changes rapidly near the QC-boundary, as shown in figure \ref{fig:mU_rmsUF}(a), and this results in a local peak of the spanwise vorticity \citep{kwon_quiescent_2014, yang_structural_2016}. In the present context, the QC-boundary is expected to influence the distribution of inertial particles. Therefore, three-dimensional volumes of particle Vorono\"i cells are calculated to quantify the particle distribution. The conditional-averaged Vorono\"i volume and conditional root-mean-square of Vorono\"i volume in the QC frame are plotted in figure \ref{fig:VoroV}. Figure \ref{fig:VoroV}(a) indicates that the number of particles in $-0.2 \leq \xi/h \leq 0.2$ are all lower than those of the initial random distribution. The particle distribution in the QC is nearly constant and independent of the location $\xi$, while an abrupt reduction appears near the boundary for intermediate $St$ particles. For particles with $St\geq 10$, the mean Vorono\"i volume $\langle V \rangle$ outside the QC is larger than that inside the QC, which is consistent with the particle accumulation in and outside the QC in figure \ref{fig:V_IO}. These particles prefer to locate in the QC rather than outside the QC in the present QC frame. The r.m.s. value of the Vorono\"i volume fluctuations in figure \ref{fig:VoroV}(b) shows the variance of Vorono\"i cells in a wall-parallel slab and thus reflects the degree of spatial particle clustering in the slab \citep{nilsen_voronoi_2013}. Except the black-dashed line ($St=0.5$), the other solid lines and the dashed lines are different from the dotted line, which represents a random particle distribution, thereby revealing that clustering exists for all these sorts of particles with $St > 0.5$. The topmost line is the black-solid line, which means that local clustering close to QC-boundary is most prominent for particles with $St = 30$. Moreover, it is observed that the r.m.s.s change only slightly across the QC-boundary. However, the tendency of local clustering of particles depends on the Stokes numbers such that the clustering degree of particles with $St \leq 10$ slightly attenuates while those with $St \geq 30$ increase slowly with $\xi$.

\begin{figure}
	\centering
	\sidesubfloat[]{\includegraphics[width=0.45\linewidth]{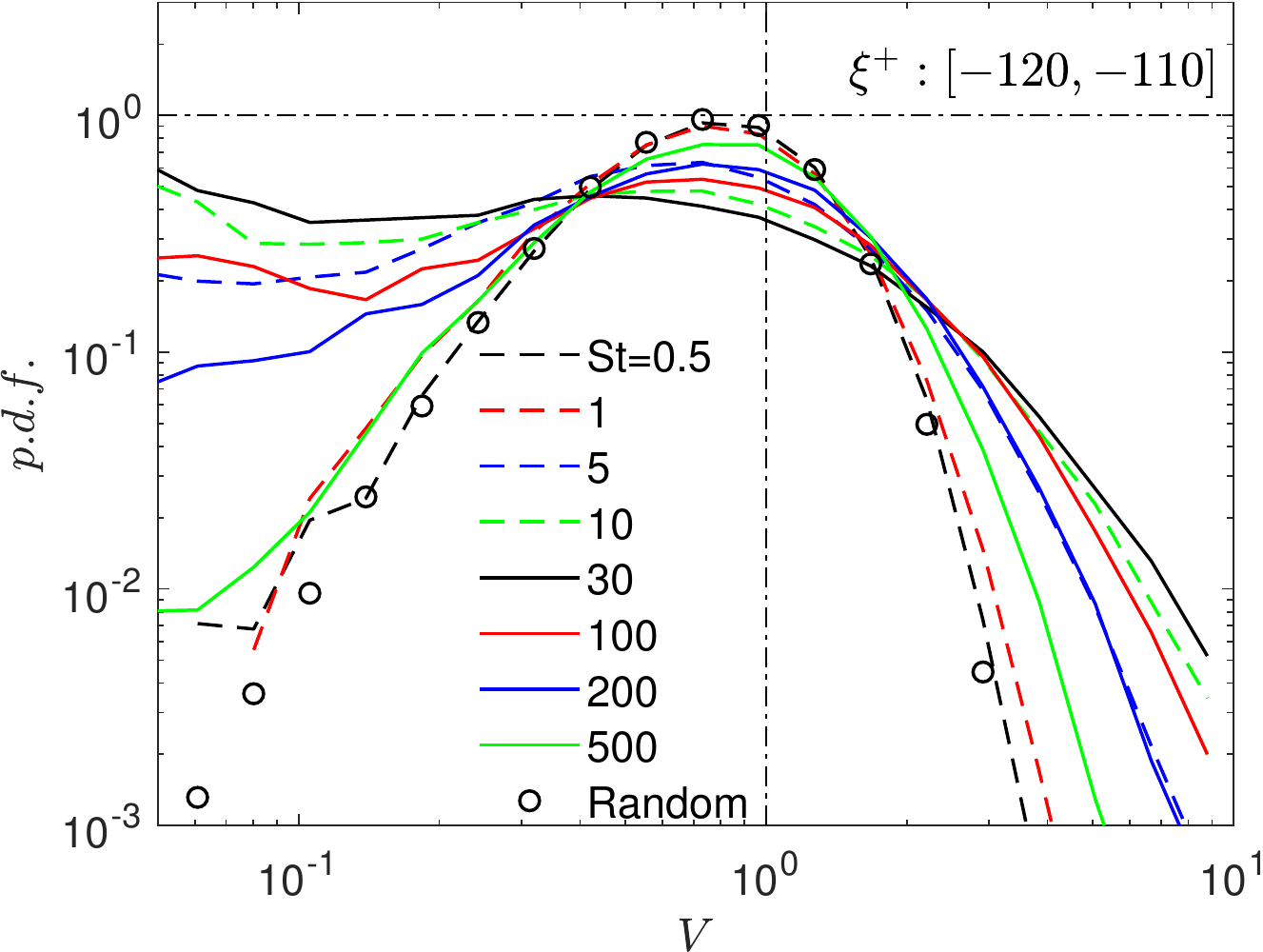}}
	\sidesubfloat[]{\includegraphics[width=0.45\linewidth]{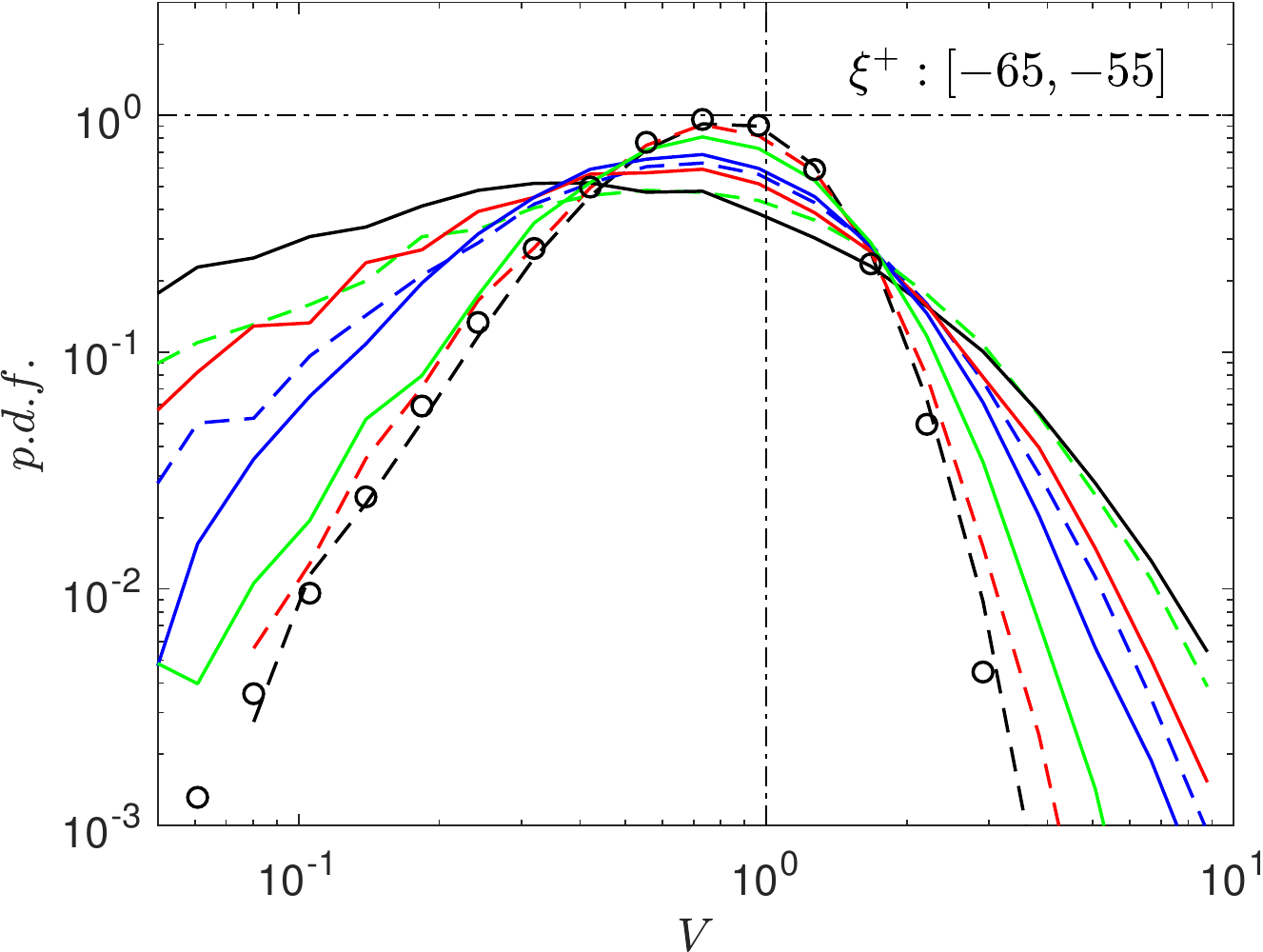}}  \hfill
	\\
	\sidesubfloat[]{\includegraphics[width=0.45\linewidth]{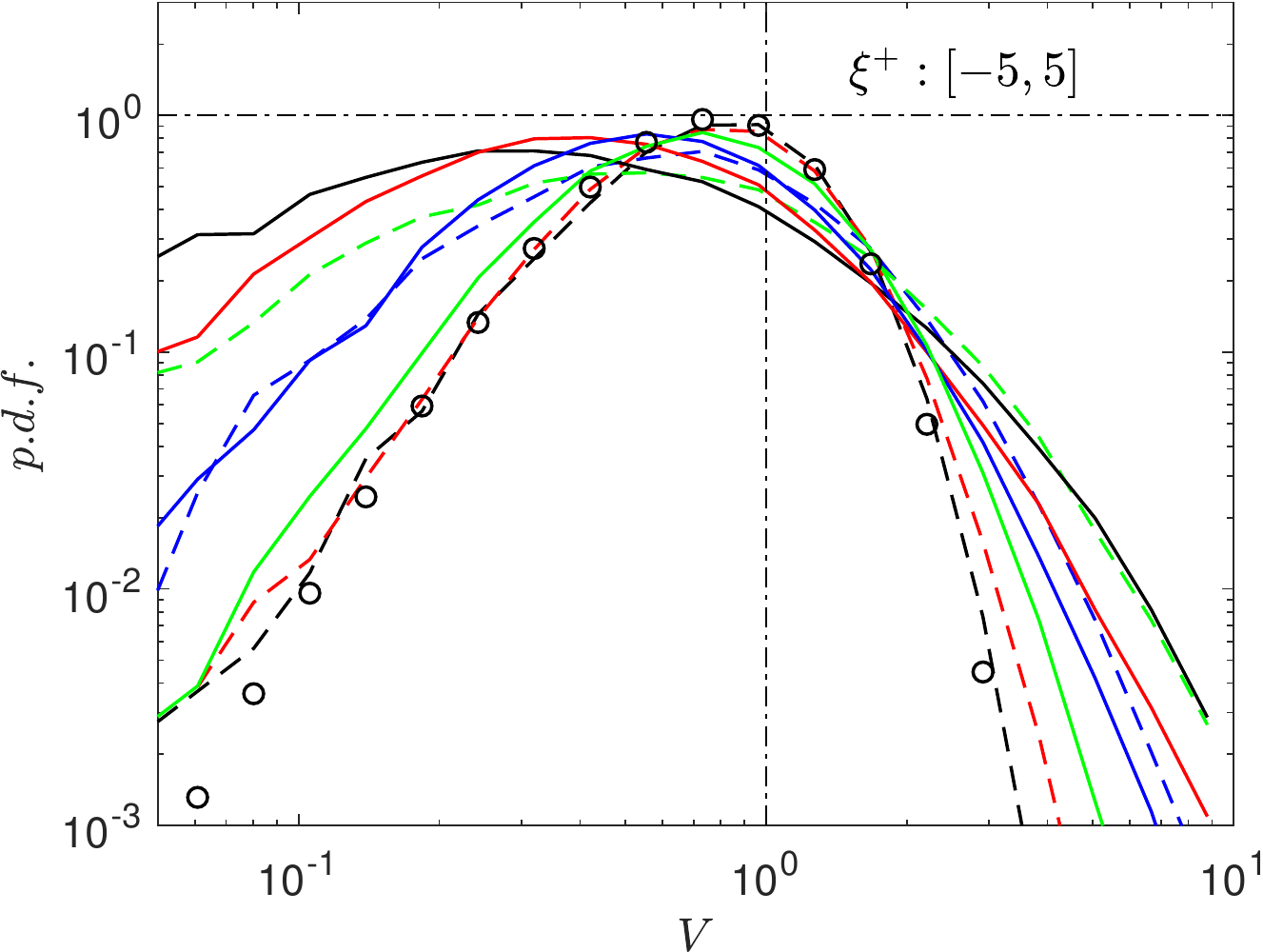}}
	\sidesubfloat[]{\includegraphics[width=0.45\linewidth]{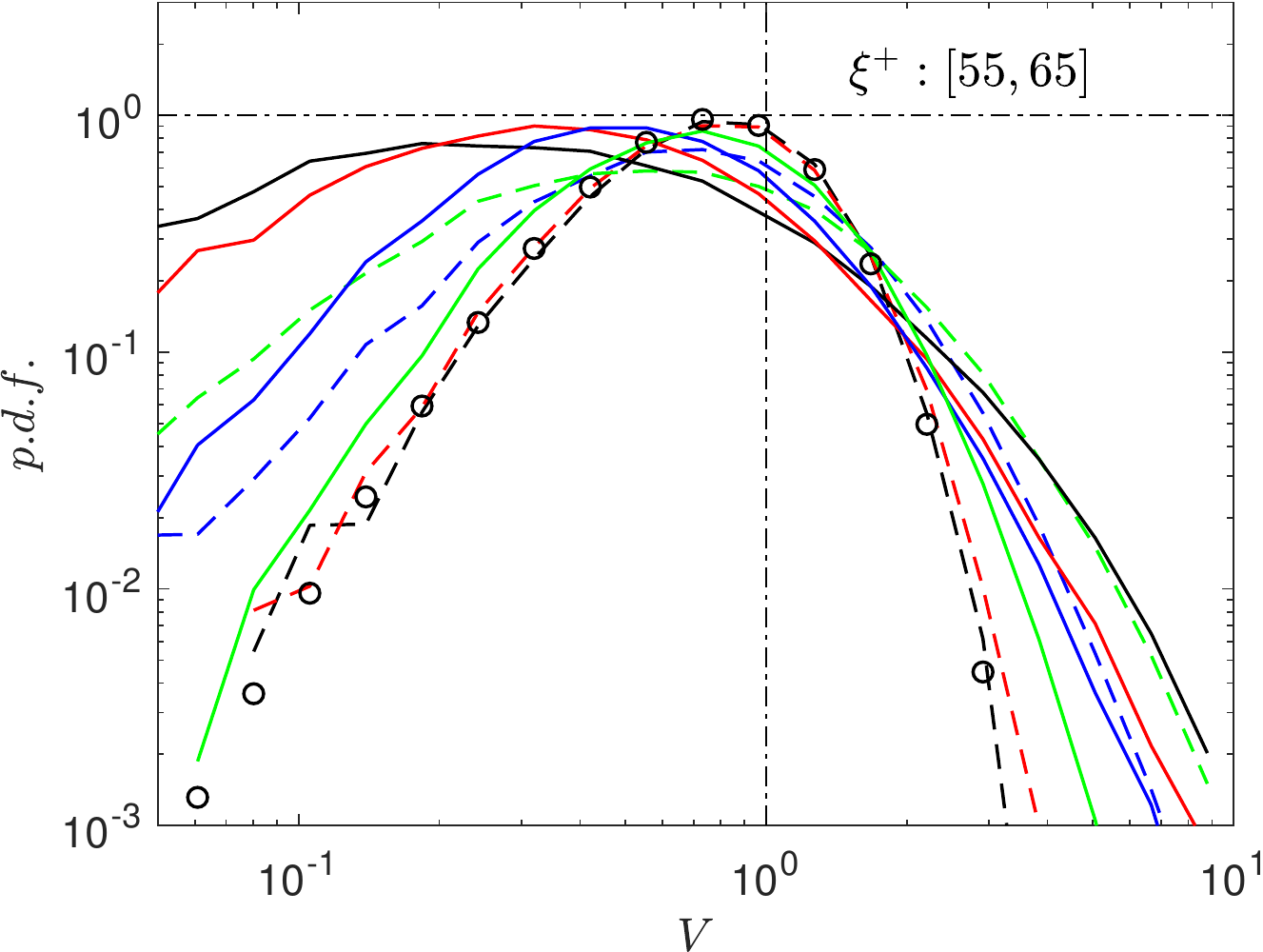}}  \hfill
	\\
	\sidesubfloat[]{\includegraphics[width=0.45\linewidth]{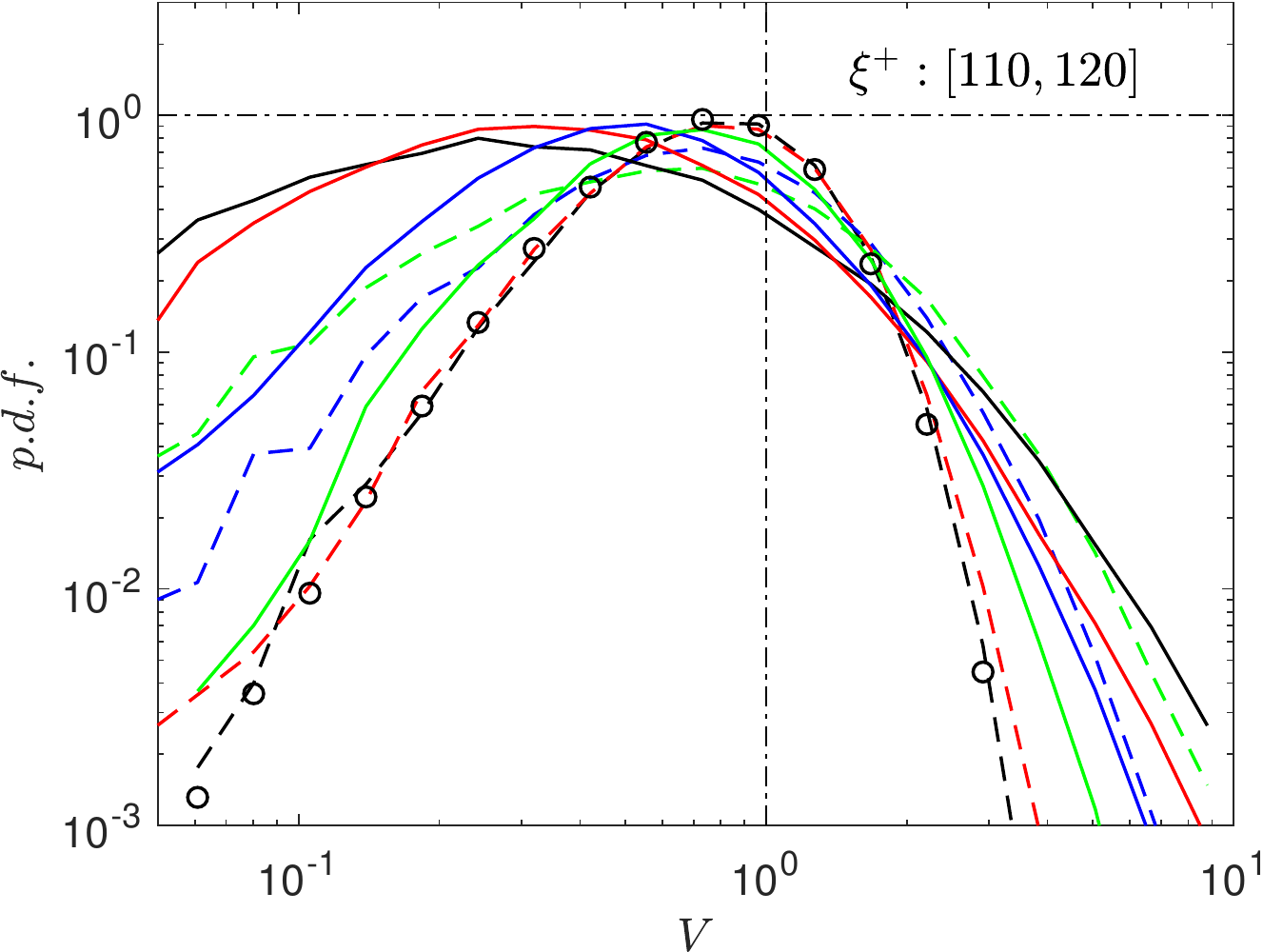}}  \hfill
	\caption{P.d.f.s of Vorono\"i volumes of particles in five different sections in the QC frame. (a) $\xi^+:[-120, -110]$; (b) $\xi^+:[-65, -55]$; (c) $\xi^+:[-5, 5]$; (d) $\xi^+:[55, 65]$; (e) $\xi^+:[110, 120]$.}
	\label{fig:pdf_V}
\end{figure}

The p.d.f.s of three-dimensional Vorono\"i volumes in different sections in the QC frame are shown in figure \ref{fig:pdf_V}, in which the black circles stand for the p.d.f. of Vorono\"i volumes of randomly distributed Vorono\"i cells \citep{ferenc_size_2007}. The differences between the circles and the lines reveal the degree of clustering in a slab of $\xi$. The clustering of $St = 30$ particles is most distinct while the clustering of $St = 100$ particles becomes significant with increasing $\xi$. The black-dashed line and the black circles almost collapse because the inertia effect becomes negligible for $St = 0.5$ particles. The green-solid line is quite close to the black circles due to the strong inertia effect of the $St = 500$ particles. These ballistic particles are more likely to ignore small-scale fluctuations of the fluid velocity. Figure \ref{fig:pdf_V}(a) and (b) show the p.d.f.s of Vorono\"i volumes outside the QC, while figure \ref{fig:pdf_V}(d) and (e) show the p.d.f.s inside the QC. The curves for the same $St$ in figure \ref{fig:pdf_V}(a) and (b) are different when $V$ is small. However, the plots in figure \ref{fig:pdf_V}(d) and (e) are almost identical, indicating that the particle distribution in the QC seems to be independent of the actual location $\xi$. In the QC, flow properties, such as mean spanwise vorticity and shear stress, remain nearly constant as $\xi$ changes \citep{kwon_quiescent_2014}. This tends to make the distribution of these p.d.f.s in the QC almost identical.

\begin{figure}
	\centering
	\sidesubfloat[]{\includegraphics[width=1.0\linewidth]{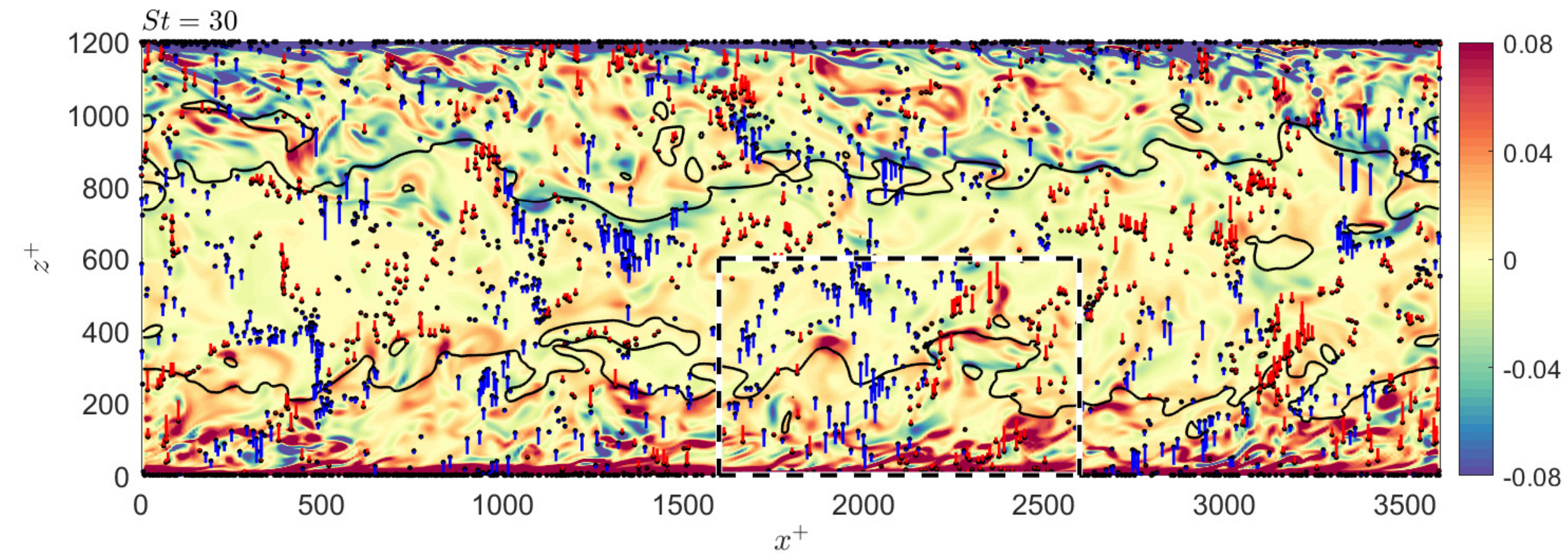}} \\
	\sidesubfloat[]{\includegraphics[width=0.49\linewidth]{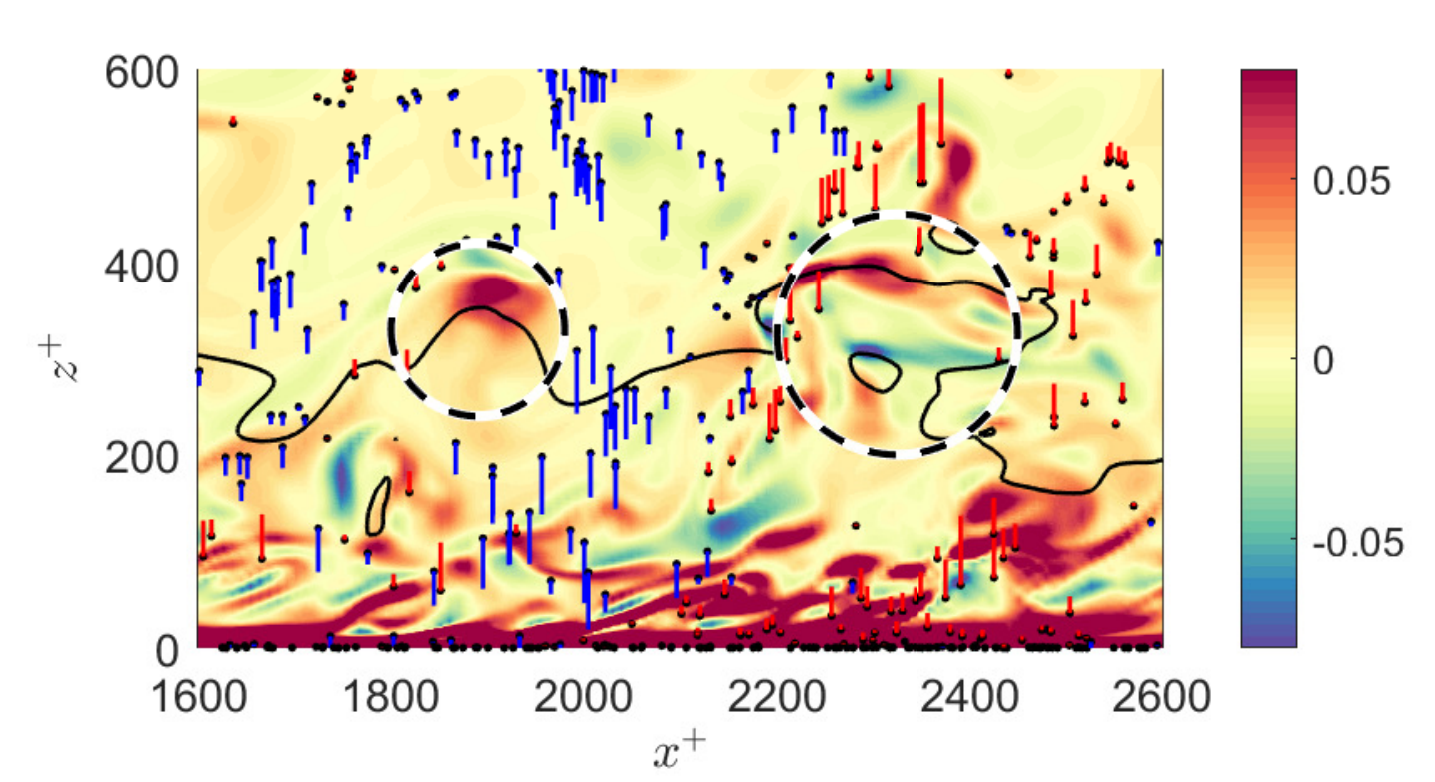}}
	\sidesubfloat[]{\includegraphics[width=0.49\linewidth]{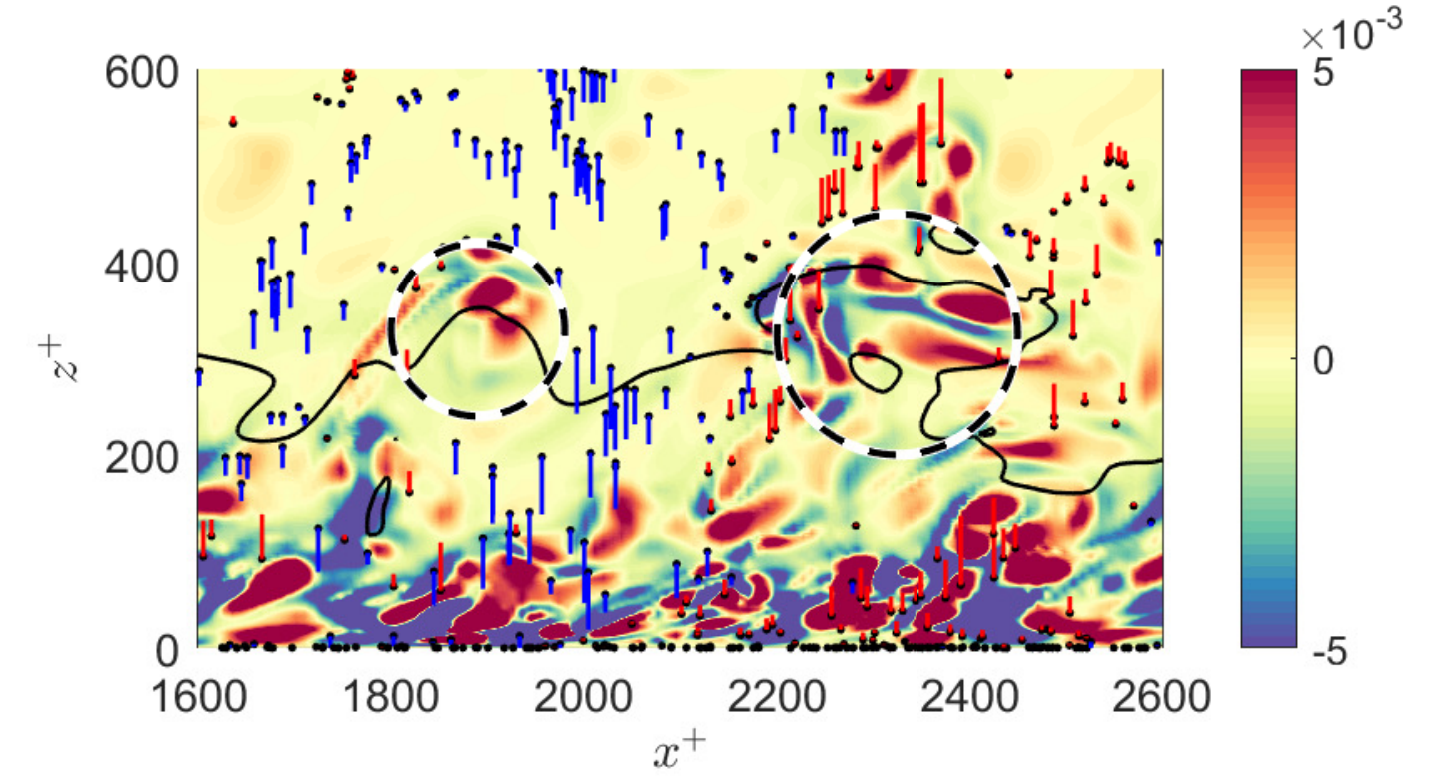}}
	\caption{Instantaneous distribution of particles with $St=30$ in a $x$-$z$ plane at $y^+\approx 703$. The background color represents (a,b) spanwise vorticity $\omega_y^+$, and (c) $Q$ value of fluid velocity field. The black lines show the QC-boundary. Both (b) and (c) are local snapshots of the black-and-white-dashed rectangular area in (a).	A black point stands for a particle and the bar connected with the point shows the magnitude of wall-normal velocity relatively. A red bar denotes positive wall-normal velocity while a blue bar represents negative one.}
	\label{fig:wyInXZ}
\end{figure}

\begin{figure}
	\centering
	\sidesubfloat[]{\includegraphics[width=0.49\linewidth]{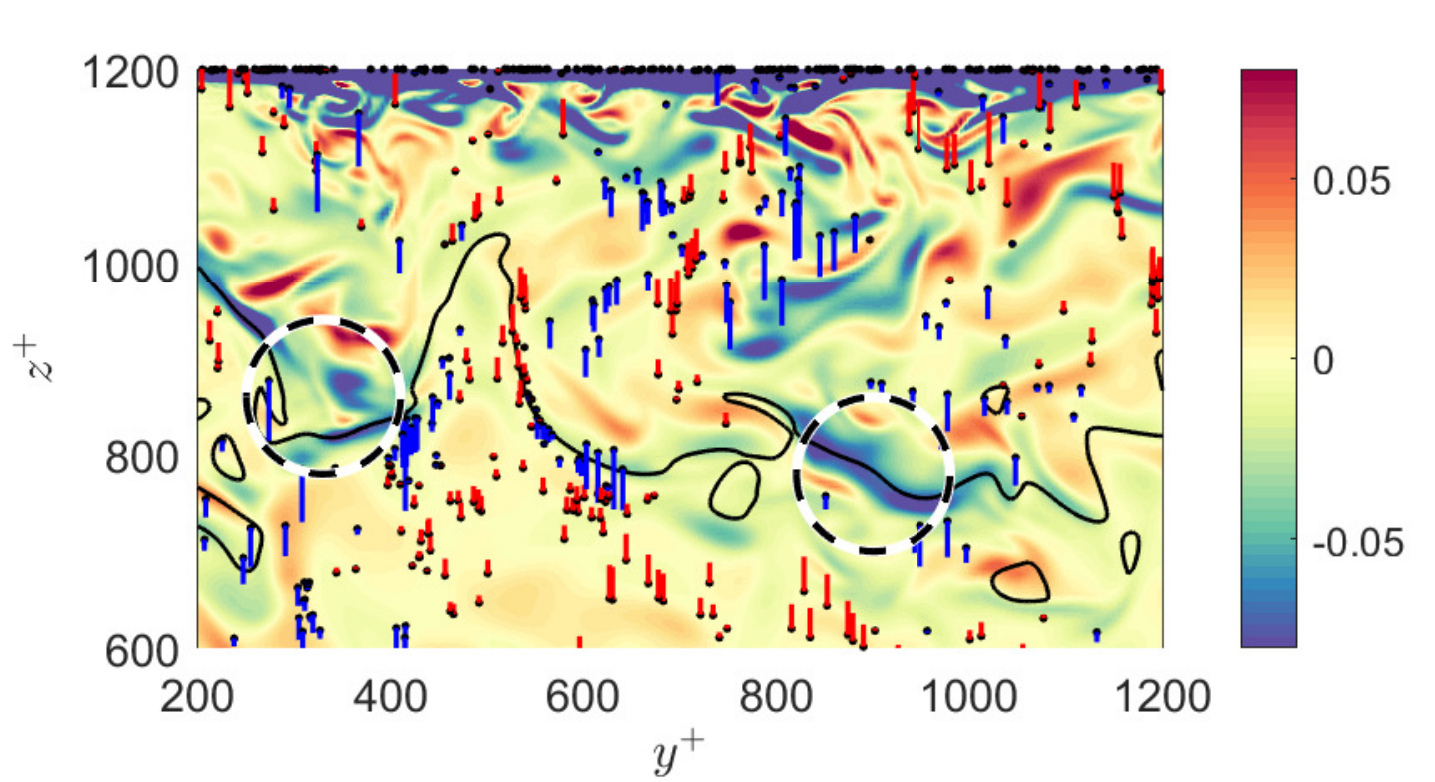}}
	\sidesubfloat[]{\includegraphics[width=0.49\linewidth]{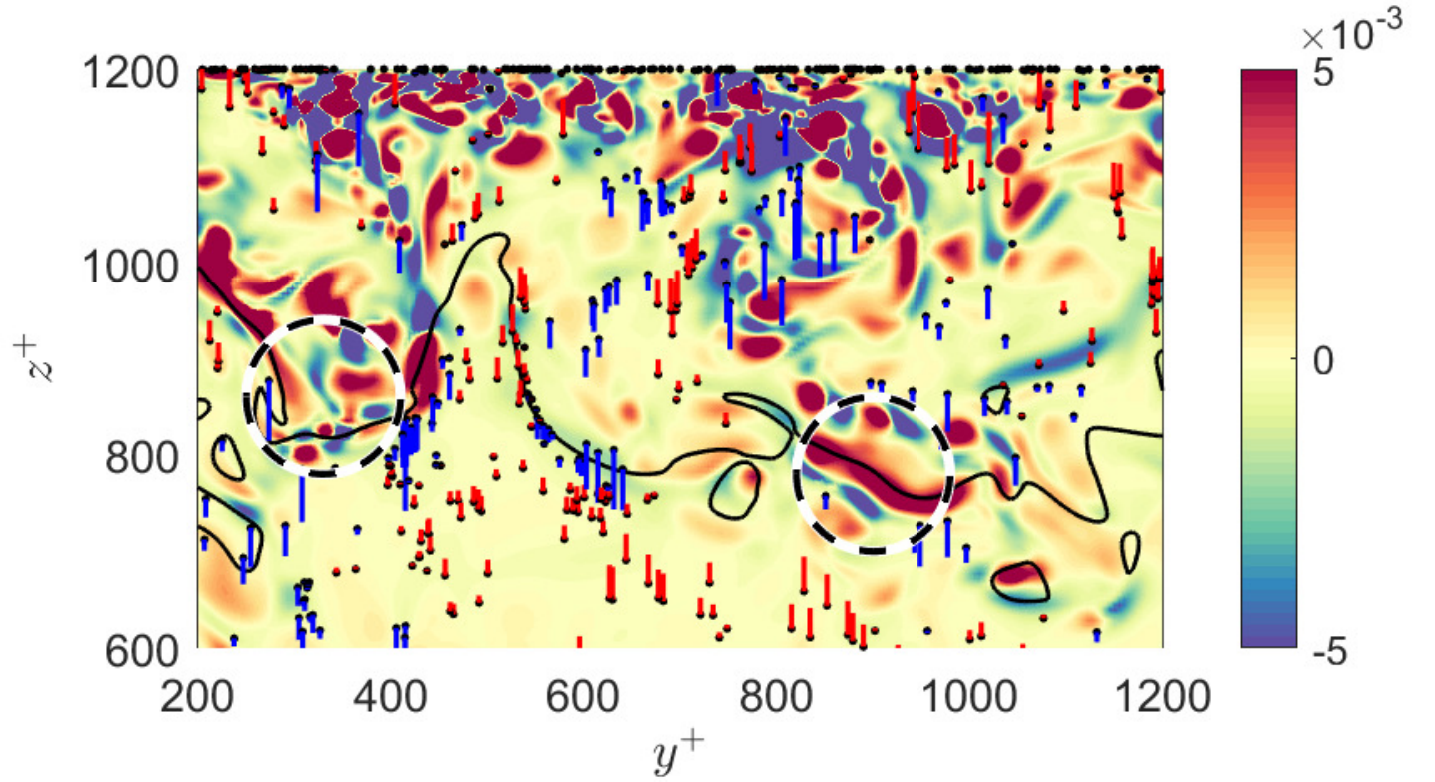}}
	\caption{Instantaneous distribution of particles with $St=30$ in a $y$-$z$ plane at $x^+ \approx 1594$. The background color represents (a) spanwise vorticity $\omega_y^+$, and (b) $Q$ value of fluid velocity field. The black lines show the QC-boundary. A black point stands for a particle and the bar connected with the point shows the magnitude of wall-normal velocity. A red bar denotes positive wall-normal velocity while a blue bar represents negative one.}
	\label{fig:wyInYZ}
\end{figure}

\begin{figure}
	\centering
	\sidesubfloat[]{\includegraphics[width=0.45\linewidth]{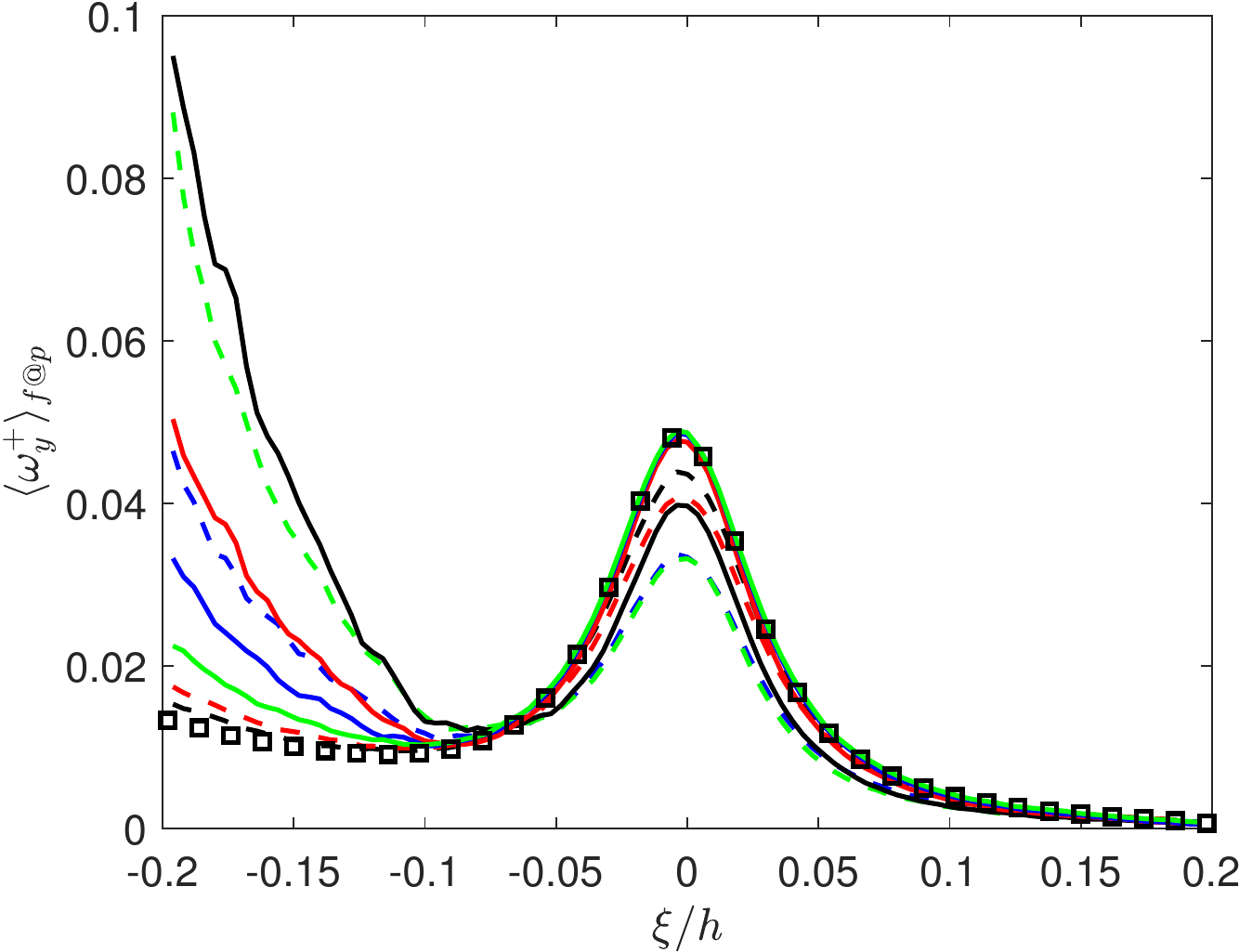}} 
	\sidesubfloat[]{\includegraphics[width=0.45\linewidth]{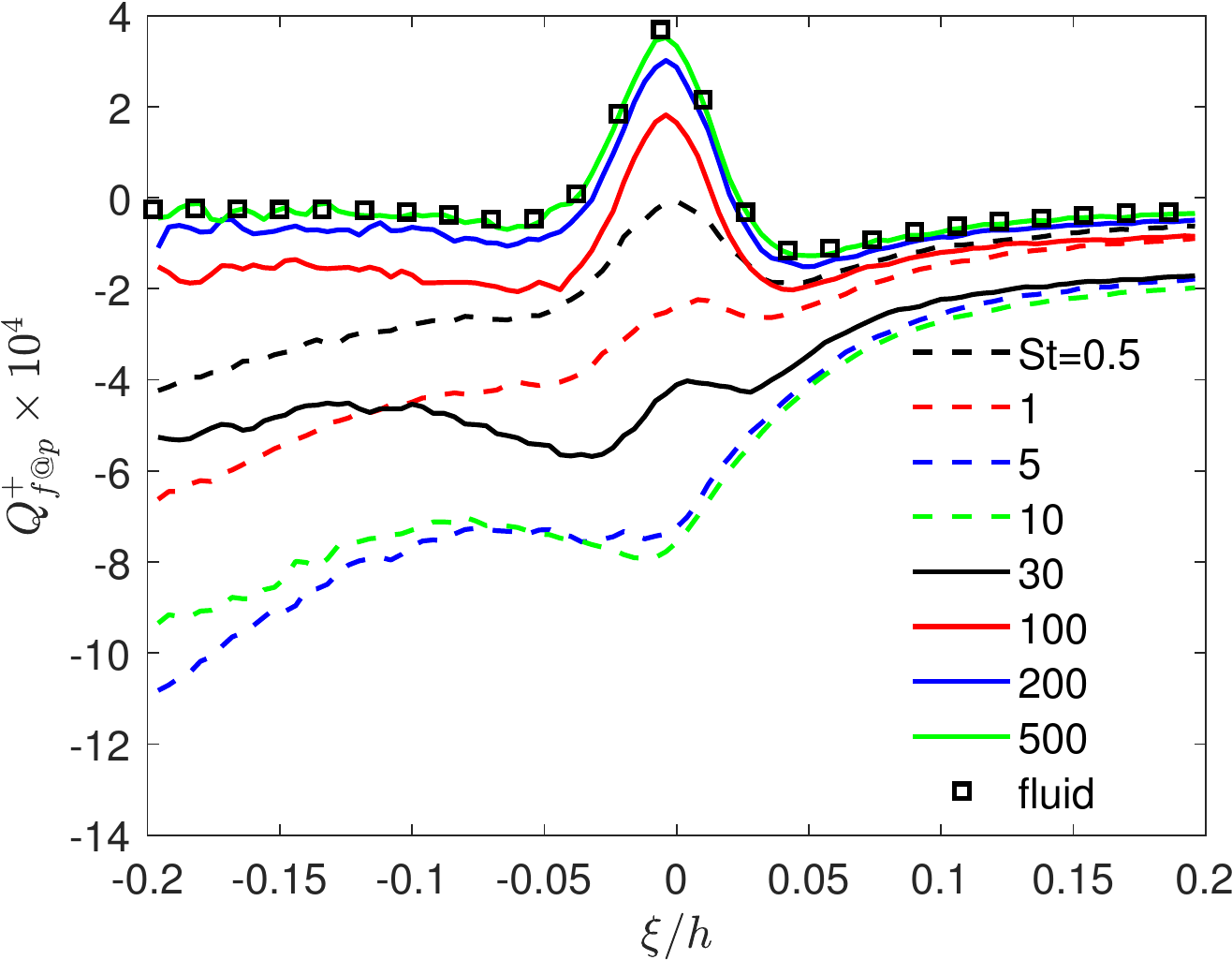}}	\hfill
	\\
	\sidesubfloat[]{\includegraphics[width=0.45\linewidth]{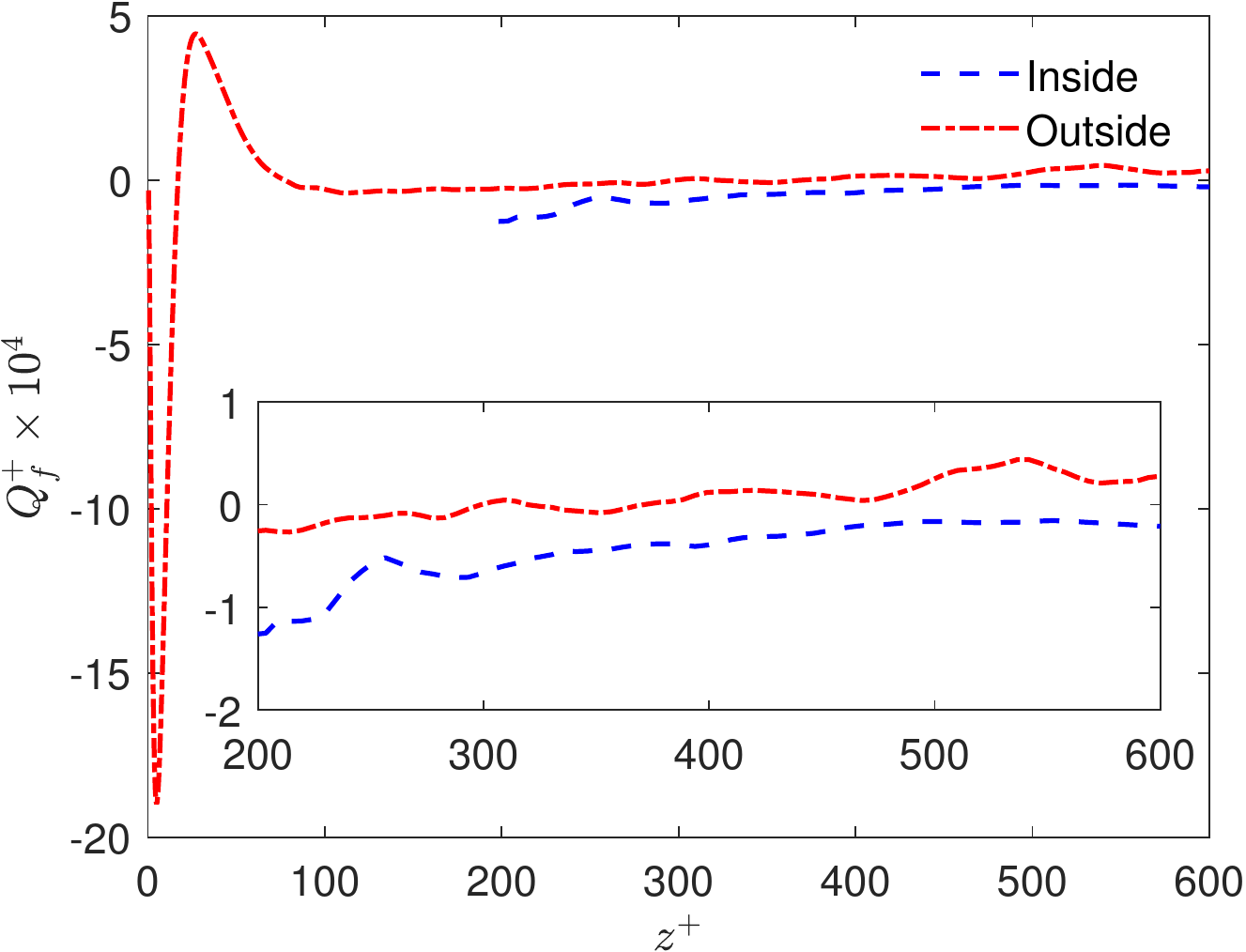}}	
	\caption{ (a) Conditional-averaged spanwise vorticity $\omega_y^+$ of fluid at particle position and $\omega_y^+$ of fluid around the QC-boundary. (b) Conditional-averaged $Q$ of fluid at particle position and $Q$ of fluid around the QC-boundary. (c) Conditional-averaged $Q$ inside the QC (the blue-dashed line) and that outside the QC (the red-dashed-dotted line) versus wall-normal distance $z^+$.}
	\label{fig:Q}
\end{figure}

According to figures \ref{fig:V_IO} and \ref{fig:VoroV} we observed that more particles accumulate in the QC than outside the QC. We suspect that the presence of the QC boundary hinders the transport of particles and induces an uneven distribution of particles in and outside of the QC. We interpret this observation as follows: In the QC-frame \citet{yang_structural_2016} demonstrated that there is a local maximum of the densities of prograde and retrograde vortices near the boundary of a QC, which is associated with large-scale flow structures and characterized by large-scale bulges and valleys \citep{yang_influence_2019}. More specifically, the QC consists of uniform momentum separated from the wall region by shear layers with sharp velocity gradients \citep{kwon_quiescent_2014}, which, we anticipate, may affect the transport of particles because of a centrifuging effect induced by the shear.

Therefore, we plot an instantaneous distribution of particles with $St=30$ in a $x$-$z$ plane, background colored by the spanwise vorticity $\omega_y^+$ (see figures \ref{fig:wyInXZ}a and \ref{fig:wyInXZ}b). Prominent spanwise vorticity is shown inside the two black-and-white-dashed circles in figure \ref{fig:wyInXZ}(b), and particles are rarely located inside the circles. Outside of the circles, however, there are a few particles locate around the QC-boundary. The particles apparently prefer to avoid regions with high spanwise vorticity. On the other hand, the local snapshot of the particles' location and movement is also shown in figure \ref{fig:wyInXZ}(c) but the background color shows the instantaneous $Q$-value of the fluid velocity field. Red regions show the flow areas with high vorticity but low strain rate, while blue regions represent the flow with high strain rate but low vorticity. High $Q$-values in the red regions in figure \ref{fig:wyInXZ}(c) is mainly caused by high spanwise vorticity in figure \ref{fig:wyInXZ}(b). As a consequence, particles tend to avoid regions with vortices, namely the regions with high $Q$, around the QC-boundary. The presence of strong vortices around the QC-boundary hinders the wall-normal transport of particles and the QC-boundary acts as a barrier. 

Figure \ref{fig:wyInYZ} shows the instantaneous distribution of particles with $St = 30$ in a $y$-$z$ plane. The two circles in panel (a) display regions with high negative spanwise vorticity, which contributes to the red regions in panel (b). Vorticity rather than strain rate is dominant in these red regions. Particles are rarely found inside these two black-and-white-dashed circles, which suggests that inertial particles with $St = 30$ are trying to avoid vortices, as observed also in figure \ref{fig:wyInXZ}.

To quantitatively examine the barrier effect of the QC boundary, conditional-averaged spanwise vorticity and $Q$ of the fluid at the particle positions around the QC-boundary are computed and presented in figure \ref{fig:Q}(a) and (b), respectively. As shown in figure \ref{fig:Q}(a), there is a distinct peak of fluid spanwise vortictiy (the black squares) at the QC-boundary, which is expected since there is an abrupt change of streamwise velocity in figure \ref{fig:mU_rmsUF}(a). The lines with $St < 100$ are below the black squares around the QC-boundary, suggesting a local preferential concentration of these particles in regions with lower vorticity. These phenomena can also be seen in figure \ref{fig:Q}(b). The green solid line, representing $St = 500$ particles, and the black squares almost collapse because of the nearly random distribution of particles at ballistic Stokes number. All other lines are lower than the black squares, revealing that particles tend to accumulate in regions with relatively high strain rate and low vorticity, inducing that the local mean value of $Q$ at particle position is lower than the mean value of the unconditioned $Q$. The local peaks of $Q$ and $\omega_y^+$ of the fluid flow are consistent with the instantaneous $Q$-field in figures \ref{fig:wyInXZ} and \ref{fig:wyInYZ}. Around the QC-boundary, the lines of particles with intermediate inertia (e.g. $St = 5, 10, 30$) are quite lower than the black squares, thereby suggesting that particles tend to avoid vortices around the QC-boundary, as visualized in figures \ref{fig:wyInXZ} and \ref{fig:wyInYZ}.

Furthermore, the conditional-averaged $Q$ of the fluid velocity within and outside of the QC are compared (see figure \ref{fig:Q}c). Since the channel center occasionally can be out of the QC, the red-dash-dotted line is not overlapping with blue-dashed line. From $z^+=200$ to $z^+=600$, the red-dash-dotted line is always above the blue-dashed line, suggesting that there are more vortices outside the QC than within the QC even at the same wall-normal position. The blue-dashed line is consistently below zero, revealing that strain rate rather than rotation rate is dominant in the QC. Since inertial particles tend to accumulate in regions with low vorticity but high strain rate, more particles are found in the QC rather than outside of it at a given wall-normal distance, as shown in figures \ref{fig:V_IO} and \ref{fig:VoroV}.

\begin{figure}
	\centering
	\sidesubfloat[]{\includegraphics[width=0.45\linewidth]{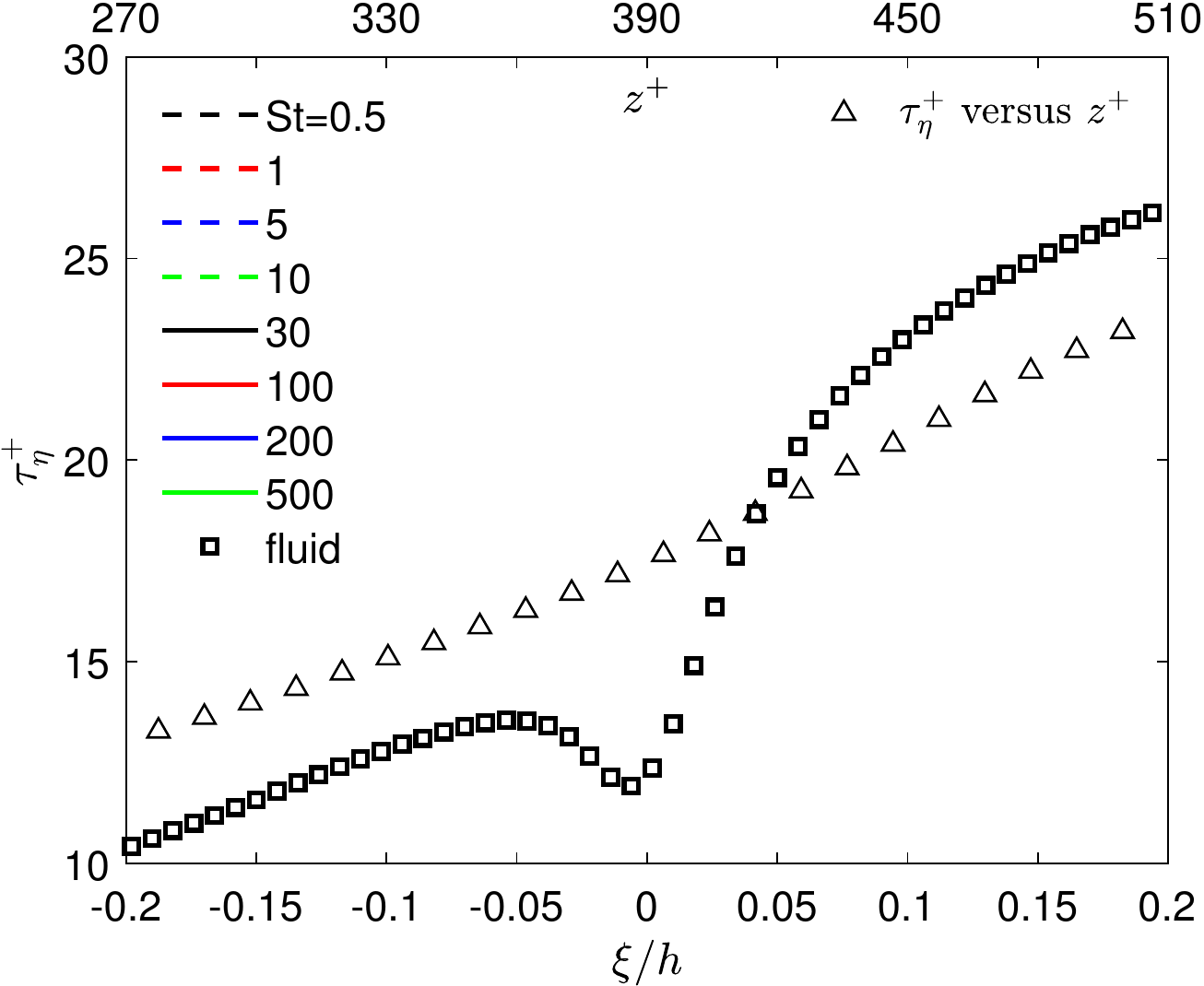}}
	\sidesubfloat[]{\includegraphics[width=0.45\linewidth]{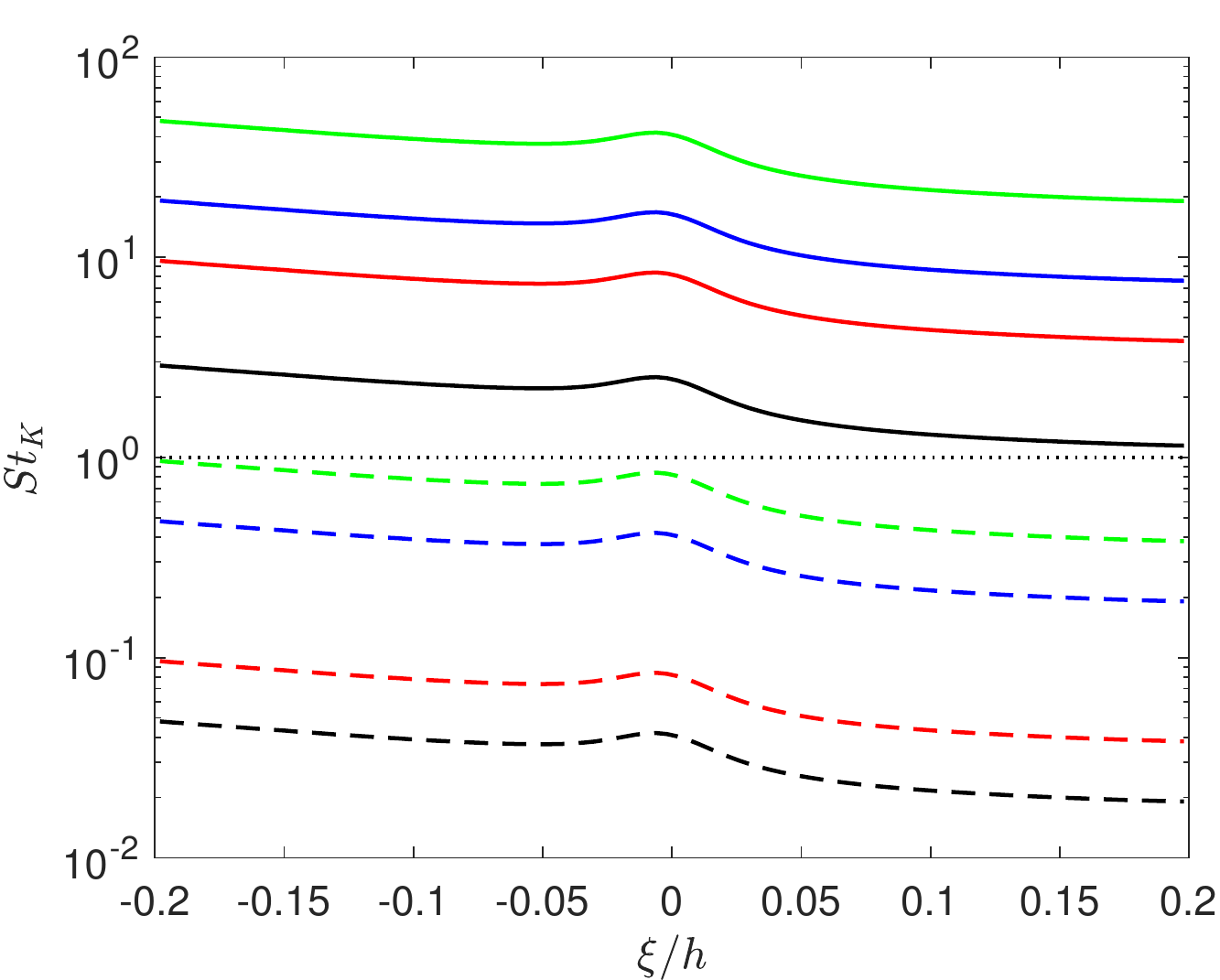}}  \hfill
	\caption{(a) Conditional-averaged Kolmogorov time scale $\tau_\eta^+$ versus $\xi/h$ around the QC-boundary (the squares) and the unconditioned Kolmogorov time scale versus wall-normal distance $z^+$ (the black triangles). (b) Conditional-averaged local $St_K$ based on the local Kolmogorov time scale around the QC-boundary.}
	\label{fig:tau}
\end{figure}

The $Q$-value of the fluid velocity field at particle positions is the smallest for $St = 5$ or $10$ particles (see figure \ref{fig:Q}b). The spanwise fluid vorticity $\omega_y$ at particle positions for particles with $St = 5$ and $St = 10$ is also lowest. This is related to the local $St_K$, which is the Stokes number based on the local Kolmogorov time scale $\tau_\eta = \sqrt{\nu/\epsilon} $, where $\epsilon$ denotes the energy dissipation rate. Note that $St_K = \tau_p / \tau_\eta = St / \tau_\eta^+$ shows the relationship between the two Stokes numbers. The conditional-averaged Kolmogorov time scale $\tau_\eta^+$ versus $\xi/h$ around the QC-boundary is plotted in figure \ref{fig:tau}(a), together with the unconditioned Kolmogorov time scale versus the wall-normal distance $z^+$. The conditional-averaged $St_K$ for different particles are shown in figure \ref{fig:tau}(b). As indicated in table \ref{tab:t}, the mean wall-normal distance of the lower QC-boundary is $\mu^+(z_{lower}) \approx 390$. Hence, the center of the upper abscissa is set as $z^+=390$ in figure \ref{fig:tau}(a). The time scale versus $z^+$ increases monotonically and its value is not much different from the conditional-averaged $\tau_\eta^+$ versus $\xi/h$. As shown in the figure, $\xi/h=0.2$ or $(\xi/h)^+=120$ locates close to channel center. The local dip of the squares around $\xi=0$ in figure \ref{fig:tau}(a) corresponds to the local maximum of the turbulent dissipation rate $\epsilon$ around the QC-boundary, which results in the modest local peak in figure \ref{fig:tau}(b). $St_K$ of particles with $St = 5$ and $10$ are of order 1, which means that the response time of these two classes of particles are close to the local Kolmogorov time scale, namely the scale of the smallest turbulent eddies.

It should be noted that clustering near QC-boundary is most prominent for particles with $St = 30$, corresponding to $St_K \approx 1\sim 3$, as shown in figure \ref{fig:tau}(b). This is consistent with previous findings that clustering of particles is most prominent when $St_K\approx 1$ in HIT \citep{eaton_preferential_1994, monchaux_preferential_2010}. However, the Stokes number of particles with the lowest $Q_{f@p}^+$ is about $5$ or $10$ according to figure \ref{fig:Q}, i.e. relatively different from the case of $St = 30$. This may suggest that one can not just use the local $St_K$ and mean $Q_{f@p}^+$ to distinguish the clustering degree of particles in regions where the mean shear is nonnegligible outside of QC.

\subsection{Translation of particles in the QC frame}
\begin{figure}
	\centering
	\sidesubfloat[]{\includegraphics[width=0.45\linewidth]{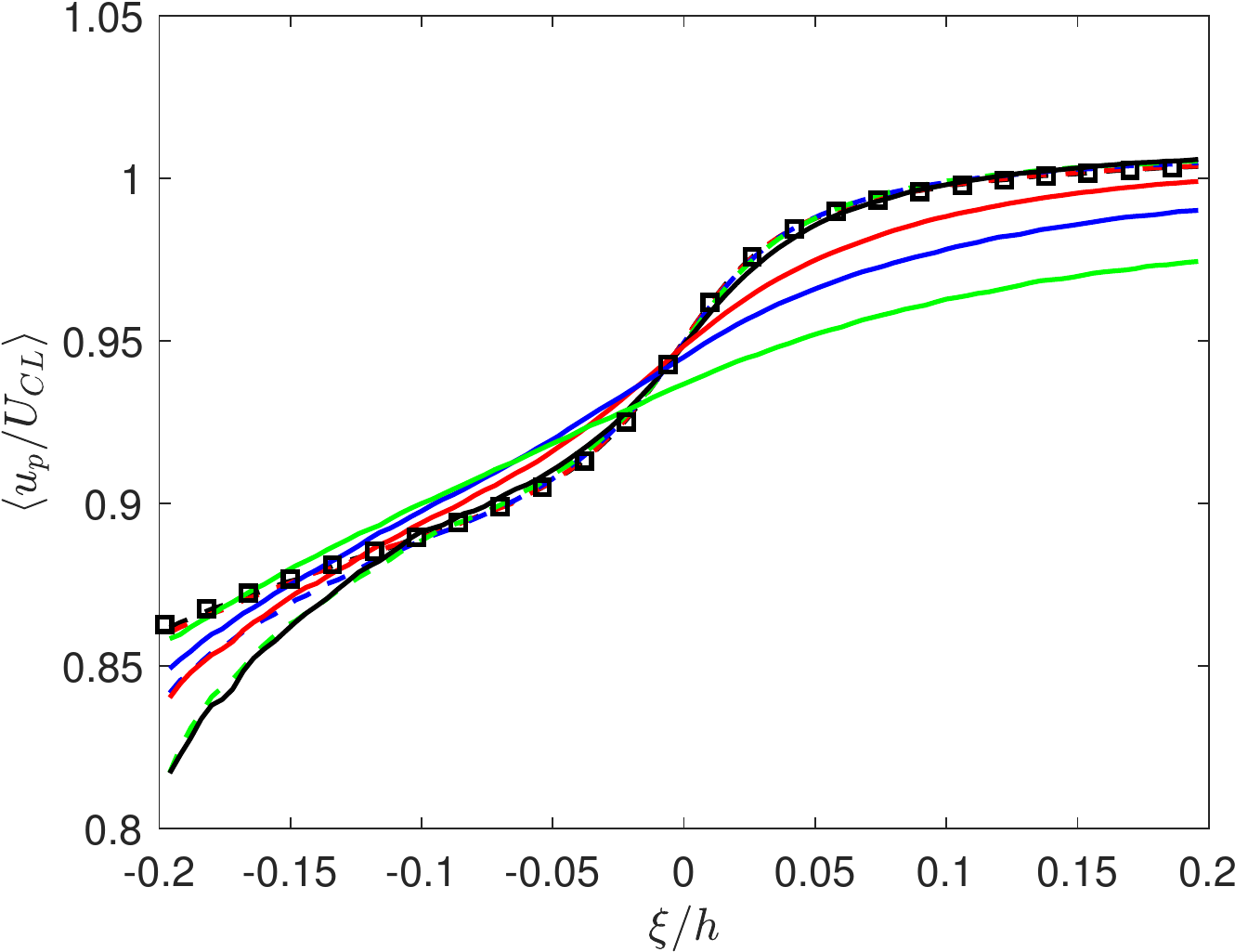}}
	\sidesubfloat[]{\includegraphics[width=0.45\linewidth]{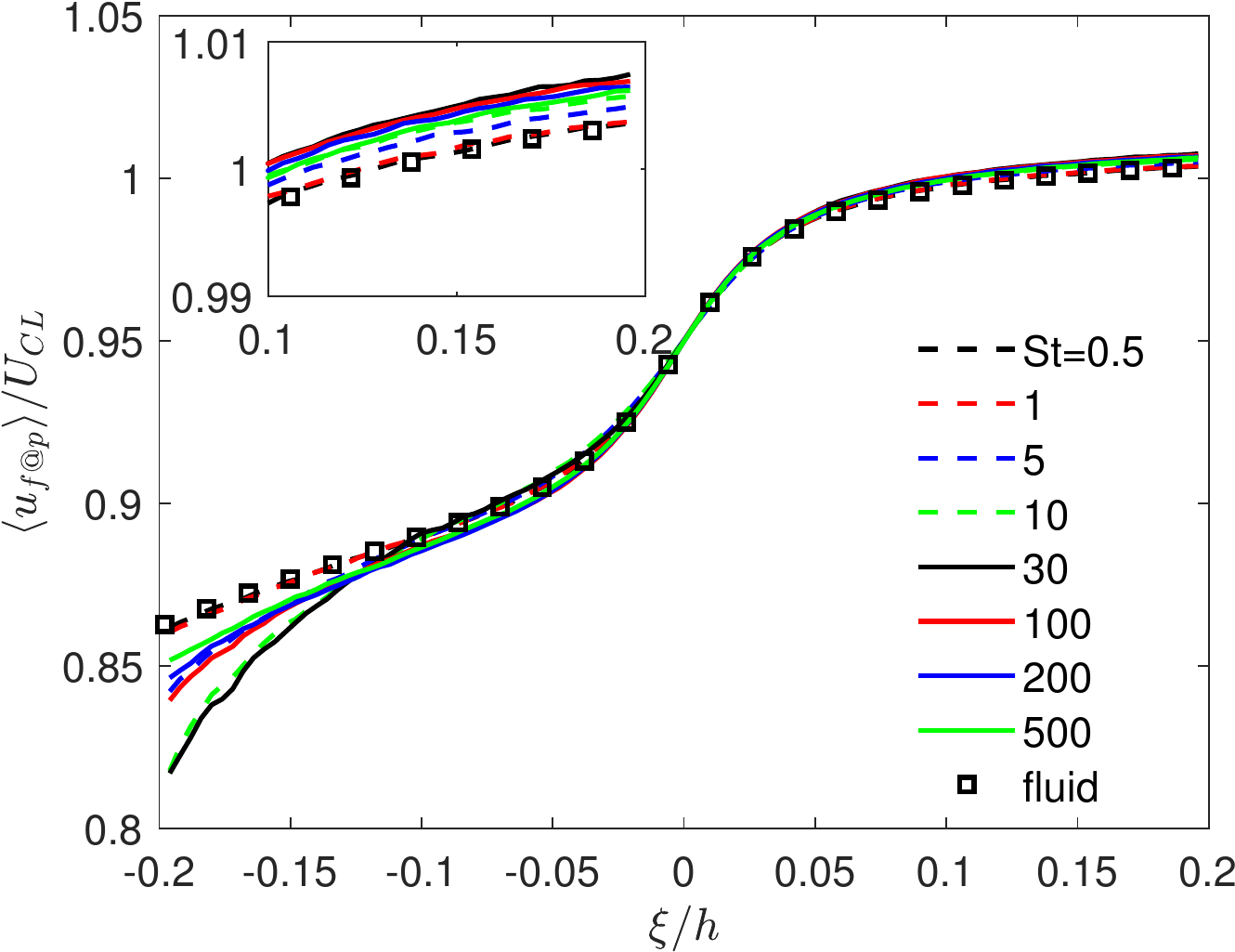}}  \hfill
	\caption{Conditional-averaged streamwise velocity of (a) particles, and (b) fluid at particle positions in the QC frame. Note that the squares in both figures represent the conditional-averaged streamwise velocity of the flow.}
	\label{fig:up}
\end{figure}
As mentioned earlier, the unique characteristics of the flow around the QC-boundary can influence both particle distribution and transport. The latter is discussed in this subsection.

First of all, the conditional-averaged streamwise velocity of particles and fluid at particle location around the QC-boundary are computed. As shown in figure \ref{fig:up}(a), an abrupt jump of the fluid velocity appears near the QC-boundary, while inside the QC the streamwise momentum remains relatively uniform \citep{kwon_quiescent_2014, yang_structural_2016}. The streamwise velocities of particles with $St \leq 30$ are close to that of the fluid, although small discrepancies of the velocity around the QC-boundary exist. However, for particles with $St > 30$, an obvious inertia effect is present and leads to differences of profiles, revealing that heavy particles do not respond to the presence of the QC-boundary and their streamwise velocity increases smoothly with increasing $\xi$. Ballistic particles $St\geq 100$ simply ignore the local shear layer at $\xi=0$ and the velocity gradient $\partial{u_p}/\partial{z}$ is smeared out. The streamwise fluid velocity at particle position and that of fluid are compared in figure \ref{fig:up}(b). There are slight differences which reflect the particle preferential accumulation. For example, in the QC ($\xi>0$), the solid lines and the dashed lines are slightly higher than the black squares, indicating that particles in the QC tend to reside in regions with positive streamwise velocity fluctuations. However, this tendency is quite modest in figure \ref{fig:up}(b).

\begin{figure}
	\centering
	\sidesubfloat[]{\includegraphics[width=0.45\linewidth]{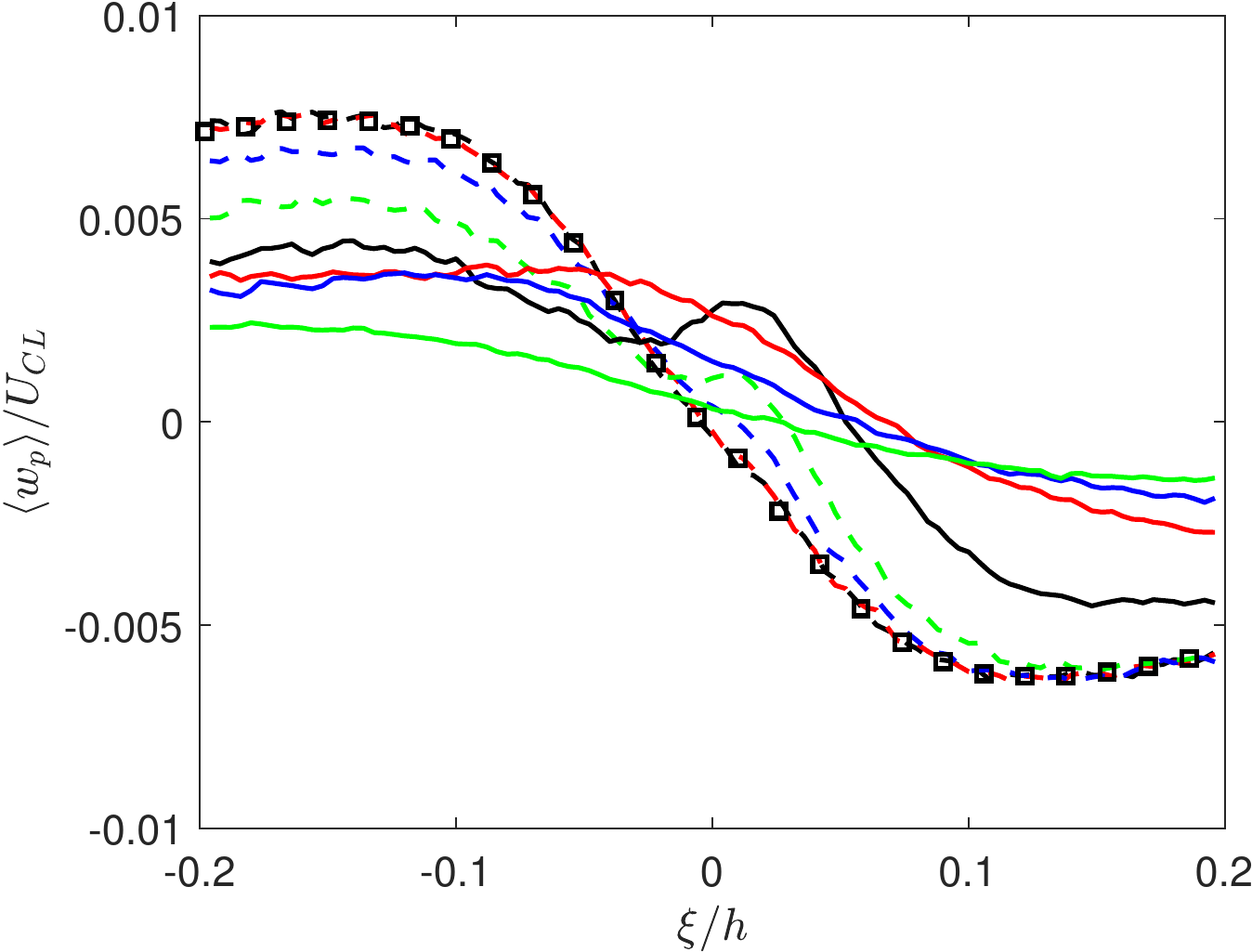}}
	\sidesubfloat[]{\includegraphics[width=0.45\linewidth]{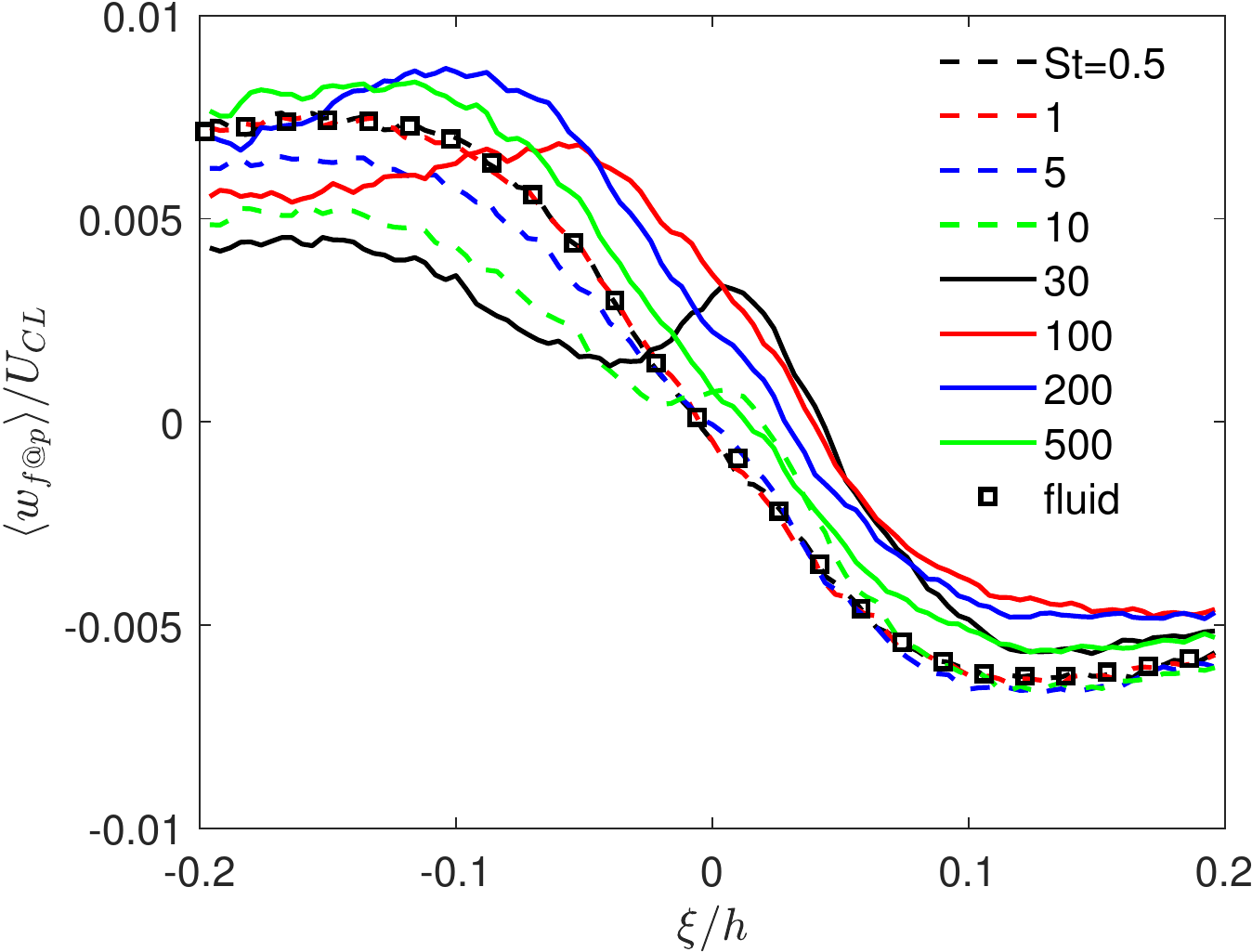}}  \hfill
	\caption{Conditional-averaged wall-normal velocity of (a) particles, and (b) fluid at particle positions in the QC frame.}
	\label{fig:wp}
\end{figure}

The conditional-averaged wall-normal velocity of particles and fluid at particle position around the QC-boundary are presented in figure \ref{fig:wp}. Figure \ref{fig:wp}(a) shows that the wall-normal velocity of the fluid, represented by the black squares, is greater than zero when $\xi<0$ while less than zero when $\xi>0$, which indicates that on average fluid outside the QC prefers to move away from the wall while fluid above QC-boundary moves toward the wall. The same phenomenon of upward and downward motions (referred to as in the lower half of channel) has been observed by \citet{kwon_quiescent_2014} and \citet{yang_influence_2019}. The upward motion taking place outside the QC may be caused by the flow structures originating from the wall, while the downward motion may be owing to the core flow drifting towards the wall \citep{hunt_eddy_2000}. Similar upward motions outside the QC and downward motions inside the QC are observed for inertial particles, but the magnitudes of particles' wall-normal velocity are all lower than that of the fluid, due to the effect of inertia. Particles with smaller Stokes number move faster in the QC frame because they behave more like tracers. In most part of the QC-frame, the absolute value of wall-normal fluid velocity is greater than that of particles. Particles tend to be entrained by fluid upward and downward motions. It should be pointed out that particles with intermediate inertia at $St=10$ and $30$ show local peaks near $\xi = 0$. In figure \ref{fig:wp}(b), the wall-normal velocities of the fluid at the particle position vary non-monotonically with Stokes number, showing different degree of particle preferential clustering. One should note that these findings of particle behavior are on the basis of conditional averages in the QC-frame. Because of the dynamic feature of the meandering QC boundary, the upward motion outside the QC and downward motion inside the QC do not lead to a local maximum particle concentration in the conventional statistics in the inertial frame.

\begin{figure}
	\centering
	\sidesubfloat[]{\includegraphics[width=0.45\linewidth]{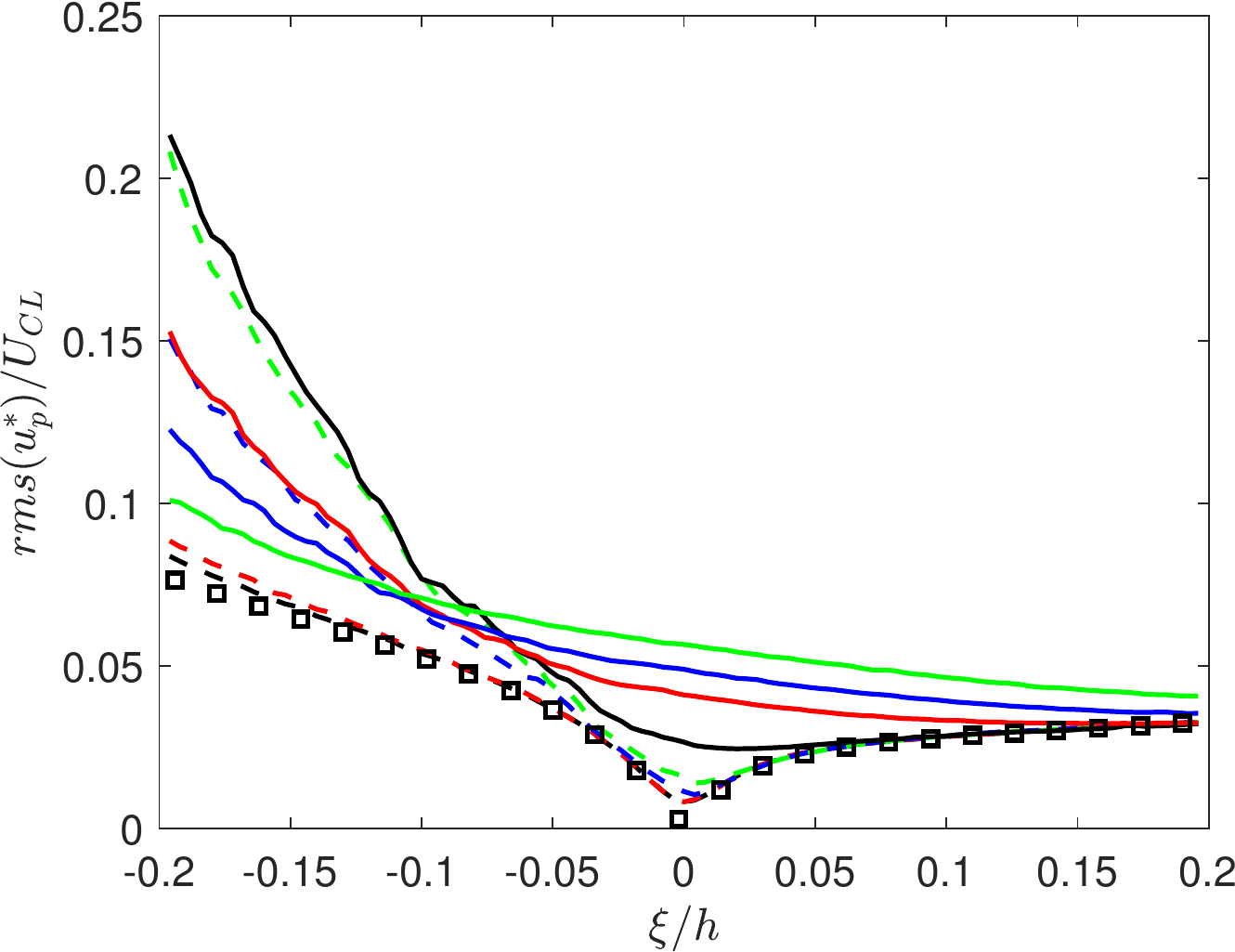}}
	\sidesubfloat[]{\includegraphics[width=0.45\linewidth]{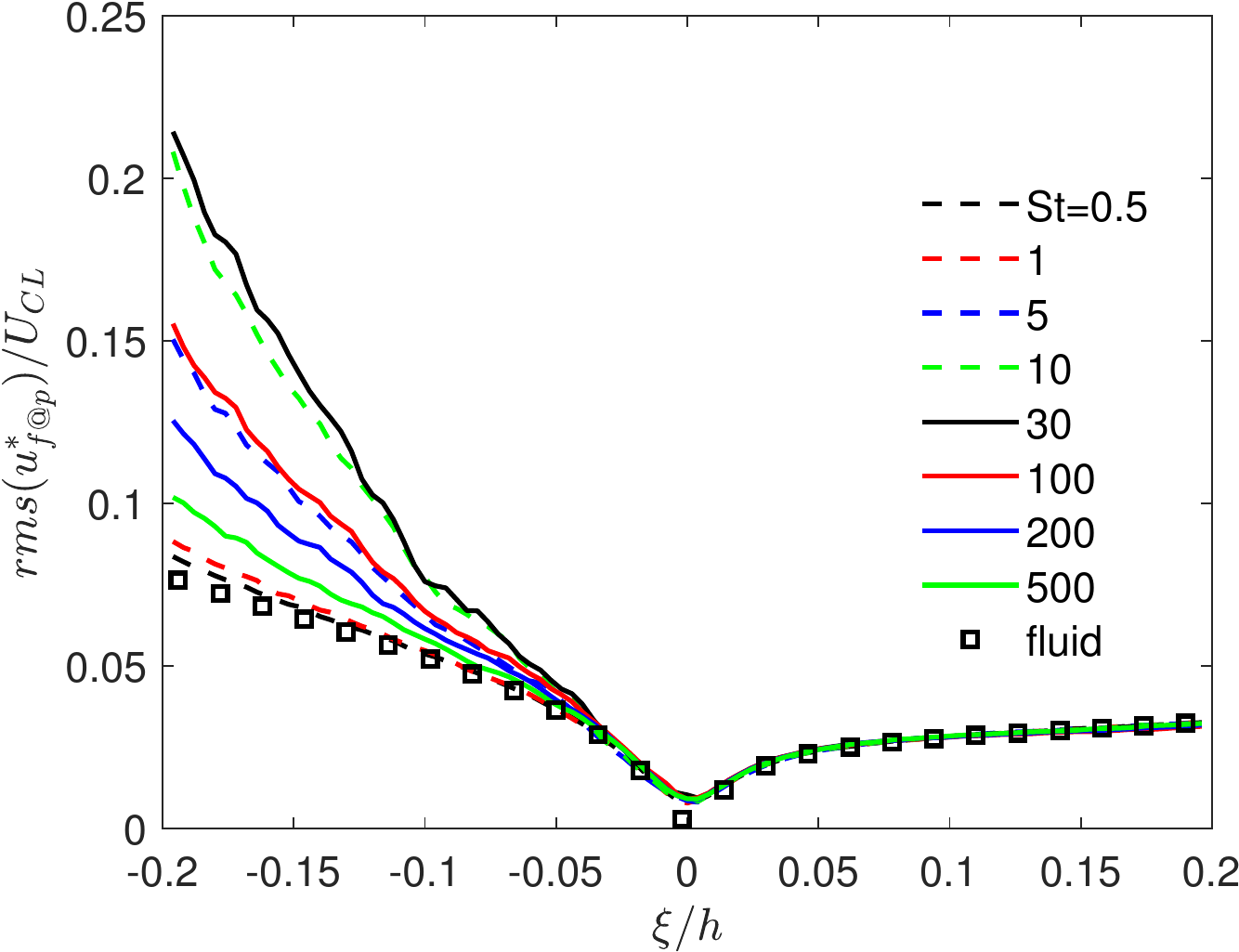}}  \hfill
	\caption{Conditional r.m.s. of streamwise velocity fluctuations of (a) particles, and (b) fluid at particle position in the QC frame.}
	\label{fig:rmsU}
\end{figure}
	
Furthermore, figure \ref{fig:rmsU} presents conditional r.m.s. of streamwise velocity fluctuations of particles and fluid at particle position around the QC-boundary. The superscript $*$ denotes fluctuations of variables in the QC frame. It is apparent that the streamwise velocity fluctuations of all particles are greater than that of fluid due to inertia. The streamwise velocity fluctuations are similar for low Stokes number particles and for the fluid. With increasing Stokes number, the discrepancy between particle and fluid streamwise velocity fluctuations appears firstly outside the QC and subsequently around the QC-boundary and within the QC. In figure \ref{fig:rmsU}(b), the streamwise fluid velocity fluctuations at particle position remain almost the same as that of the fluid. However, the fluid velocity fluctuations at particle position are greater than those of the fluid, especially for intermediate Stokes number particles, revealing a high-level clustering of particles outside the QC. As a result, particles outside the QC tend to accumulate in regions with negative streamwise fluid velocity fluctuations.

\begin{figure}
	\centering
	\sidesubfloat[]{\includegraphics[width=0.45\linewidth]{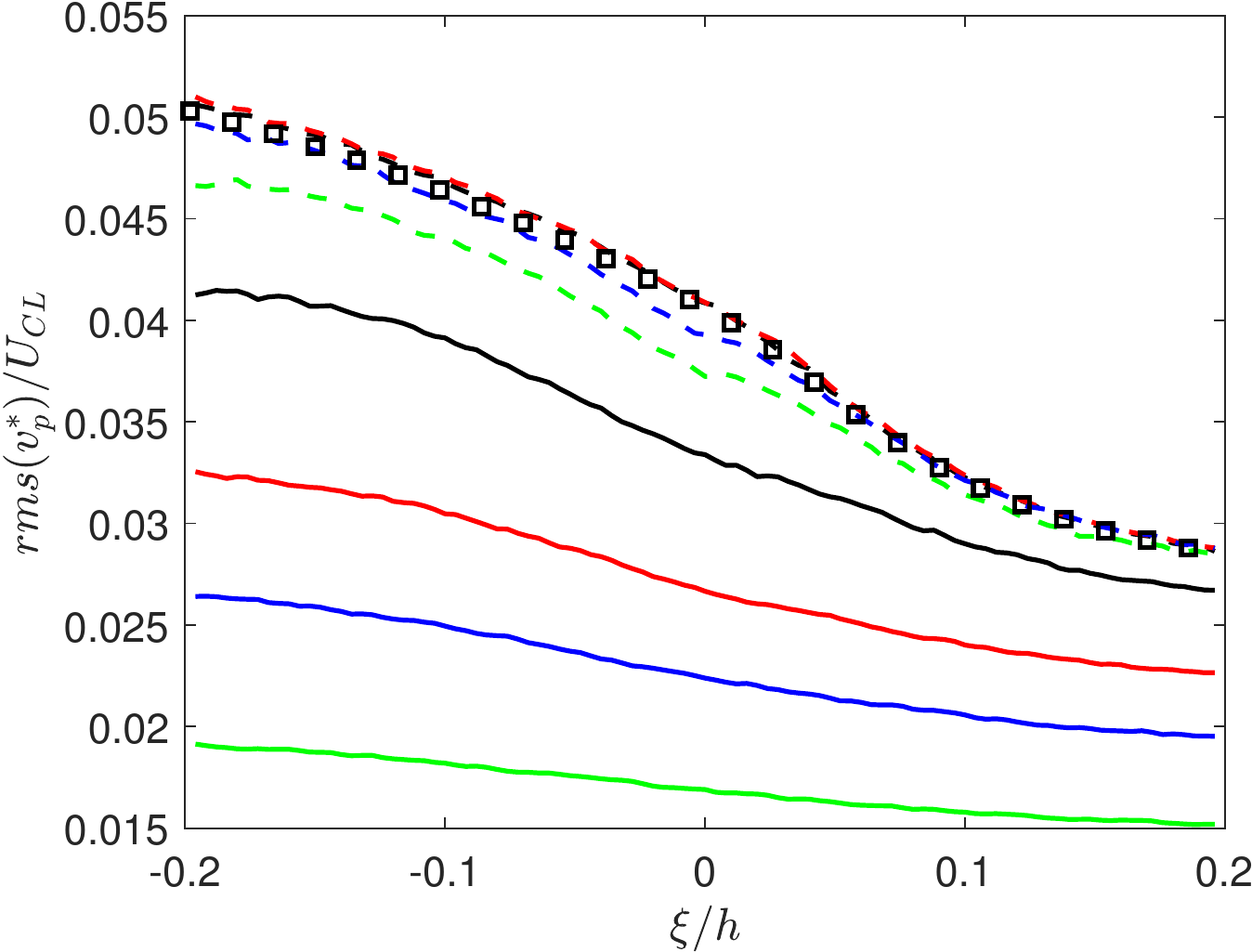}}
	\sidesubfloat[]{\includegraphics[width=0.45\linewidth]{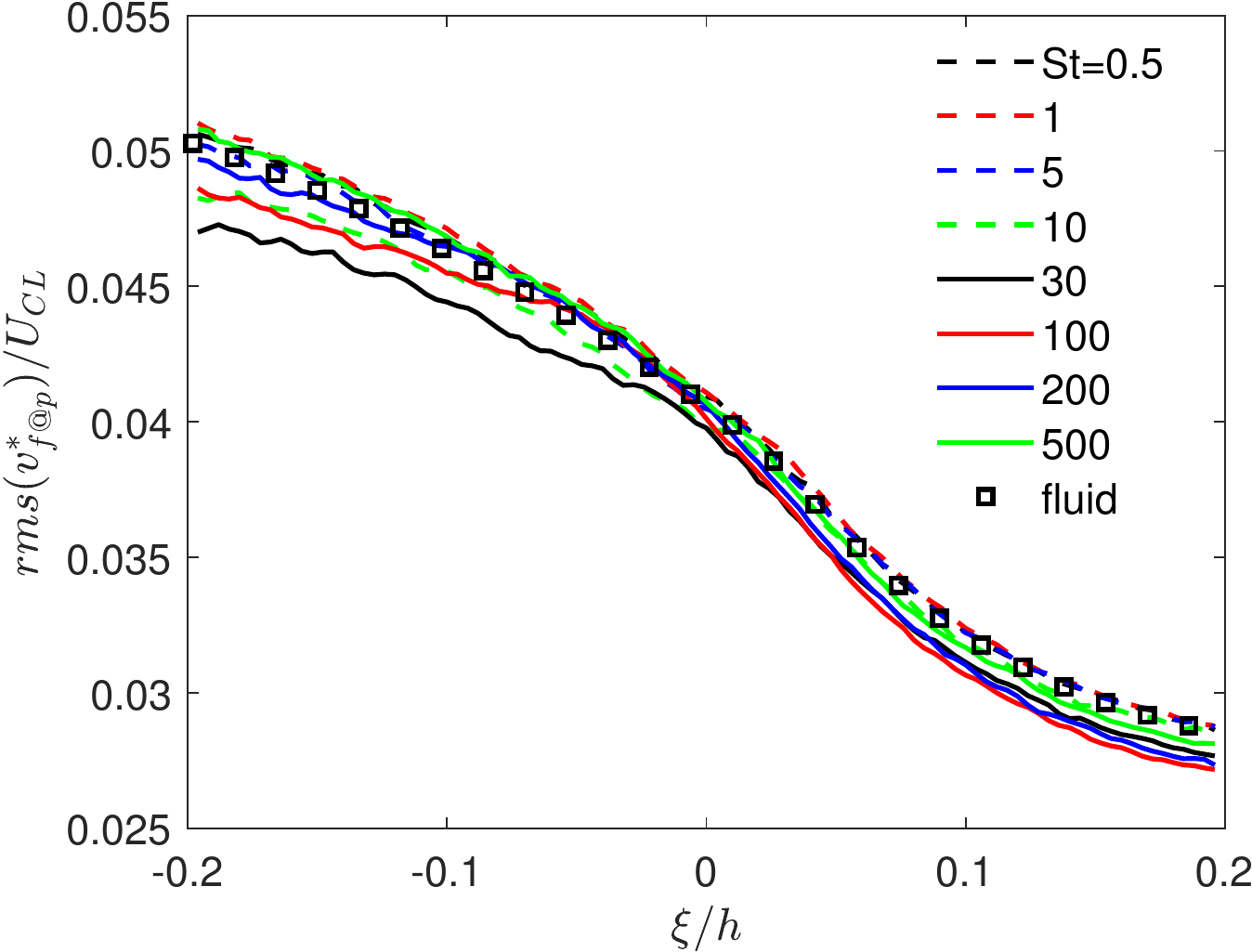}}  \hfill
	\caption{Conditional r.m.s. of spanwise velocity fluctuations of (a) particles, and (b) fluid at particle position in the QC frame.}
	\label{fig:rmsV}
\end{figure}

\begin{figure}
	\centering
	\sidesubfloat[]{\includegraphics[width=0.45\linewidth]{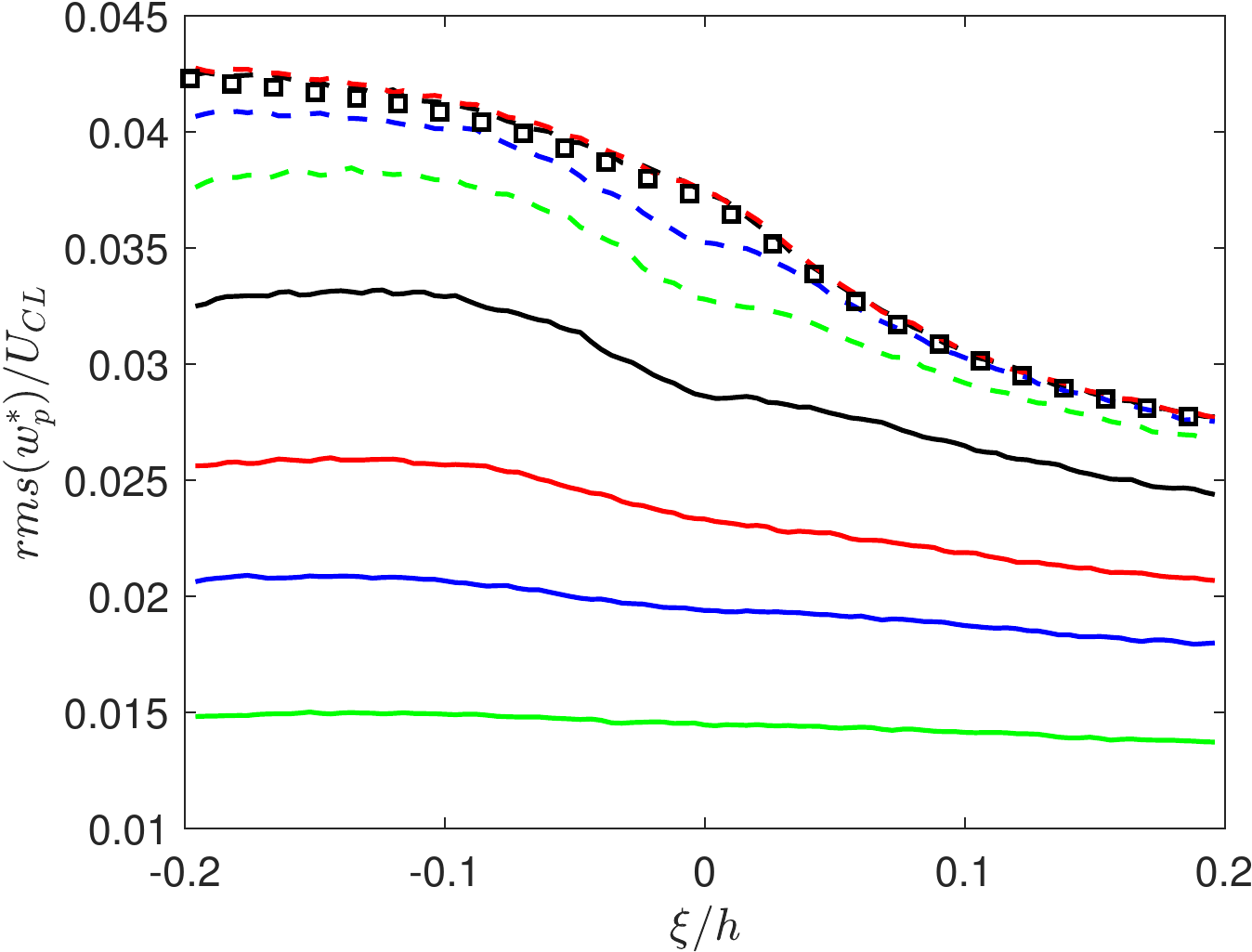}}
	\sidesubfloat[]{\includegraphics[width=0.45\linewidth]{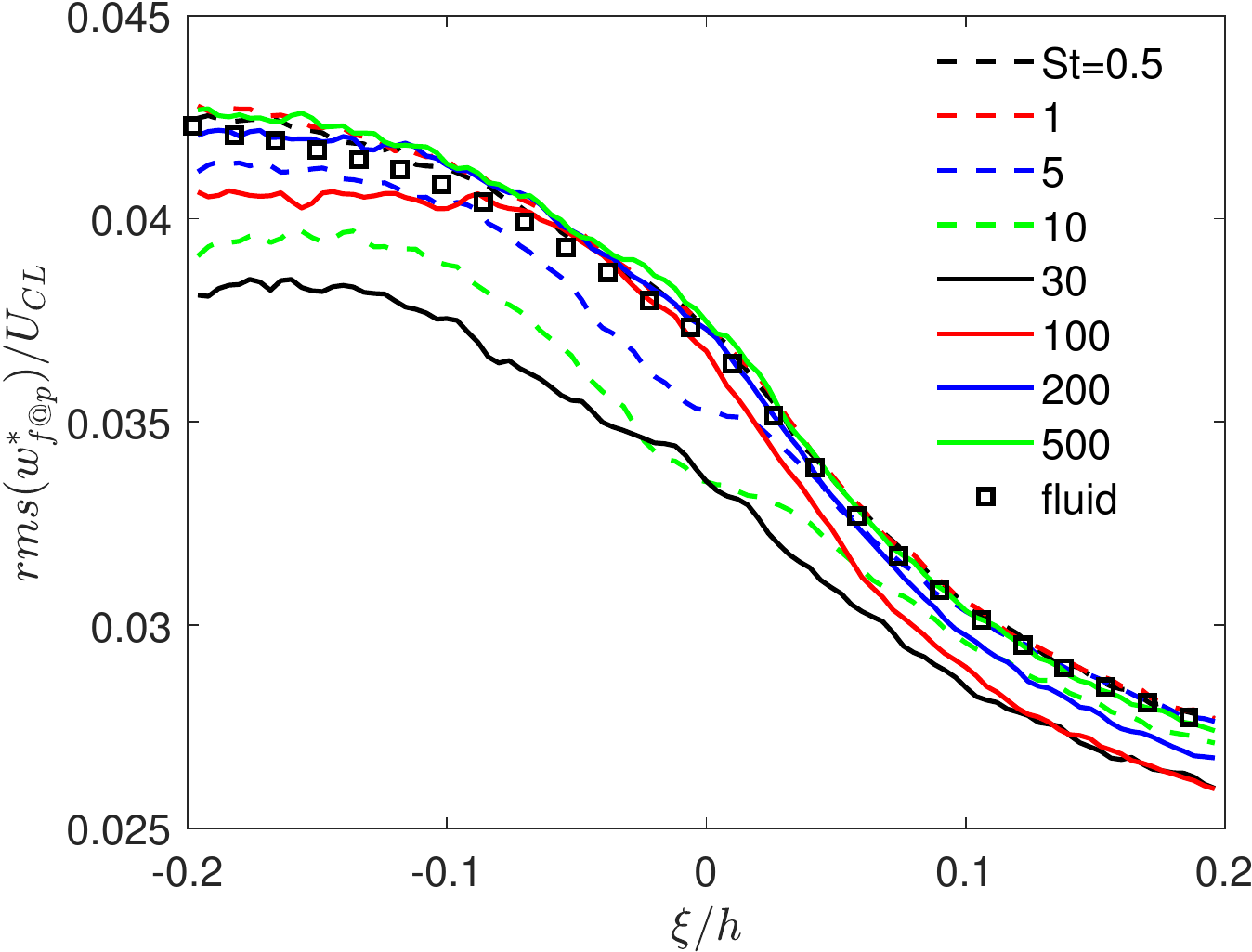}}  \hfill
	\caption{Conditional r.m.s. of wall-normal velocity fluctuations of (a) particles, and (b) fluid at particle position in the QC frame.}
	\label{fig:rmsW}
\end{figure}

Figure \ref{fig:rmsV} and \ref{fig:rmsW} display the conditional r.m.s. of the spanwise and the wall-normal velocity fluctuations, respectively, of particles and fluid at particle position around the QC-boundary. It can be seen that the spanwise and the wall-normal velocity fluctuations of all particles are smaller than those of the fluid, which is different from the fluctuations in the streamwise direction. The inertia effect attenuates fluctuations in the spanwise and the wall-normal direction monotonically with increasing Stokes number. However, there is not an abrupt jump of fluctuations across the QC-boundary. The discrepancy between the fluid and fluid at particle position velocity in figures \ref{fig:rmsV}(b) and \ref{fig:rmsW}(b) is slight, but the black line ($St=30$) suggests a higher clustering degree for intermediate Stokes number particles, since the difference between the r.m.s. of fluid velocity and the r.m.s. of fluid velocity at particle position is most prominent at $St=30$.

\begin{figure}
	\centering
	\sidesubfloat[]{\includegraphics[width=0.45\linewidth]{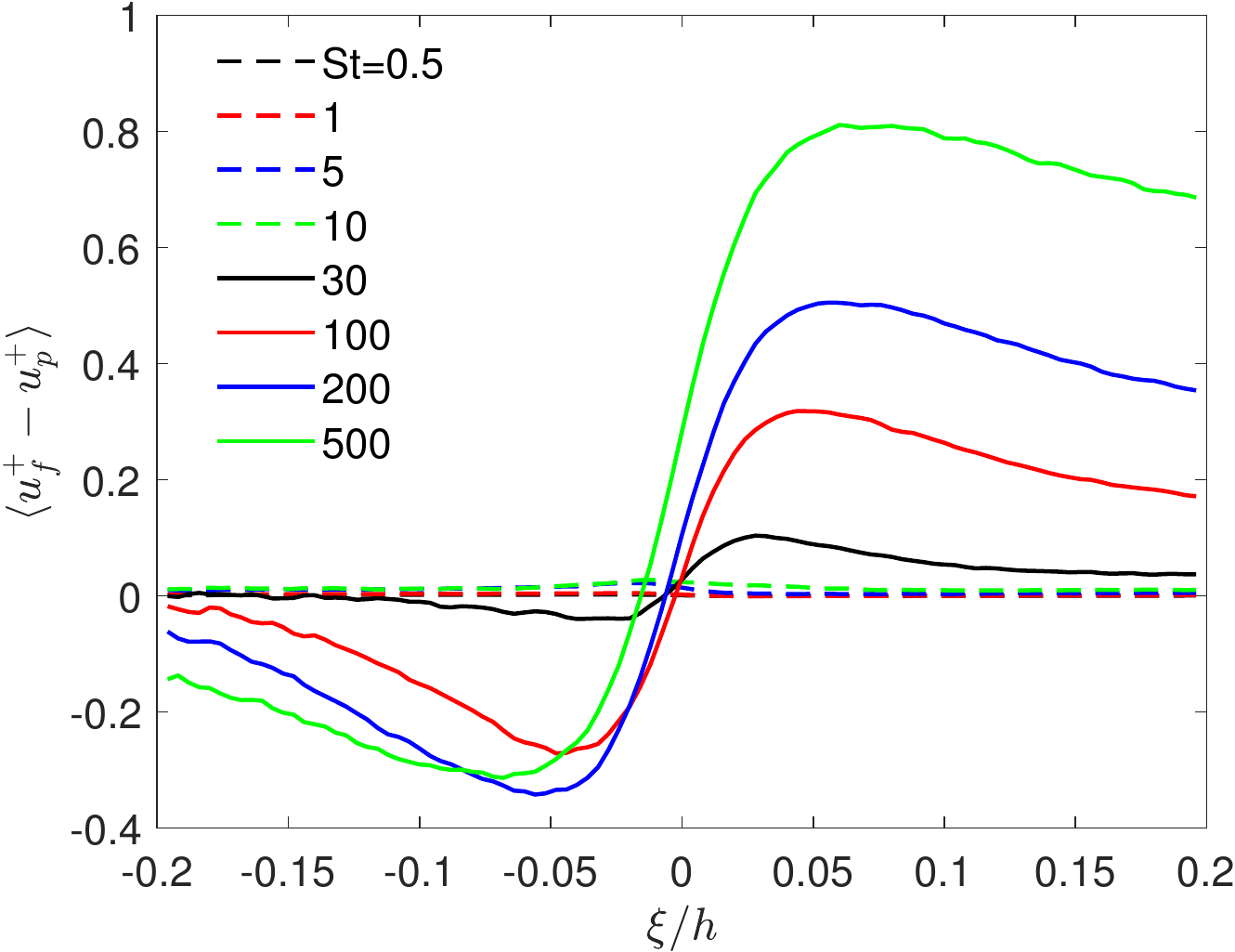}}
	\sidesubfloat[]{\includegraphics[width=0.45\linewidth]{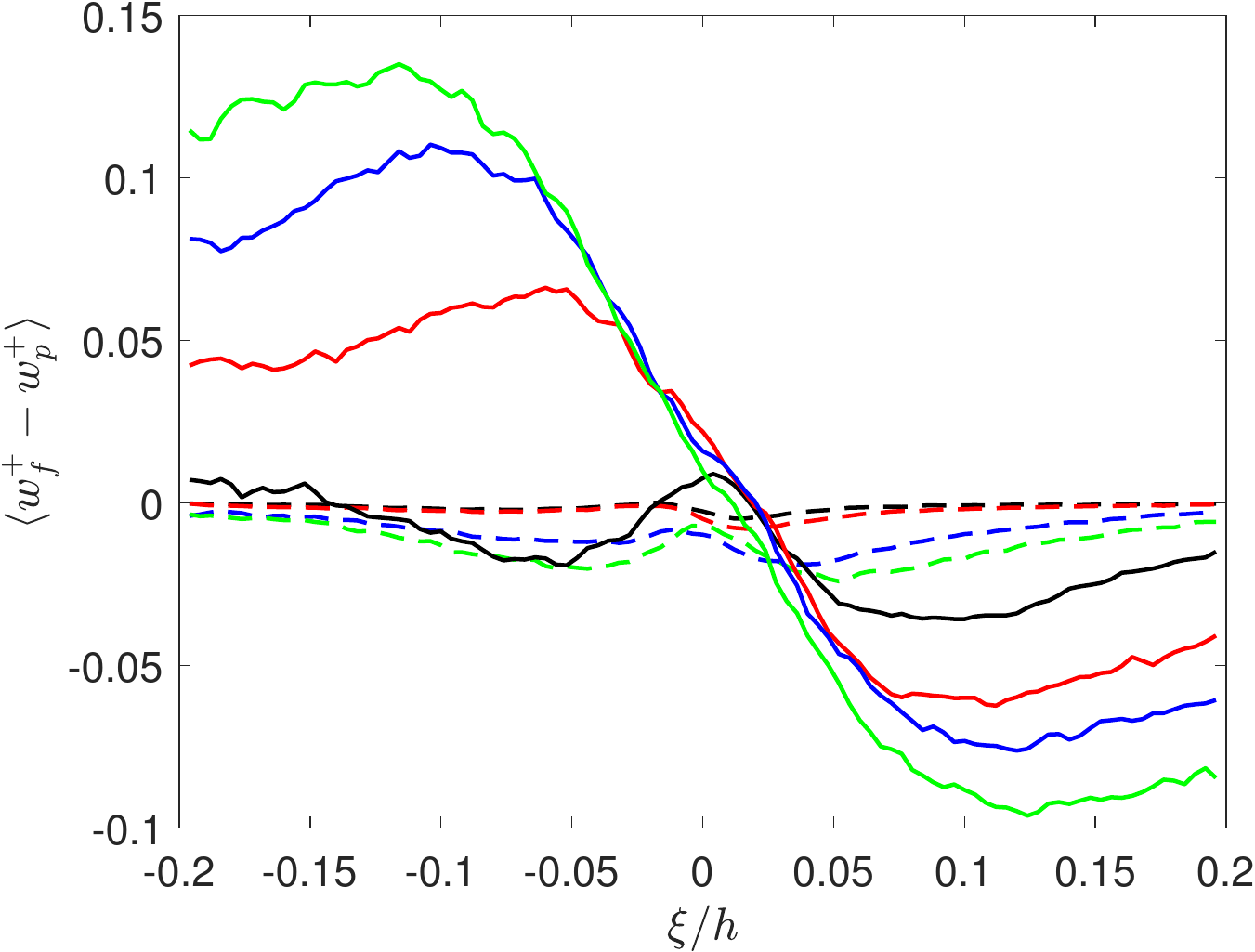}}  \hfill
	\caption{Conditional-averaged (a) streamwise, and (b) wall-normal slip velocity around the QC-boundary.}
	\label{fig:slipV}
\end{figure}

\begin{figure}
	\centering
	\sidesubfloat[]{\includegraphics[width=0.45\linewidth]{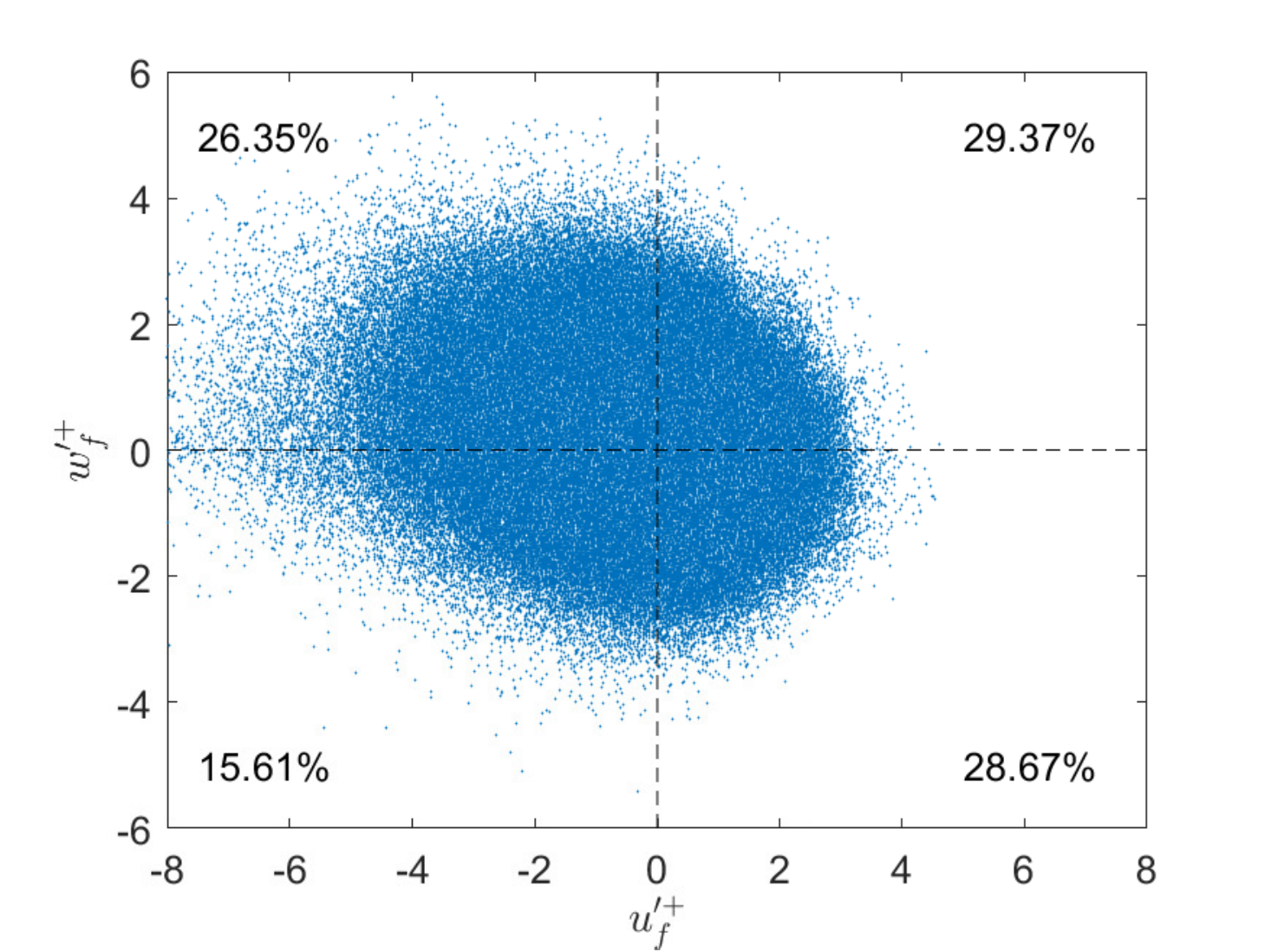}}
	\sidesubfloat[]{\includegraphics[width=0.45\linewidth]{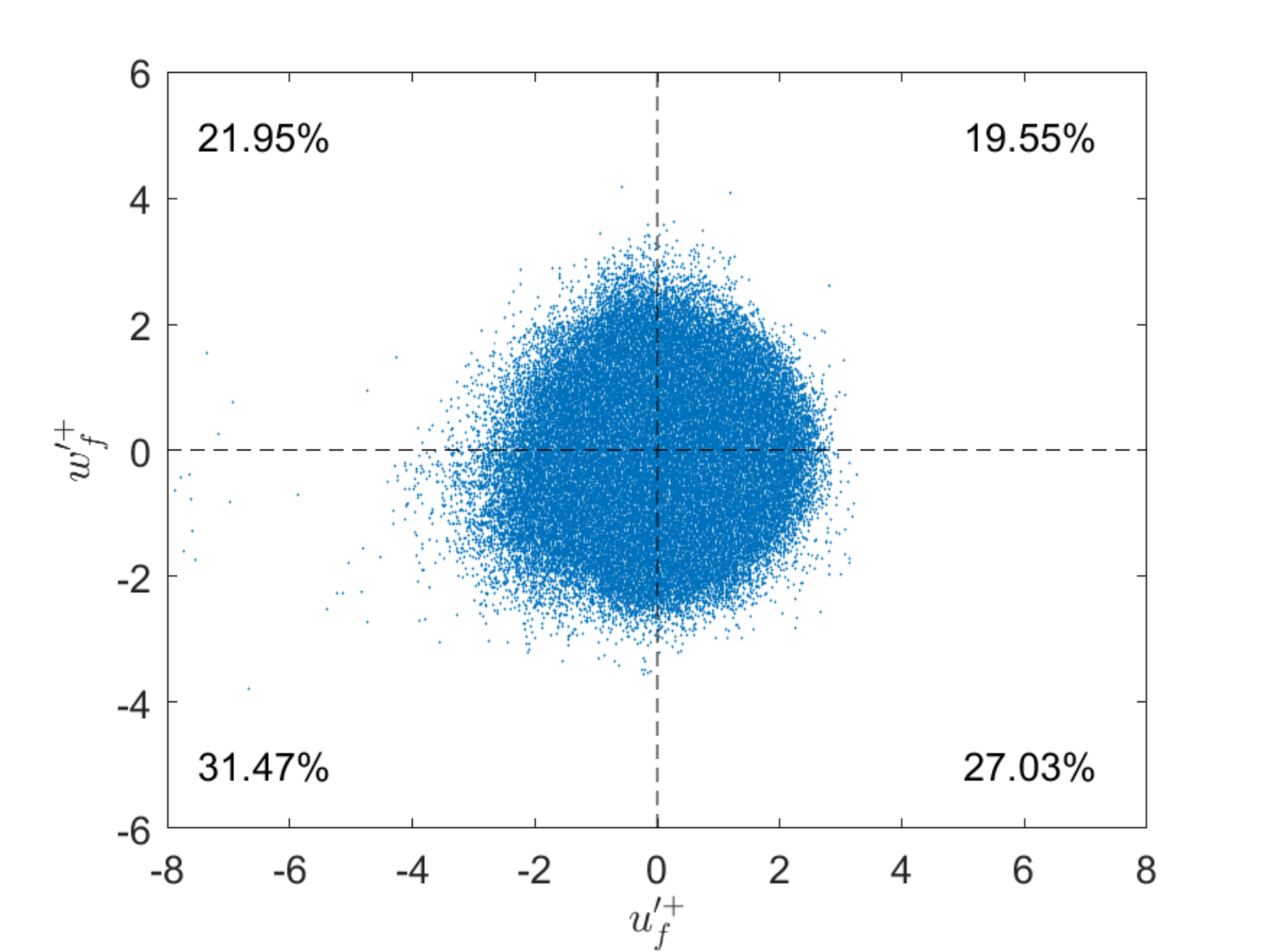}}  \hfill
	\\
	\sidesubfloat[]{\includegraphics[width=0.45\linewidth]{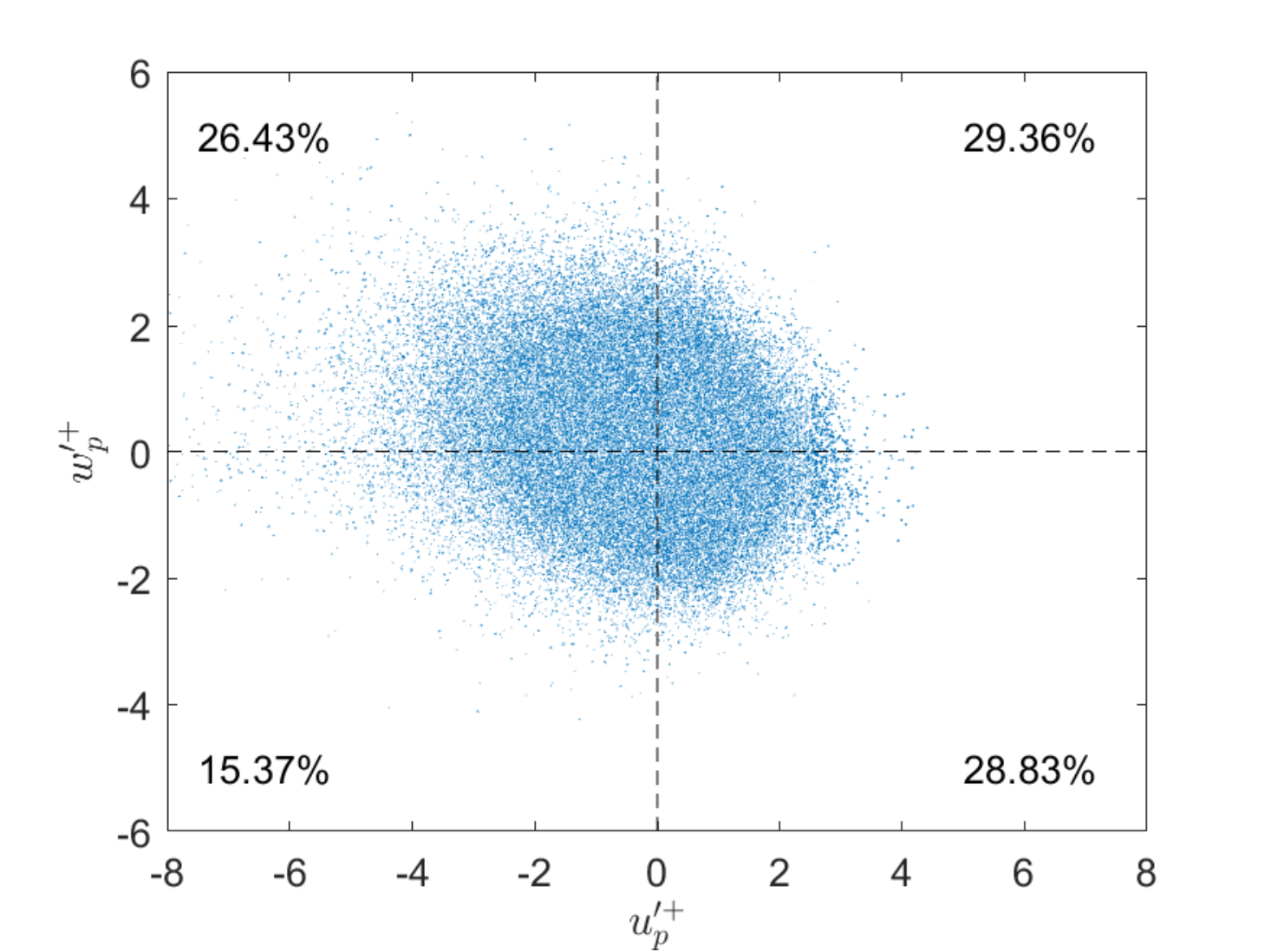}}
	\sidesubfloat[]{\includegraphics[width=0.45\linewidth]{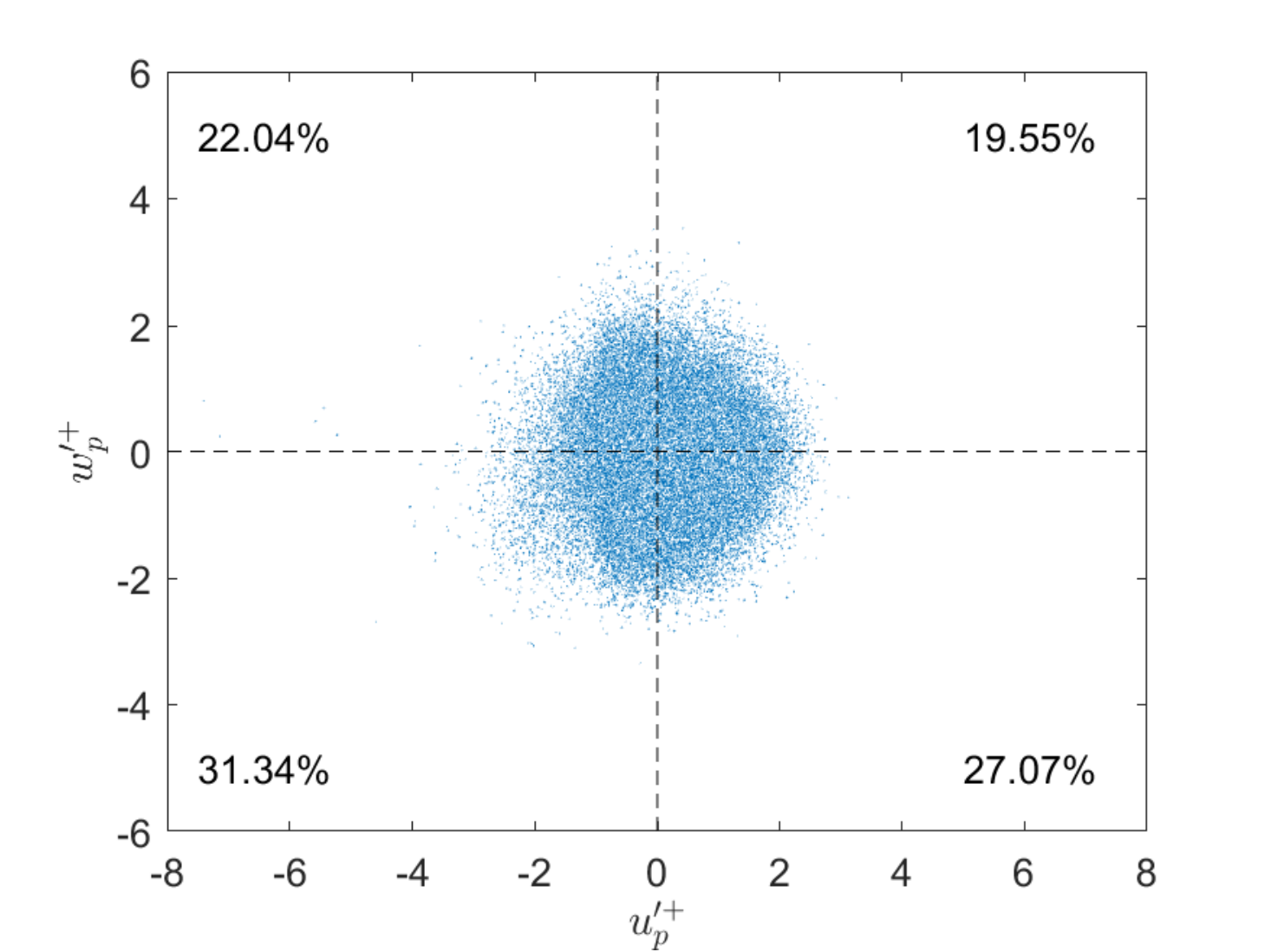}}  \hfill
	\caption{Scatter plots of fluid velocity fluctuations at (a) $\xi/h=-0.1$, and (b) $\xi/h=0.1$; and the scatter plots of particle velocity fluctuations with $St=200$ at (c) $\xi/h \in [-0.108,-0.092]$, and (d) $\xi/h \in [0.092,0.108]$.}
	\label{fig:scatter}
\end{figure}

Figure \ref{fig:slipV} shows the conditional-averaged streamwise and wall-normal slip velocity in the QC frame. It is striking that the sign of streamwise slip velocity change from negative to positive across the QC-boundary for $St \geq 30$, as shown in figure \ref{fig:slipV}(a). Statistically, this suggests that particles with large inertia are accelerated by fluid in the QC while those outside the QC are likely to be decelerated by the surrounding fluid in the streamwise direction. Moreover, the results in figure \ref{fig:slipV}(a) indicate that the QC plays an important role on the inertial particle transport in the flow direction and it tends to contribute to $\partial{u_p} / \partial{z}$. This trend is most prominent for $St=500$, whereas the effect is small for $St < 30$.

A similar situation occurs in the wall-normal direction as well. Particles with large Stokes numbers are driven by the fluid toward the QC-boundary in the wall-normal direction while the particles located in QC are pushed toward the wall due to the negative slip velocity. Here, again, the QC is found to be of importance on the particle migration in the wall-normal direction. Low inertia particles behave more like tracers so that the streamwise and wall-normal slip velocities are relatively small. We deduced that the upward and downward fluid motions are the underlying reason for these phenomena.

The scatter of streamwise and wall-normal fluid velocity fluctuations around the QC-boundary is shown in figures \ref{fig:scatter}(a) and (b). The scatter plot in figure \ref{fig:scatter}(a) is broader than that in figure \ref{fig:scatter}(b), which means that the velocity fluctuations outside the QC is more intensive than inside the QC, i.e. consistent with figures \ref{fig:rmsU}(b) and \ref{fig:rmsW}(b). The percentages in these figures show the proportion of fluid fluctuations in the different quadrants. The distinct asymmetry of the scatter plot in figure \ref{fig:scatter}(a) suggests a substantially negative correlation between $w'_f$ and $u'_f$, i.e. a significant Reynolds shear stress. The proportion of flow with positive wall-normal velocity fluctuations in figure \ref{fig:scatter}(a) is about $56\%$, i.e. greater than $50\%$, thus indicating dominance of upward motions of flow outside the QC. While the proportion of flow with negative wall-normal velocity fluctuations in figure \ref{fig:scatter}(b) is about $59\%$, reflecting downward motions of flow in the QC. As a consequence, the translation of particle shows similar upward and downward motions, as shown in both figure \ref{fig:wp}(a) and figure \ref{fig:scatter}(c,d). Interestingly, the scatter plots of fluid in the QC frame, on the other hand, show the dominance of scatters in the first quadrant ($29.37\%$) outside of QC and of the third quadrant ($31.47\%$) inside of QC, which outrules the correspondence with near-wall sweep and ejection, referred to as the fourth quadrant event and the second quadrant event, respectively \citep{kim_turbulence_1987}.

Figure \ref{fig:scatter}(c) shows the scatter of particle velocity fluctuations with $St = 200$ outside the QC while figure \ref{fig:scatter}(d) shows the same inside the QC. The spreading of the scatters in figure \ref{fig:scatter}(c) is much smaller than in figure \ref{fig:scatter}(a), revealed that the upward motion of flow is more dominant than that of the particles. This is consistent with the positive wall-normal slip velocity outside the QC in figure \ref{fig:slipV}(b). Likewise, the downward motion of the flow inside the QC is stronger than that of the particles, thereby resulting in a negative wall-normal slip velocity in the QC in figure \ref{fig:slipV}(b).

\section{Concluding remarks}
In present work we examined particle behavior influenced by the QC of a turbulent channel flow at $Re_\tau=600$ by means of one-way coupled direct numerical simulation. The QC, which plays important role in the spatial distribution and transport of particles, is only detected in medium and high Reynolds and not prominent at $Re_\tau=180$. Moreover, we believe that the observations of influence of QC on particle dynamics in the present study are also valid in turbulent pipe flows \citep{yang_influence_2019}.

The present results, first of all, show that the inertial particles tend to reside in high-speed regions in the QC, which is opposite to the preferential accumulation of inertial particles in low-speed regions in the vicinity of walls. We found that the underlying mechanism behind particle's preferential clustering in the QC is to stay in low vorticity region similar to earlier findings of inertial particles in HIT \citep{eaton_preferential_1994}. The first key message that we would like to convey is that the preferential accumulation in the high-speed region in the QC accelerates the streamwise particle transport, which becomes more significant when one considers that the QC becomes gradually dominant in realistic high-Reynolds-number flows. More importantly, we found that inertial particles tend to avoid the region of vortices around the QC-boundary, which is a shear layer with sharp gradient of velocity, due to the centrifugal mechanism. Therefore, the second key finding of the present study is that the QC-boundary can be regarded as a barrier that hinders the wall-normal transport of particles.

Furthermore, the mean velocity and velocity fluctuations of both particles and fluid at particle position are examined in the QC frame. We observed an abrupt jump of the streamwise particle velocity across the QC-boundary for particles with small and intermediate Stokes numbers. Interestingly, the particles in the QC tend to drift to the wall while those outside the QC tend to migrate towards the channel center, as shown in figure \ref{fig:wyInXZ}, which is actually due to the fluid upward and downward motions in and outside of QC that have been observed by \citet{kwon_quiescent_2014} in channel flow and \citet{yang_influence_2019} in pipe flow. Conditional-averaged slip velocities in the streamwise and wall-normal direction show that there is a sudden change of direction of Stokes drag near the QC-boundary in both directions. Particles in QC tend to be accelerated by the fluid in the streamwise direction while those outside of QC are more likely to be decelerated by fluid. Therefore, QC boundary at $\xi=0$ represents a transition region between particle acceleration and deceleration.

Considering that the QC region occupies around $50\%$ of the channel in more realistic high Reynolds number wall turbulence and that the thickness of the core increases with increasing Reynolds number \citep{kwon_quiescent_2014}, it is, therefore, of importance to advance our understanding of the role played by the QC in particle dynamics. The present work shows the influence on particle distribution and transport due to the presence of the QC in a turbulent channel flow. However, there are more remaining open questions, such as how would the QC be modulated by considering the feedback of the inertial particles? What is the Reynolds number effect on the particle behaviour in the QC region? These interesting questions are awaiting for exploration in future work.

\section*{Acknowledgments}
The work was supported by the Natural Science Foundation of China (grant Nos. 11702158 and 91752205), Tsinghua University Initiative Scientific Research Program (2019Z08QCX10) and the China Scholarship Council and the Research Council of Norway (grant No 250744). The authors acknowledge the computational resources through grant No. NN2694K and No. NN9191K (Program for Supercomputing). The second author appreciates the great hospitality offered by Applied Mechanics Lab during his research stay at Tsinghua University.


\bibliographystyle{apalike}

\bibliography{Sphere_QC_Ref}


\end{document}